\documentclass[]{aa}

\usepackage{graphicx}
\usepackage{txfonts}
\usepackage{color}
\usepackage[]{hyperref}
\begin{document}

   \title{J-VAR: the northern variable sky in 7 filters}

   \subtitle{First Data Release}

\titlerunning{J-VAR DR1}
   \authorrunning{Ederoclite et al.}

   \author{A. Ederoclite
          \inst{1},
          H. V\'azquez Rami\'o \inst{1},
          A. Alvarez-Candal \inst{2},
          B. B. Siffert \inst{3},
          V. M. Placco\inst{4} ,
          D. Morate \inst{1},
          S. Pyrzas \inst{1},
          C. López-Sanjuan \inst{1},
          M. Mahlke \inst{5},
          S. Kulkarni \inst{1}
          L. Espinosa \inst{6},
          M. J. Castro \inst{7,8},
          B. Zacarias \inst{6}
          M. Akhlaghi \inst{1},
          J. Castillo \inst{1}, 
          T. Civera \inst{1},
          J. Hernández-Fuertes \inst{1}, 
          A. Hernán-Caballero \inst{1},
          A. López-Sainz, \inst{1},
          G. Lorenzetti \inst{1},
          D. Muniesa-Gallardo \inst{1},
          A. Moreno-Signes \inst{1},
          H. Vives-Arias \inst{1},
          J. Zaragoza Cardiel \inst{1},
          M. C. Díaz-Martín \inst{1}, 
          F. Galindo-Guil \inst{1},
          R. Iglesias-Marzoa \inst{1},
          R. Infante-Sainz \inst{1},
          T. Kuutma \inst{1},
          E. Lacruz \inst{1,9},
          J. Lamadrid-Gutierrez \inst{1}, 
          F. López-Mart\'{\i}nez \inst{1},
          N. Ma\'{\i}cas-Sacristan \inst{1},
          F. Hernández Pérez \inst{9},
          J. Carvano \inst{10},
          P. Cruz \inst{11},
          F. R. Herpich \inst{12},
          E. Solano \inst{11},
          A. F. Pala \inst{13},
          R. R. R. Reis \inst{3}
          }

   \institute{
        Centro de Estudios de F\'isica del Cosmos de Arag\'on\\
              \email{aederocl@cefca.es}
        \and
            Instituto de Astrof\'isica de Andaluc\'ia, CSIC, Apt 3004, E18080 Granada, Spain
        \and
            Universidade Federal de Rio de Janeiro
        \and
            NSF NOIRLab, Tucson, AZ 85719, USA
        \and
            Université Marie et Louis Pasteur, CNRS, Institut UTINAM (UMR 6213), équipe Astro, F-25000 Besançon, France
        \and
            Department of Astronomy, Instituto de Astronomia, Geofisica e Ciencias Atosfericas, Universidade de Sao Paulo
        \and
            Fakultät für Physik und Astronomie, Universität Heidelberg, Im Neuenheimer Feld 226, D-69120 Heidelberg, Germany
        \and
            Landessternwarte, Zentrum für Astronomie der Universität Heidelberg, Königstuhl 12, D-69117 Heidelberg, Germany
        \and
            Universidad Internacional de Valencia (VIU), C/Pintor Sorolla 21, 46002 Valencia, Spain
        \and
            Observatorio Nacional
        \and
            Centro de Astrobiología (CAB), CSIC-INTA, Camino Bajo del Castillo s/n, E-28692, Villanueva de la Cañada, Madrid, Spain.
        \and 
            Laboratório Nacional de Astrofísica (LNA/MCTI), Rua Estados Unidos, 154, Itajubá 37504-364, Brazil
        \and
            European Southern Observatory
             }

   \date{Received \ldots; accepted \ldots}

  \abstract
  {}
   {The analysis of variability of astronomical sources is of extraordinary
   interest, as it
   allows the study of astrophysical phenomena in real time. This paper presents 
   the Javalambre Variability Survey (J-VAR) which leverages the 
   narrow band filters available at the Javalambre Auxiliary Survey 
   Telescope (JAST80) at the Observatorio Astrofísico de Javalambre (OAJ).}
   {The JAST80 equipped with T80Cam,
   providing a field of view of 2\,square degrees and a pixel scale of 0.55\,arcsec/pixel
   has been designed for wide-field studies. The main characteristic is the availability of a variety of narrow band filters strategically located
   on stellar spectral features (the $J0395$ in correspondence of the
   Ca H+K doublet, the $J0515$ of the Mg $b$ triplet, the $J0660$ of the H$\alpha$ line,
   and the $J0861$ of the Ca~triplet).
   This project combines, for the first time, the wide-field with a variety
   of narrow band filters for a unique variability survey, observing each field 
   11~times with a standardised observing sequence.
   The median limiting magnitude for individual exposures are 19.1 mag in $J0395$ and $J0515$, $19.6$ mag in $J0660$ and $J0861$, $19.8$ mag in $i$, and $20.2$ in $g$ and $r$. The typical FWHM of the $r$-band images is $1.5$ arcsec.
   }
   {This article introduces the first data release of J-VAR
   including 
   more than 6000 individual asteroids, 10\,detected optical transients (4\,discovered 
   supernovae), and 1.3 million light curves of point-sources. 
   On average, J-VAR delivers
   an unprecedented $\sim$5000 light curves per square degree of 11\,epochs in 7\,bands, opening
   research opportunities for theoretical studies and new discoveries alike.
   }
   {}


   \maketitle

\section{Introduction}
\label{sec:introduction}

Variability is a fundamental property of the Universe, offering key insights across all spatial and temporal scales. From the pioneering use of Cepheid pulsations to establish the extragalactic distance scale (e.g., \citealt{leavitt, hubble}), to the discovery of exoplanets through periodic variations in stellar brightness (e.g., \citealt{51Pegb}), the study of time-dependent phenomena has continually transformed the field of astrophysics.

In recent years, the field of time-domain astrophysics has entered a golden era, driven by the start of several high-cadence (one day) photometric surveys. Ground-based projects such as All-Sky Automated Survey for Supernovae \citep[ASAS-SN;][]{ASASSN_ref}, Catalina Real-time Transient Survey \citep[CRTS;][]{CRTS_ref}, and Zwicky Transient Facility \citep[ZTF;][]{ZTF_ref}, operating mostly in up to three broadband filters, have mapped vast swathes of the sky, revealing countless new transients and variable sources. In parallel, space missions like the COnvection ROtation et Transits satellite \citep[CoRoT;][]{corot_ref}, {\it Kepler} \citep{kepler_ref}, the Transiting Exoplanet Survey Satellite \citep[TESS;][]{tess_ref} , and \textit{Gaia} \citep{GaiaDR3} have provided light curves of exquisite precision for millions of sources, albeit typically in a single passband or pseudo-filter. Currently, the Vera C. Rubin Observatory
is starting the Large Survey of Space and Time which is expected to deliver millions
of alerts of transients per night covering the full sky every few nights with a
dedicated 8\,m telescope with 6\,filters \citep{lsst}.

Photometric monitoring in a single filter enables the detection and characterization of variability — amplitudes, timescales, and periodicities — but limits our understanding of the physical mechanisms at play. Spectroscopic time-domain follow-up adds the much-needed context of temperature, velocity, and chemical composition changes, but is observationally expensive and limited to small samples.
An intermediate approach, increasingly explored in recent years, is time-domain photometry in multiple filters. This strategy combines the temporal cadence of imaging surveys with the spectral diagnostic power of multi-band photometry, allowing variability to be tracked not just in brightness, but across the spectral energy distribution (SED). Large-scale projects such as the Javalambre Photometric Local Universe Survey  \citep[J-PLUS;][]{jplus}  and the Javalambre-Physics of the Accelerating Universe Astrophysical Survey \citep[J-PAS;][]{jpas,Bonoli2021} have demonstrated the power of this approach for static sky studies, using a combination of broad and narrow-band filters to recover low-resolution SEDs for millions of sources.

The Javalambre Variability Survey (J-VAR) builds on this legacy by bringing the 
multi-filter strategy into the time domain. 
Carried out with the JAST80, J-VAR employs a unique seven-filter system — three broad-band ($g$, $r$, $i$) and four narrow-band ($J0395$, $J0515$, $J0660$, and $J0861$) — to repeatedly observe each field 11 times. Designed to operate during conditions not suitable for the J-PLUS survey  (e.g. poor seeing, bright sky background), J-VAR is unveiling the temporal evolution of stars, compact binaries, Solar System objects and optical transients with both cadence and spectral coverage.

The goal of this paper is to introduce J-VAR, its motivation, and to demonstrate its scientific potential. This paper is outlined as follows: Sect.~\ref{sec:observations} describes in detail the technical aspects and observing strategy of J-VAR, followed by an overview of the data reduction in Sect.~\ref{sec:datareduction}. The First Data Release (DR1) of J-VAR is described in Sect.~\ref{sec:DR1} and its main scientific drivers (including early results) in Sect.~\ref{sec:science}. Our conclusions and perspectives are outlined in Sect.~\ref{sec:conclusions}.

A series of companion papers will offer further insights into different aspects of the survey and the data release, including the construction of calibrated light curves and variability indices (Pyrzas et al., submitted), the catalogue of minor bodies (Morate et al., submitted), and the light-curve fitting of RR Lyrae stars (Kulkarni et al., submitted).

Magnitudes are in the AB system \citep{abmag_definition}, unless noted otherwise.

\section{Observations}
\label{sec:observations}

J-VAR is a multi-band time-domain survey of the northern sky designed to study variability in sky regions previously observed by the J-PLUS survey. By targeting the same fields using identical instrumentation, J-VAR leverages J-PLUS's robust photometric calibration \citep{Lopez-Sanjuan2024} and other survey products that have matured over time, including source classifications \citep{Lopez-Sanjuan2019,delPino2024} and photometric redshifts for extragalactic objects. The current J-VAR photometric calibration (Pyrzas et al. submitted; Morate et al. submitted) directly depends on the one of the J-PLUS third data relase \citep[J-PLUS DR3;][]{Lopez-Sanjuan2024}. For this reason the J-VAR pointings follow the J-PLUS field definitions.

\begin{figure}
   \includegraphics[width=\linewidth]{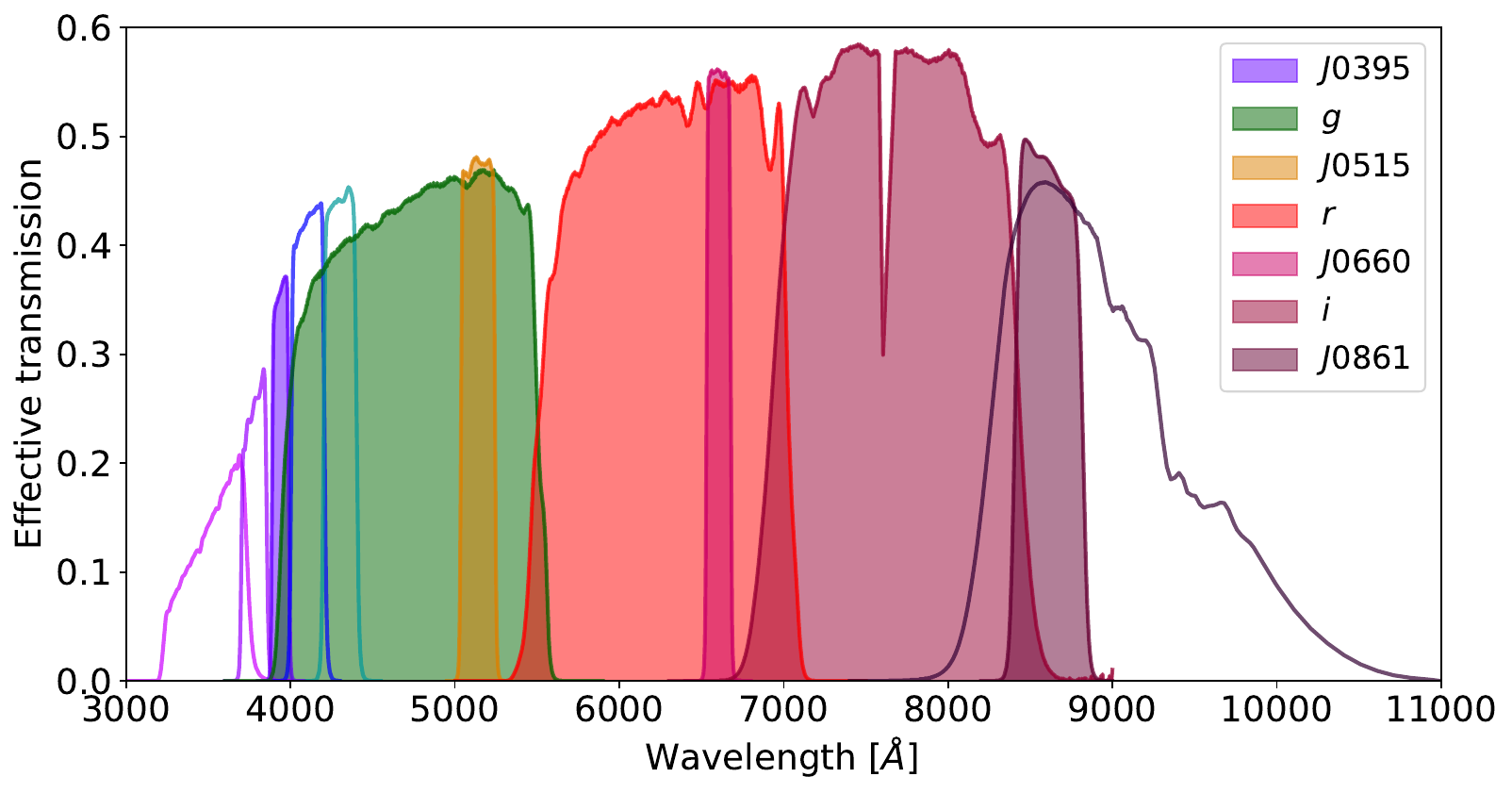}
      \caption{The J-VAR filter set (shown as filled curves) is a subset of the J-PLUS photometric system (empty and filled curves). The effective transmission accounts for the CCD quantum efficiency, the filters measured transmission and the reflectively of the primary and secondary mirror of JAST80.}
         \label{Fig:filters}
   \end{figure}

The main driver behind the choice of the filters of J-VAR is efficiency. 
Six filters ($g$, $r$, $i$, $J0515$, $J0660$, and $J0861$) have been
chosen for being the most efficient filters 
(i.e. highest effective transmission of the
whole system)
of the J-PLUS filter-set, as it can 
be seen in Fig.~\ref{Fig:filters}. The $J0395$ has also been 
included to study the variation of the Ca~H+K doublet, which 
is specially sensible to e.g. stellar chromospheric activity. 

Observations are carried out at the OAJ 
\citep{oaj}
with the T80Cam panoramic camera attached to the Javalambre Auxiliary Survey Telescope (JAST80) \citep{Marin-Franch2012,Marin-Franch2015}. JAST80 is an 83 cm, f/4.5 modified Ritchey-Chr\'{e}tien reflector on a German equatorial mount. The telescope and camera provide a pixel scale of $0.556\,\textrm{arcsec}/\textrm{pix}$ and a $1.4\,\textrm{deg}\times1.4\,\textrm{deg}$ ($2\,\textrm{deg}^2$) filed-of-view (FoV)
 that, on average, allows the detection of a number of sources of the order of $10^4$ per individual image. J-VAR uses a subset of the J-PLUS photometric system (see \mbox{Fig.~\ref{Fig:filters}}). The latter was devised to measure some of the most prominent atmospheric stellar features with the aim of studying a wide variety of phenomena in the local Universe (for more details see \citealt{jplus}).

\begin{figure}
    \centering
    \includegraphics[width=\linewidth]{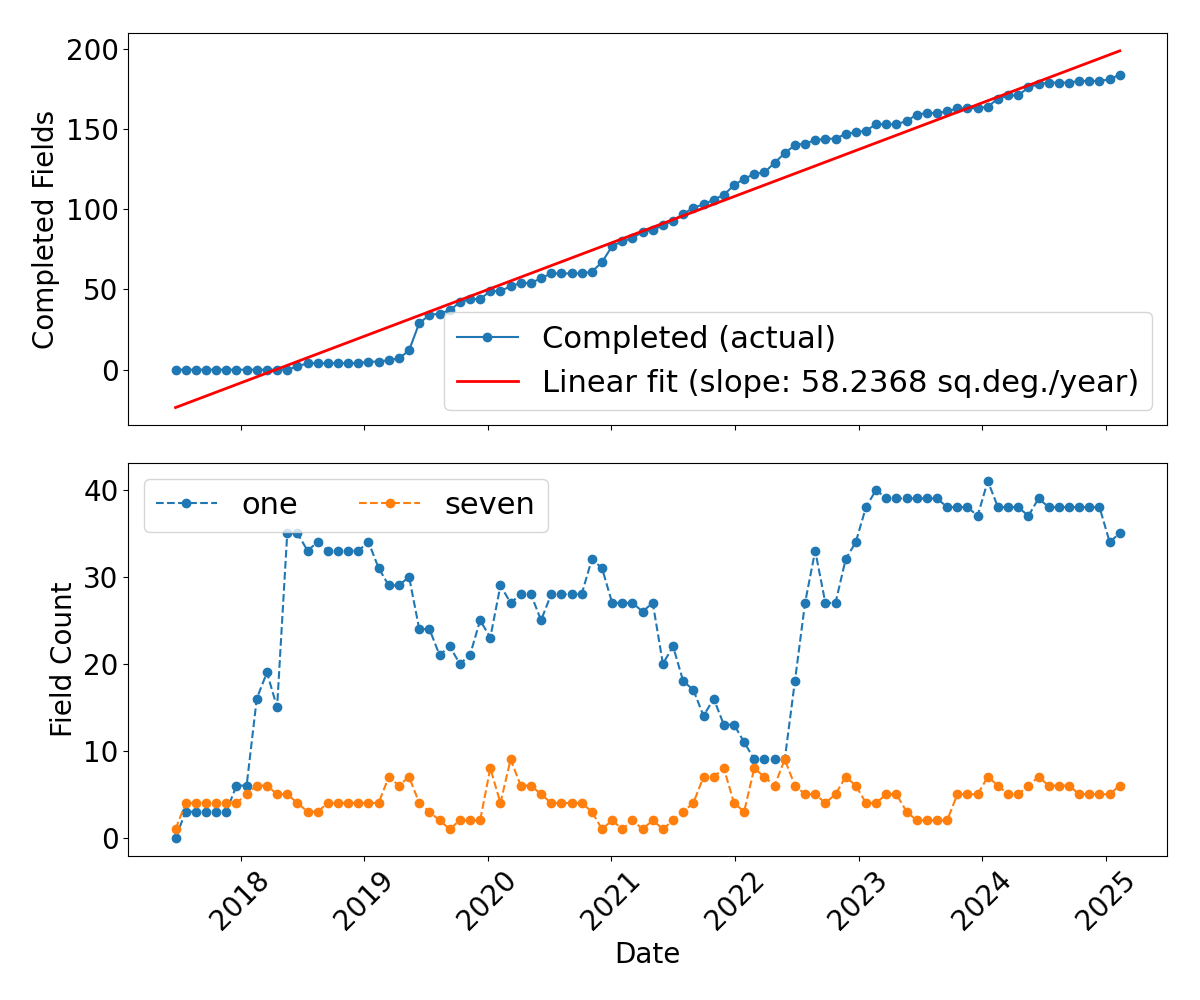}
    \caption{Evolution of the observations of J-VAR. The upper panel shows the number of
    completed fields (i.e. fields with at least 11 observed epochs). The bottom panel shows
    the number of fields which have been observed one or seven times (the rest has been omitted
    for clarity). The number of fields with one observed epoch shows that there is a 
    continuous incoming stream of new fields to 
    be observed and the number of fields with seven epochs shows that the strategy prevents an accumulation of fields not observed.}
    \label{fig:evolution}
\end{figure}

J-VAR's observations started in May 2017, shortly after the beginning of the scientific operation of the JAST80/T80Cam, as a filler program running on non-photometric nights, whenever J-PLUS or other competitive open time programs demanding photometric conditions could not be executed. From 2019 to the end of 2022, J-VAR has been granted with competitive open time offered every semester by the OAJ as a large program. Since 2023 the survey continued, again as a filler program, as it did in the beginning. The evolution of the
survey is shown in Fig.~\ref{fig:evolution}.

\begin{table}[]
\begin{center}
\caption{
Filter names, limiting magnitudes used to establish the exposure times and the corresponding exposures times, calculated using the exposure time calculator for bright nights. The ''1exp'' and ''3exp'' refer to the
magnitude limits for an individual image and three combined images respectively.
Mean and standard deviation estimated from the individual catalogues. J-VAR values are compared with those of J-PLUS DR1 \citep{jplus} from coadded images. All values referred to $SNR>5$ within a circular aperture of 3\,arcsec.
}
\label{tab:depths}
\begin{tabular}{l|ccc|cc|c}
\hline
Filter & $m_{\textrm{lim}}^{\textrm{1exp}}$ & $m_{\textrm{lim}}^{\textrm{3exp}}$ & $t_{\textrm{exp}}$ & $\langle m_{\textrm{lim}}^{\textrm{J-VAR}}\rangle$ & $\sigma_{ m_{\textrm{lim}}}^{\textrm{}}$ &  $m_{\textrm{lim}}^{\textrm{J-PLUS}}$ \\
\hline\hline
$J0395$  & 20.2 & 20.5 & 87  & 19.1  & 0.5 & 20.5 \\
$g$  & 21.2 & 21.5 & 33  & 20.3 & 1.6 & 21.5 \\
$J0515$  & 20.2 & 20.5 & 40  & 19.1 & 0.6  & 20.7 \\
$r$  & 21.2 & 21.5 & 40  & 20.2 & 0.7 & 21.5 \\
$J0660$  & 20.8 & 21.0 & 135 &19.6 & 0.7  & 20.7 \\
$i$  & 21.3 & 21.5 & 34  & 19.8 & 0.7  & 21.2 \\
$J0861$  & 20.9 & 21.0 & 160 & 19.6 & 0.6  & 20.0 \\
\hline
\end{tabular}
\end{center}
\end{table}

\subsection{Observing strategy\label{sec:strategy}}

The filler program nature of J-VAR necessitates consideration of potential cloud cover during observations. J-VAR targets a limiting magnitude $\sim$1~mag brighter than J-PLUS, conducted with identical instrumentation including telescope, camera and filters.
Non-photometric conditions cannot be modelled reliably, so bright time is used as the closest operational proxy in the exposure time calculator\footnote{\url{https://www.cefca.es/jplusetc/}}
and assume that the system will deliver observations 0.5-1\,magnitudes shallower than
the input value. The input limiting magnitudes and the
exposure times for each band are shown in Table~\ref{tab:depths}, together with actual average measures per filter from a sample of completed fields. As it can be seen, the individual exposures of J-VAR are between $0.4$ ($J0861$) and $1.4$ ($J0395$) magnitudes shallower than the stacked images of J-PLUS (in most of the cases, three images are combined). So it can be considered that the goal of the individual images of J-VAR being as deep as J-PLUS is overall fulfilled by the use of increased exposure times. It is worth noting that, in the case of J-PLUS, the observing exposure times are modulated according to the Moon phase and distance \citep{jplus}. This is possible because J-PLUS is carried out under photometric conditions. That approach was discarded for J-VAR both for simplicity, and because the survey is by construction carried out with varying atmospheric conditions that are difficult to track and model.

\begin{figure}
    \centering
    \begin{tabular}{c}
    \includegraphics[width=0.9\linewidth]{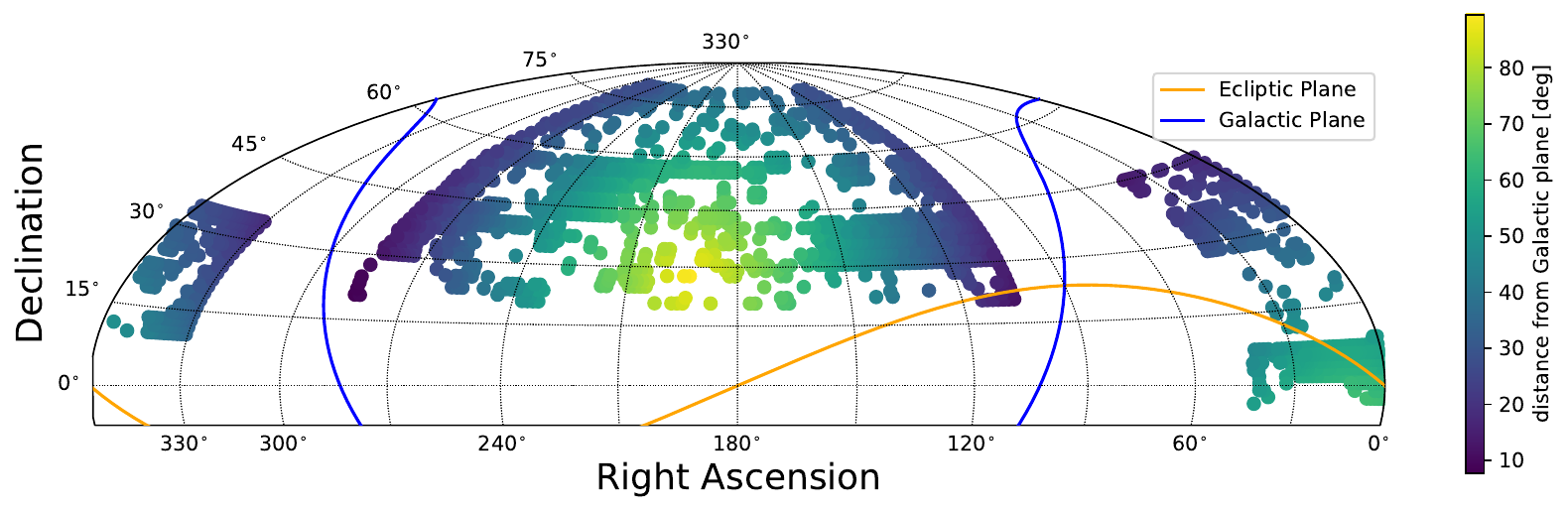} \\
    \includegraphics[width=0.9\linewidth]{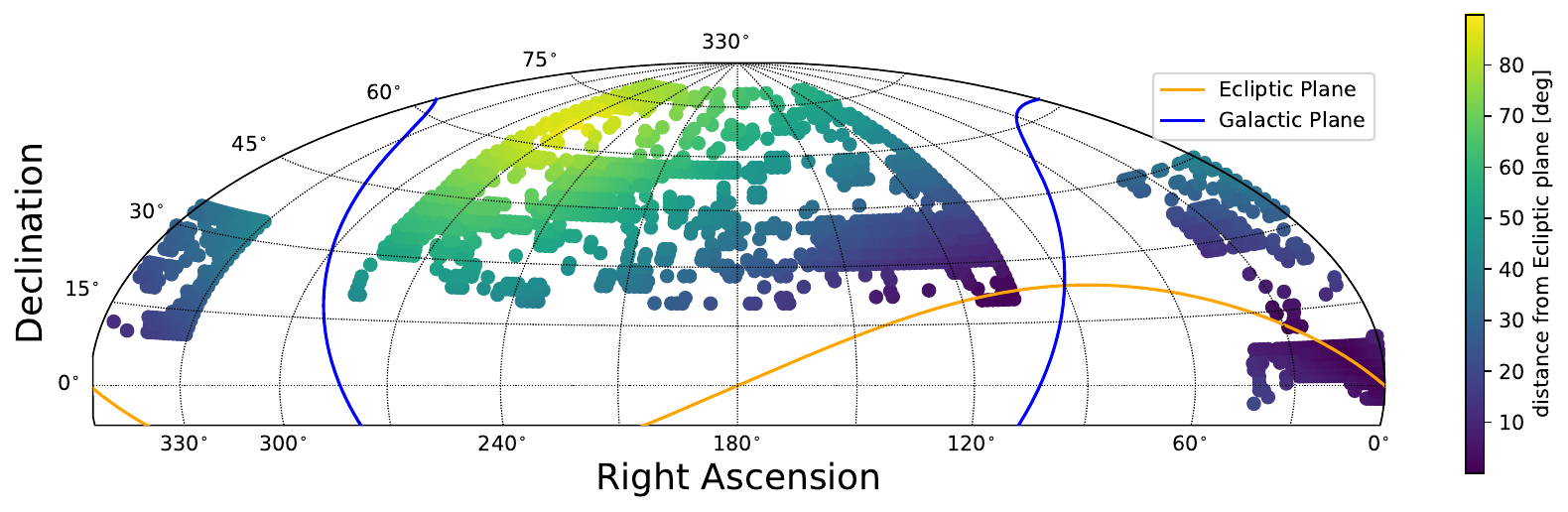} \\
    \includegraphics[width=0.9\linewidth]{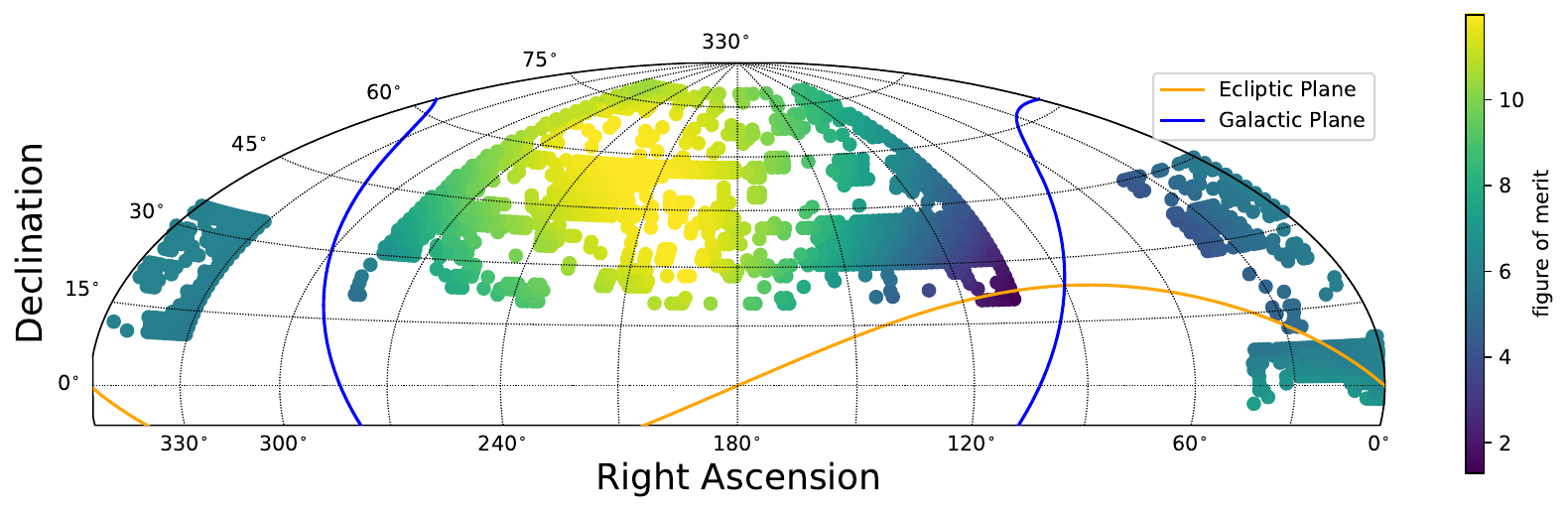} \\
    \end{tabular}
    \caption{
    The selection of the pointings to be observed is based on a trade off
    between the distance from the Galactic plane (top panel) and the distance
    from the Ecliptic plane (middle panel). The bottom panel shows the figure of
    merit given by the sum of the two previous panels (divided by 10 for convenience). 
    At each right ascension, this provides the priority for the pointing to be observed.
    }
    \label{fig:strategy}
\end{figure}

Regarding the selection of sky regions to be observed, two main considerations are taken into account. First, to ensure accurate photometric calibration, J-VAR targets J-PLUS fields that have already been published and properly calibrated. Second, the choice of fields is driven by the three key science cases of the survey: small bodies of the Solar System (Sect. ~\ref{sec:small_bodies}), optical transients—particularly supernovae (Sect.~\ref{sec:transients}), and variable stars (Sect.~\ref{sec:variable_stars}). The supernova case requires fields with a relatively low stellar density, which is best achieved by observing far from the Galactic plane. In contrast, the Solar System Objects case benefits from fields located as close as possible to the Galactic plane. This complementarity is illustrated in Fig.~\ref{fig:strategy}.

\begin{figure}
    \centering
    \begin{tabular}{c}
    \includegraphics[width=0.9\linewidth]{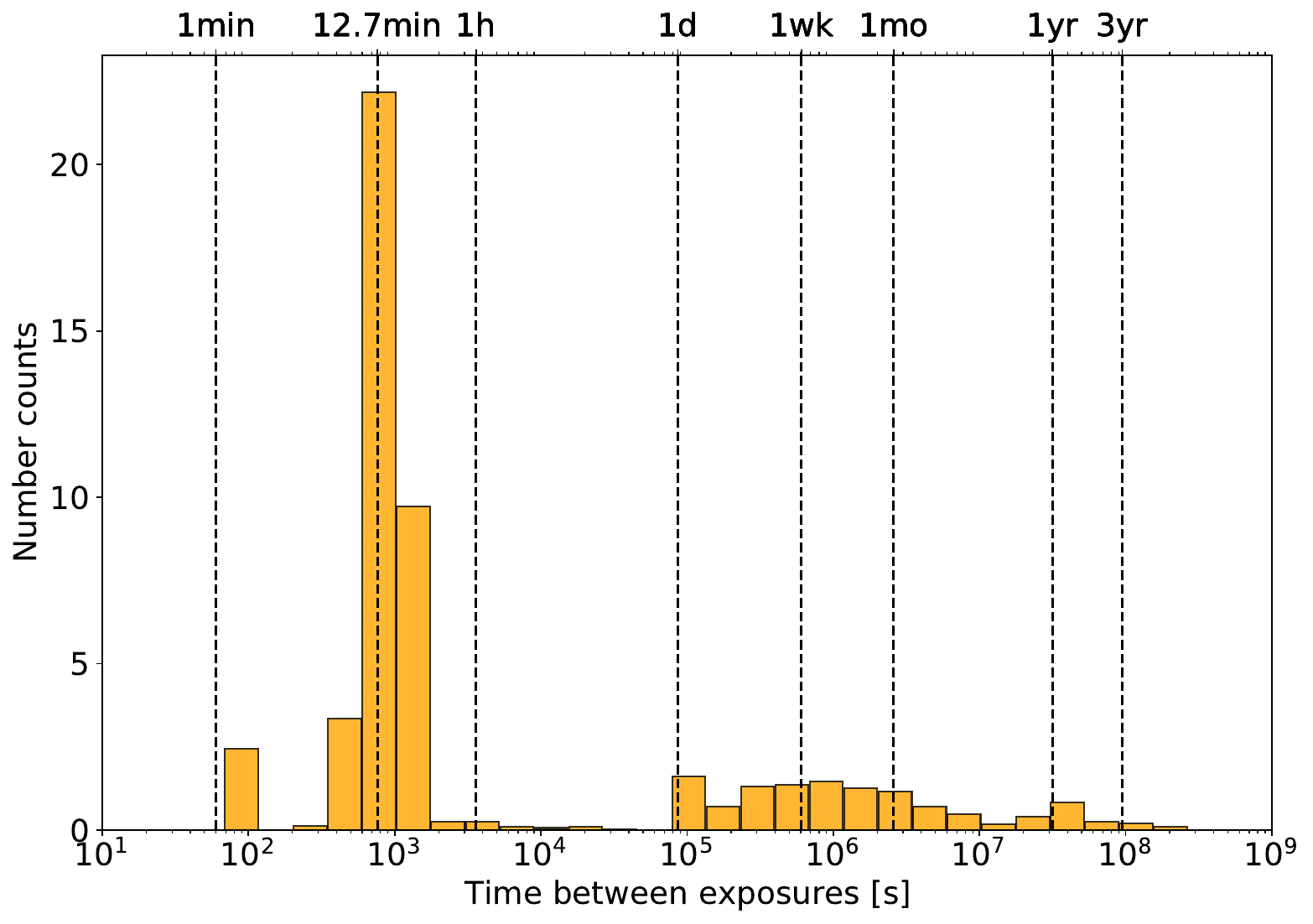} \\
    \includegraphics[width=0.9\linewidth]{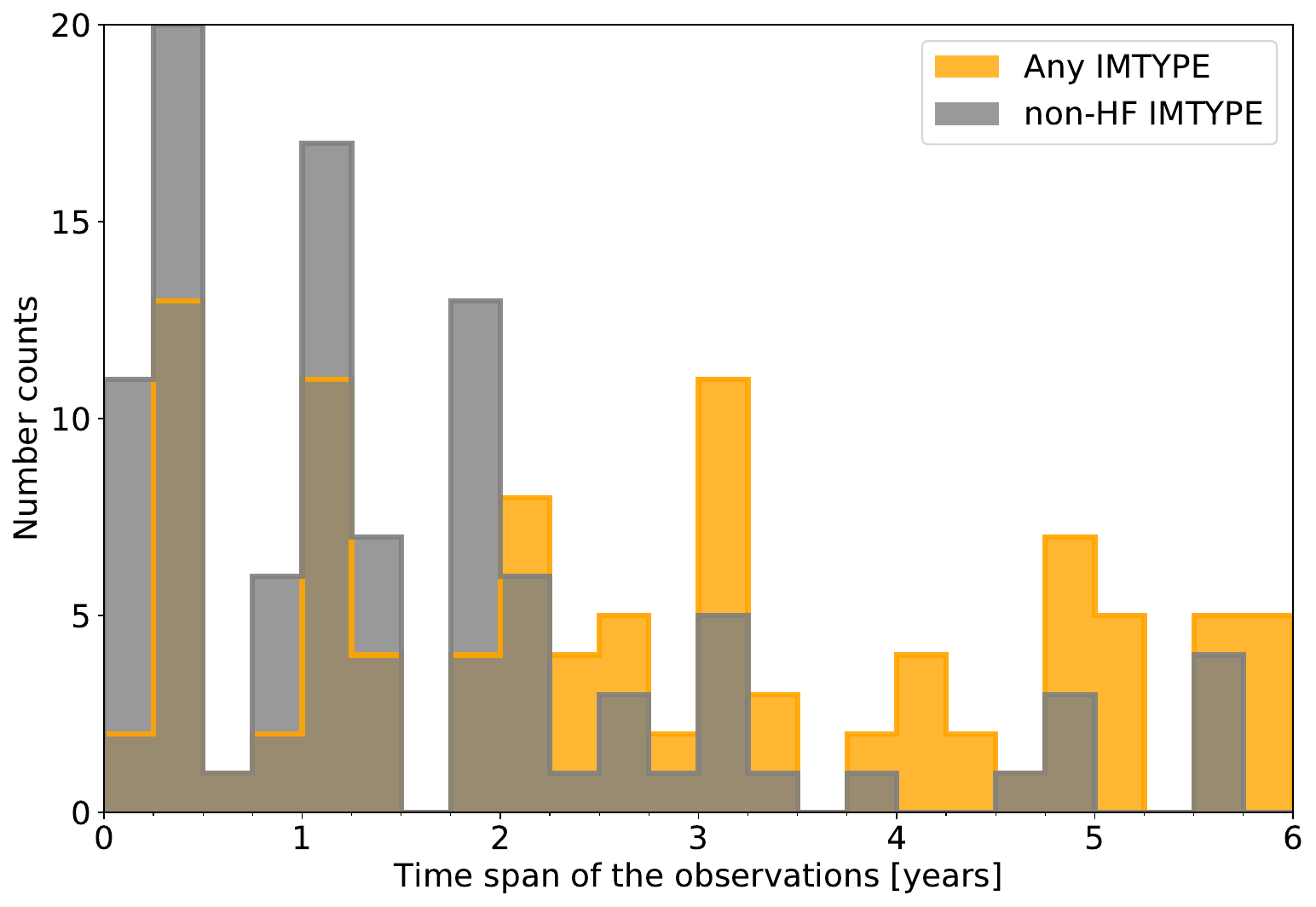} \\
    \end{tabular}
    \caption{Cadence and time span. \emph{Top:} Histogram of the times between consecutive individual exposures in $g$-band along the observation of each field. The median of all fields is shown. The median time interval between exposures is $12.7$ minutes.
    \emph{Bottom:} Time span distribution of all J-VAR DR1 fields. In \emph{gray}, nominal observations (JVN), while in \emph{yellow}, any type of observation is shown (including also high-frequency observations).}
    \label{fig:sampling}
\end{figure}

\subsubsection{Time-domain design: cadence and observing sequence}

Given the filler nature of the project, the cadence is primarily dictated by atmospheric conditions and the availability of other observing programs at any given time. 
The resulting cadence and sampling for J-VAR DR1 are shown in Fig.~\ref{fig:sampling}.

\begin{figure}
    \centering
    \includegraphics[width=\linewidth]{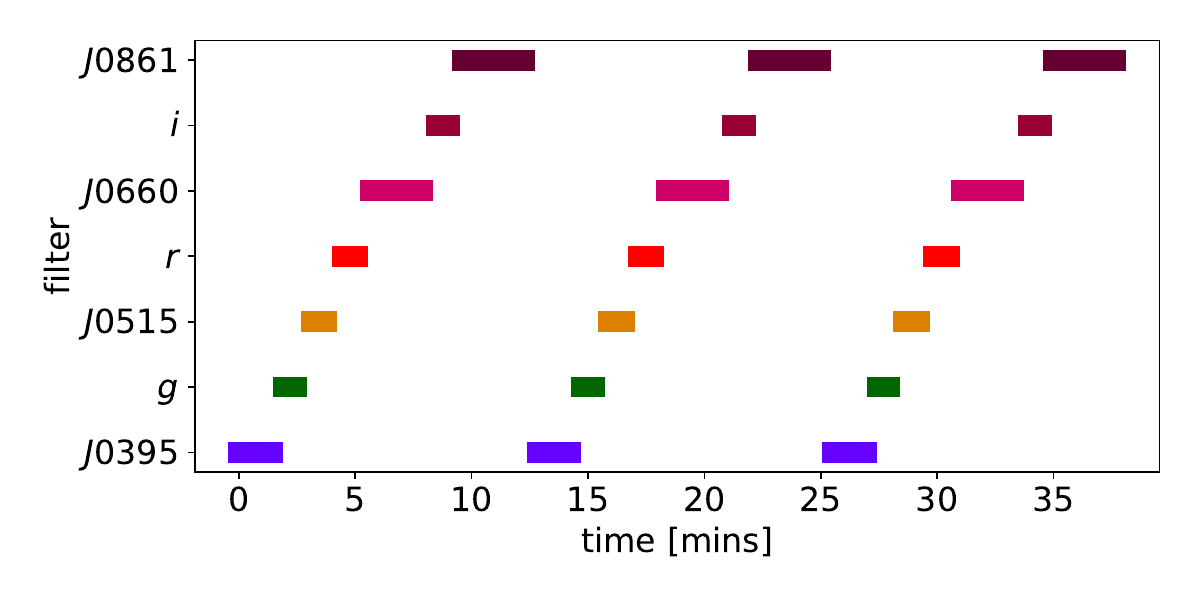}
    \caption{Structure of an observing block. First an exposure in the seven filters is obtained, then
    a small dithering is applied, then a second sequence of exposures is executed and then a third one after
    a second dithering. The two dithering directions are orthogonal. 
  }
    \label{fig:ob_scheme}
\end{figure}

For supernovae, maximizing the time span of observations increases the likelihood of detecting events in different phases of their evolution. Therefore, a goal was set to span the campaign in each field for at least one year. Additionally, to facilitate the creation of template images from any observing epoch—defined as all observations taken within the same night—the basic observing block (BBO) was structured to consist of three exposures per filter across all seven bands. These triplets allow for stacking when needed, improving signal-to-noise and enabling high-quality template image generation. The exposures are not taken sequentially in the same band; instead, three full cycles through the seven filters are executed, distributing the observations in time. This strategy, illustrated in Fig.~\ref{fig:ob_scheme}, ensures that each filter is sampled at different time intervals within an observing night. The full BBO takes approximately 40 minutes to complete, with a median gap of 12.7 minutes between exposures in the same filter.

This approach benefits not only the detection of transient events but also minor bodies in the Solar System, where repeated observations over short timescales help constrain orbital trajectories and facilitate rapid follow-up. Whenever possible, the same field is observed on consecutive nights to improve trajectory predictions. On longer timescales, sampling on the scale of weeks is crucial for tracking the brightness evolution of supernovae. The final distribution of visit frequencies for a given field is illustrated in Fig.~\ref{fig:sampling}.

Beyond transients and Solar System objects, J-VAR was also designed to provide multi-band time-domain information for variable stars, with RR Lyrae stars serving as a reference case for defining the total number of epochs. Unlike transient detections, where the absolute timing of observations is critical, RR Lyrae characterization primarily depends on adequately sampling the pulsation cycle. Given the somewhat randomized observing cadence, this condition is generally fulfilled naturally over time.

To determine the optimal number of epochs required, a simulation was conducted using real RR Lyrae light curves from the OGLE III Catalog of Variable Stars \citep{Soszynski2010,Soszynski2011} as a reference. The test focused on a single OGLE band (I), where different scenarios with varying numbers of randomly selected epochs were simulated. Each simulated epoch consisted of three data points, mirroring the J-VAR observing sequence. The derived periods were then compared to the true values provided by OGLE, allowing us to assess the period recovery success rate. The study found that at least 11 epochs were necessary to retrieve the correct period in $\sim$$75\%$ of cases when using a single band and a signal-to-noise ratio (SNR) above 50. Given that J-VAR includes observations in seven filters, this success rate was expected to improve when considering multi-band data. Consequently, 11 epochs were adopted as a reasonable trade-off between ensuring period determination accuracy and maximizing the survey area. This choice is consistent with other surveys, such as Pan-STARRS \citep{Sesar2017}. The actual period recovery success rate for J-VAR DR1 is analyzed in the accompanying RR Lyrae catalogue paper (Kulkarni et al. submitted). Using only the J-VAR $r$ band, \emph{Gaia} DR3 \citep{GaiaDR3,Clementini2023} periods of RR Lyrae stars in common are recovered within $1\%$ relative difference in $\gtrsim70\%$ of the cases for $G$ magnitude from 14 to 19. Some examples of J-VAR RR Lyrae light curves are shown in Sect.~\ref{sec:variable_stars}.

\subsubsection{The High Frequency (HF) Fields}

During periods around full Moon, one of the JAST80 filter wheels may be temporarily 
replaced to host alternative filters required by other projects 
(e.g. GALANTE,  \citealt{galante}).
When atmospheric conditions prevent the execution of those programs, we implement a dedicated J-VAR observing strategy, hereafter the “high-frequency mode” (HF). In this mode, a single field is monitored continuously for $\sim$3 hours, repeating the standard sequence of exposures with the set of available J-VAR filters
J-VAR sequence of seven filters (Fig.~\ref{fig:ob_scheme}).
This configuration delivers intra-night light curves with much denser sampling than in the nominal survey, while maintaining full compatibility with the standard observing setup.

\section{Data Reduction}
\label{sec:datareduction}

The image reduction is performed using \texttt{jype}, a custom-built pipeline developed at the \emph{Centro de Estudios de F\'{\i}sica del Cosmos de Arag\'on} (CEFCA) for processing data from OAJ surveys
\citep[see, e.g.][]{jplus,Bonoli2021}. Written primarily in \texttt{Python}, \texttt{jype} incorporates \texttt{SExtractor} \citep{Bertin1996} for source extraction and initial photometry. 
The pipeline has been continuously developed since 2010, and its most recent version before the release, \texttt{jype-3.1.9}, has been used 
for the processing of J-VAR DR1 images.

A key aspect of \texttt{jype} is its ability to handle the particular readout structure of the T80Cam CCD, which consists of 16 amplifiers. Each amplifier includes overscan and prescan regions along both the $x$ and $y$ axes, allowing for precise bias level removal. Immediately after observation, scientific images undergo an initial calibration using the latest valid flat field and illumination correction (see below). These pre-processed images are typically available within minutes and enable time-sensitive analyses, such as asteroid and transient detection, which are usually conducted the day after observation.

Beyond this real-time processing, a more refined reduction is performed at the end of each observing run, defined as a period during which the optical system remains stable (i.e., no interventions affecting the filter wheel or main mirror cleaning). Observing runs typically last about a month, at which point master flat fields and illumination corrections are generated to ensure optimal photometric accuracy.

In addition to standard image reduction steps, including bias subtraction, flat-fielding, fringing correction (when necessary), and cosmic ray and satellite trail masking, \texttt{jype} features an essential illumination correction step. This is particularly important for wide-field systems like JAST80, where conventional flat-fielding alone can introduce a two-dimensional photometric bias of several tens of milimagnitudes across the image due to the presence of field correctors. To mitigate this effect, an additional processing step is applied; see details in Appendix B.1 in \citet{Bonoli2021}.

Finally, \texttt{jype} computes the aperture correction for photometry performed within a $6$-arcsecond diameter integration area, which serves as a reference for total flux measurements in J-VAR DR1, particularly for point-like sources. This standardized photometric approach ensures consistency across all reduced data products.

\begin{figure*}
    \centering
    \includegraphics[width=\linewidth]{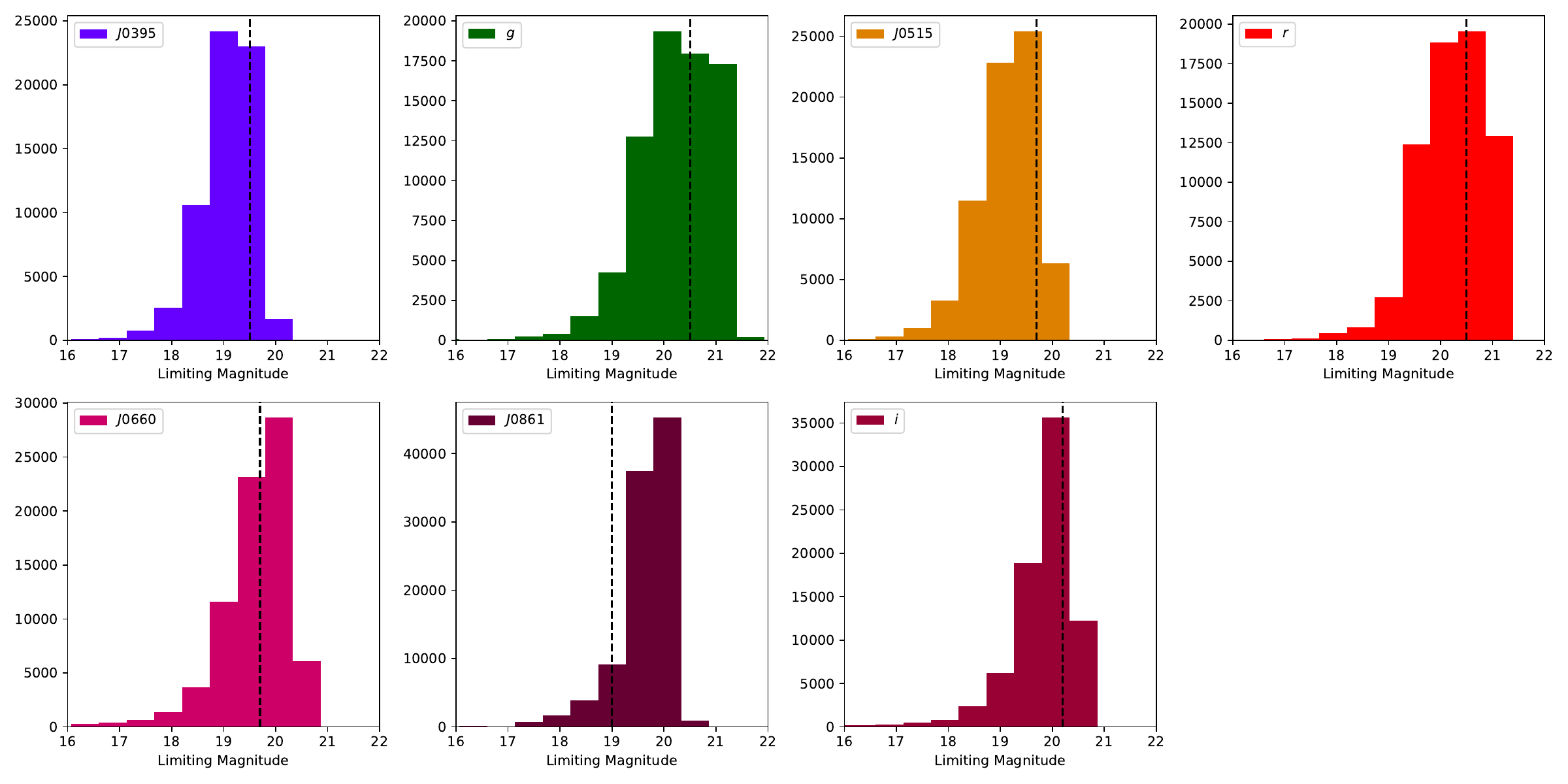}
    \caption{Distribution of the limiting magnitudes of all the images of J-VAR DR1
    in the different filters.
    The vertical line is one magnitude shallower than J-PLUS limiting magnitude (5$\sigma$ detection in 6\,arcseconds aperture) which
    was used to define the exposure times of J-VAR.
  } 
    \label{fig:limitingMags}
\end{figure*}

\begin{figure*}
    \centering
    \includegraphics[width=\linewidth]{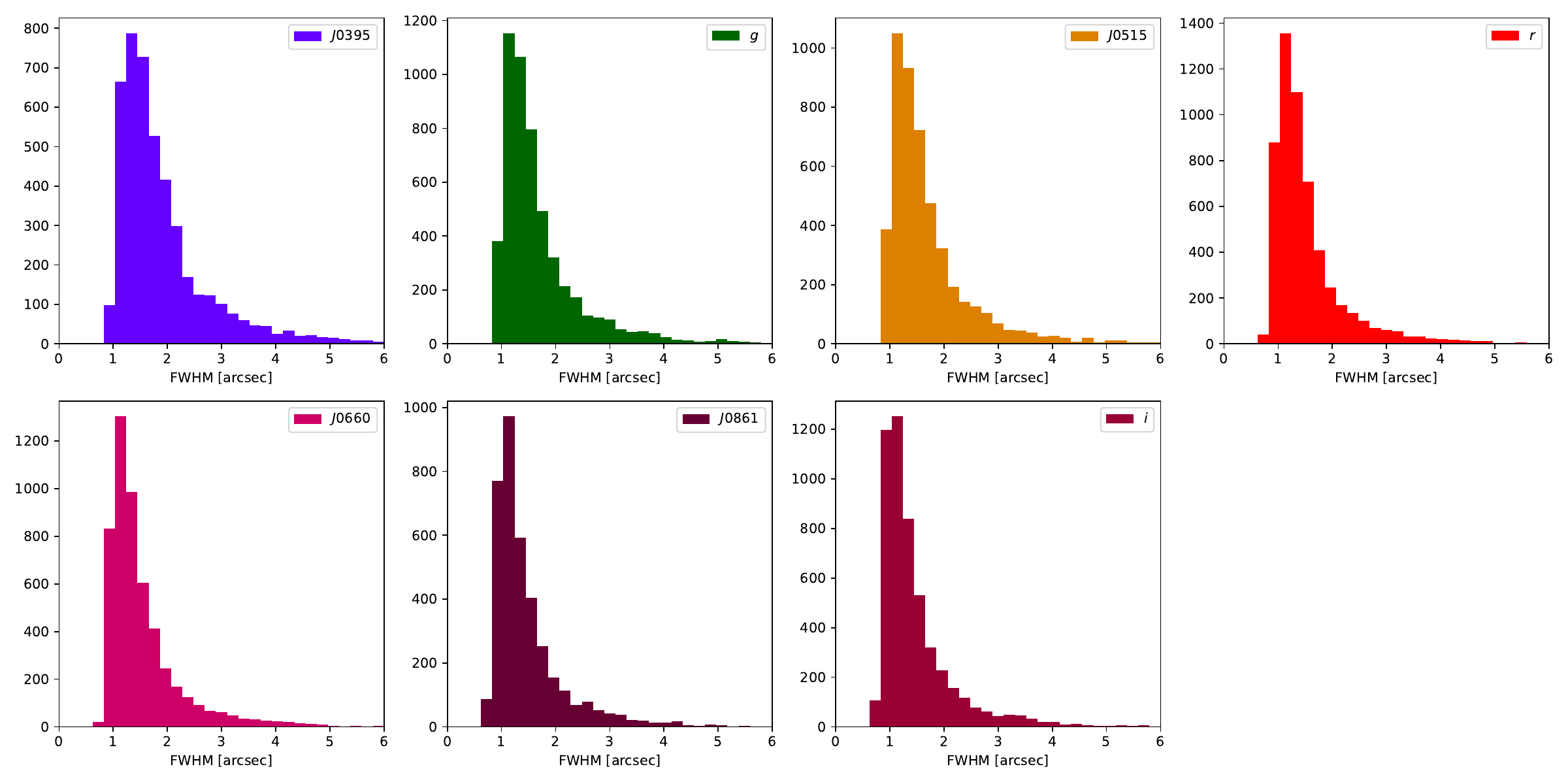}
    \caption{Distribution of the median full width at half maximum of the detected 
    sources in each image for each observed filter. The quasi-simultaneous observations
    in the different filters and the good image quality across the whole optical range
    result in very similar distributions, regardless of the central wavelength of the 
    filter. 
    }
    \label{fig:fwhm}
\end{figure*}

\section{First Data Release}\label{sec:DR1}

\begin{figure*}
    \centering
    \includegraphics[width=\linewidth]{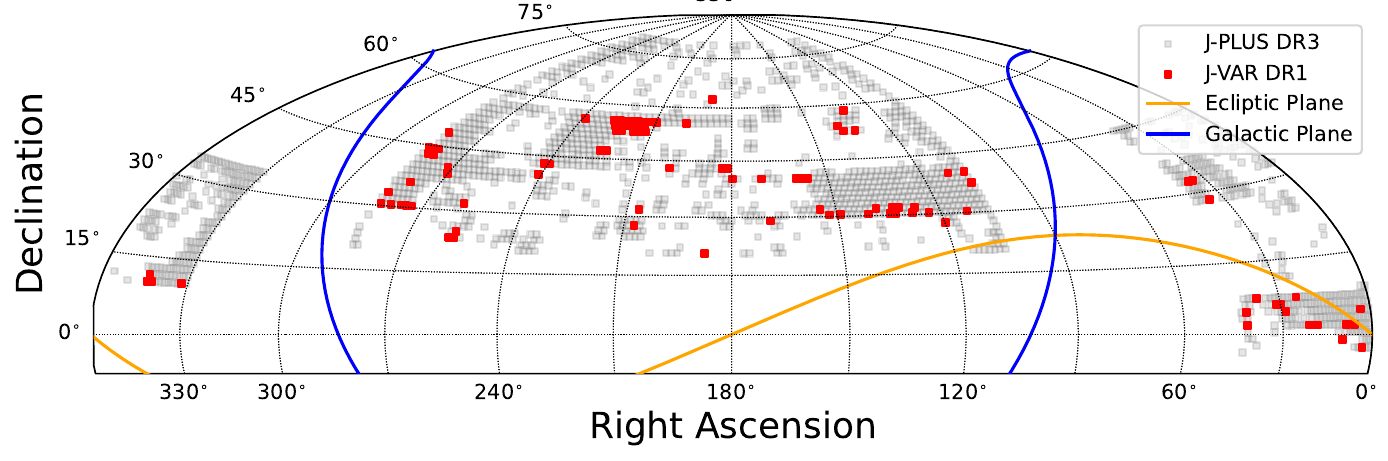}
    \caption{The red markers show the sky distribution of the 101 pointings of J-VAR DR1.
    The grey points refer to the sky distribution of J-PLUS DR3 which is used for 
    calibration. The yellow and the blue line show the Ecliptic and the Galactic plane, 
    respectively. 
    }
    \label{fig:footprint}
\end{figure*}

The First Data Release of J-VAR was made public on 19th July 2024.
The number of pointings distributed in this data release is $101$, 
providing time-domain information for $\sim$$202\,\textrm{deg}^2$ on the sky in 
seven photometric bands. The list of coordinates is in Table~\ref{tab:pointings},
together with information on the earliest and latest time of 
observation (in modified Julian date) and
if the field has high frequency observations. 
In case HF observations are available, the earliest and
latest time of observation is also available (again, as modified Julian date).
The sky distribution is shown in Fig.~\ref{fig:footprint}. 
From an analysis of each of the images in this
data release, we derived the distribution of limiting
magnitudes (Fig.\ref{fig:limitingMags}). 
The limiting magnitudes are rarely achieved in most filters
and never achieved in the reddest filters ($J0861$ and $i$).
This is a consequence of the approximations made at the time
of setting the exposure times.
Similarly, Fig.~\ref{fig:fwhm} shows the distribution of the FWHM. 
The lower limit of the distribution is close to 1\,arcsec
which is slightly smaller than 2\,pixels. The peak of the
distribution is consistently close to 1.5\,arcsec.
The quality of the data, in particular having a 
shallower survey in the red, does not affect the main 
goals of the project, as the multi-filter light curves are
still deliverable.

The data of J-VAR DR1 are accessible through the CEFCA 
archive\footnote{\url{https://archive.cefca.es/catalogues/jvar-dr1}}. 
The principal way to access to the data is through the public catalogue,
whose structure is related to the main objectives of the project.
The time of observation of each field in a given filter is in the {\tt Timestamps} table (its
contents are shown in Table~\ref{tab:timestamps}). In this table,
the observing time of each image is provided. Each image is identifiable via the identifier of the 
field and the identifier of the filter. The time is provided as coordinated universal time (UTC). 
For convenience, the modified Julian date (MJD) is also given.

Each image is also tagged according to the type of observation: regular or high frequency (see above).
The regular J-VAR observations are tagged as ``JVN''. The high frequency observations are marked 
differently, as described hereafter. The narrow-band images are marked as ``HFN''. The broad-band images are marked ``HFJ'' if executed with the same filters as the main ``JVN'' survey. If they are acquired using, not the same physical filters, but the  $g$, $r$, and $i$ filters initially devoted to the Pathfinder camera to conduct of the miniJPAS survey \citep{Bonoli2021}, they are named ``HFG''. Those $gri$ filters were usually mounted in the auxiliary filter wheel of T80Cam, and were manufactured following the same optical design.

The photometry of each object is provided in the {\tt Photometry} table (a description is
made available in Table~\ref{tab:photometry}).
In this table, each observation of an object is identified by an ID of the object. Both the J-PLUS 
identifier (the ID of the J-PLUS DR3 tile and the identifier of the object within the tile) and the
J-VAR identifier (the string {\tt jvar} followed by the right ascension and the declination 
of the object) are provided. The magnitudes and associated errors are offered as arrays. 
The full-width at half maximum of the objects is also provided as array, as well as a list of photometric 
flags which can be used for quality assessment. Finally, the photometric calibration is represented
by an array containing the values used to go from instrumental magnitude to calibrated magnitude. 
The number of comparison stars and the maximum radius within which these stars are 
located are also shown. For details, see Pyrzas et al. (submitted).
For convenience, the previous two tables are combined in a {\tt light\_curves} table  whose columns are described in Table~\ref{tab:lightcurves}. 

The table {\tt Objects} conveniently gathers useful 
information from J-PLUS DR3 (magnitudes in 3 and 6\,arcsec,
different star/galaxy separation indicators, FWMM),
the interstellar extinction, the type and period of
variability according to the International Variable Stars 
Index and the parallax, G magnitude, variability class
and score from Gaia DR3.
The columns of this table are summarised in Tab.\ref{tab:objects}.

Two tables related to the Solar System Objects are also available. The first one ({\tt SSOs\_DETECTIONS},
the columns are shown in Table~\ref{tab:ssos_detections}) shows the detections by the SSOS pipeline,
described in Sect.~\ref{sec:small_bodies} and in Morate et al. (submitted), while the second
one ({\tt SSOs\_MAGNITUDES}, whose columns are shown in Table~\ref{tab:ssos_magnitudes}) shows
the magnitudes for each object.

Following the common practice, the detections of optical transients are reported through 
the Transient Name Service\footnote{\url{https://www.wis-tns.org/search}} and therefore 
there is no associated table in the database.

All the tables are accessible via a web interface and via the Table Access Protocol of the
Virtual Observatory, both in a programmatic way (e.g. with Python) or via programs like 
TOPCAT \citep{topcat}. Similarly, The reduced images are accessible through the archive and through the
Simple Image Access Protocol (SIAP) of the Virtual Observatory.
An example of how to access the data of J-VAR is available at \url{https://github.com/aederocl/J-VAR_paper1}.

\section{Science Cases}
\label{sec:science}

\subsection{Small bodies of the Solar System}
\label{sec:small_bodies}

Small bodies (SB) in the Solar System are not usually the drivers of observational surveys, with the 
noticeable exceptions of the Asteroid Terrestrial-impact Last Alert System, ATLAS, \citep{atlas2018PASP}, the Panoramic Survey Telescope and Rapid Response System, PAN-STARRS \citep{pansta2004AN}, or the Rubin Observatory's Legacy Survey for Space and Time, LSST \citep{ivezic2019ApJ} surveys. SBs are usually regarded as contamination for the main science objectives of most large surveys. Nevertheless, the SB community has exploited this "contamination" by developing pipelines that detect and extract their information.
  The Sloan Digital Sky Survey has been largely exploited for SB \citep[e.g.,][]{sso_sdss,demeo+carry}
  and so was the first data release of J-PLUS \citep[see][]{morate2021}.
\citet{sso_2mass},  \citet{Popescu+2016} and \citet{Mahlke+2018} have extended this search to 
infrared wavelengths.
SBs information, such as their colors, is used for mineralogical analysis, which places strong constraints on how different materials are distributed in the Solar System and, together with dynamical and collisional models, may shed light into how the Solar System as we know it today came to be.

J-VAR observational strategy
is  well-suited for the detection of SBs because the fields (Sect.~\ref{sec:strategy}) are selected as a balance between its three main science drivers. Within the 101 fields included in J-VAR DR1, there are 131\,900 individual detections, i.e., an asteroid was detected in one of the J-VAR filters. These detections correspond to a total of 6\,570 individual asteroids, some of them detected in two or more epochs.

The detection of the SBs is done using the SOSS pipeline \citep[see details in][]{soss-pip2019}. The pipeline detects and extracts the instrumental magnitude of the SBs, but it does not produce calibrated magnitudes because the J-VAR data is taken during non-photometric nights. Thus, no zero points are provided in the images' headers. Therefore, the calibration step is done using J-PLUS observed fields, computing the zero points for the common objects in the field and applying this secondary calibration to the SB (more details are provided in Morate et al. submitted).

Regarding the number of detections, for those J-VAR fields centered around the ecliptic plane of the Solar System ($\pm10\deg$), there are around $\sim$400 detections per epoch (which translates roughly into $\sim$30 different objects). This number decreases as one moves away from the ecliptic: for high latitudes, i.e., above $30\deg$, the number of detections is very scarce, usually below 20 (this is, one object or none at all.)
However, the number of detected objects ensures good statistics (see below).

\begin{figure}
   \includegraphics[width=\columnwidth]{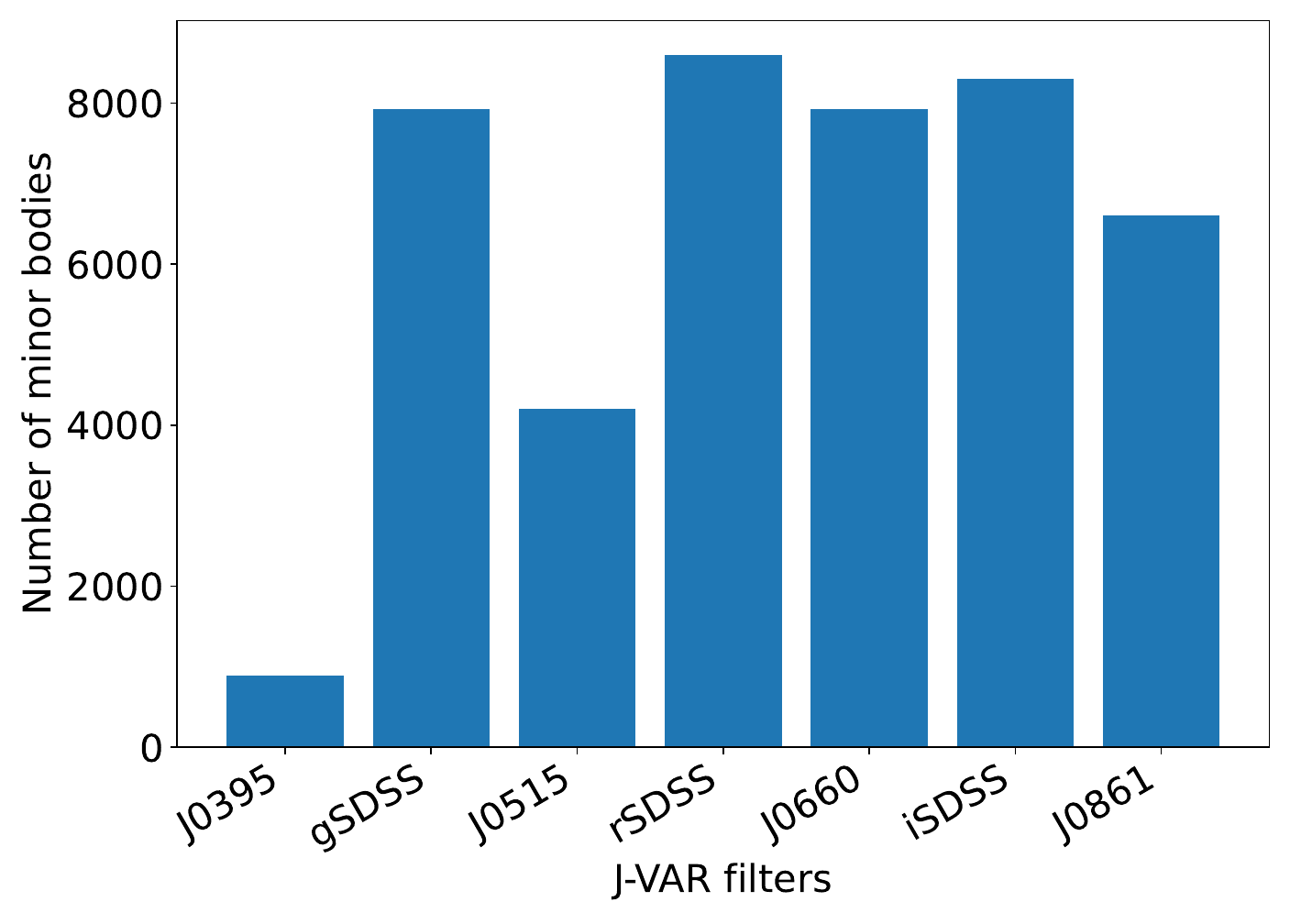}
      \caption{Number of SBs detected in each J-VAR filter.}
         \label{Fig:sbs_nfil2}
   \end{figure}

The distribution of objects observed with a given number of filters peaks at six (with about 3\,000). Almost 800 were observed with all seven filters, and only two objects were observed with one filter. 
We show the number of detections per filter in Fig.~\ref{Fig:sbs_nfil2}. The lower numbers in the $J0395$ and $J0515$ filters were expected as these are narrow-band filters, and, in the case of the $J0395$ filter, the reflectance of
SB  is known to decrease significantly at blue wavelengths.
The most efficient filters are the SDSS filters $g$, $r$, and $i$, which was expected 
because these are the widest filters in the J-VAR system.

\begin{figure}
\includegraphics[width=\columnwidth]{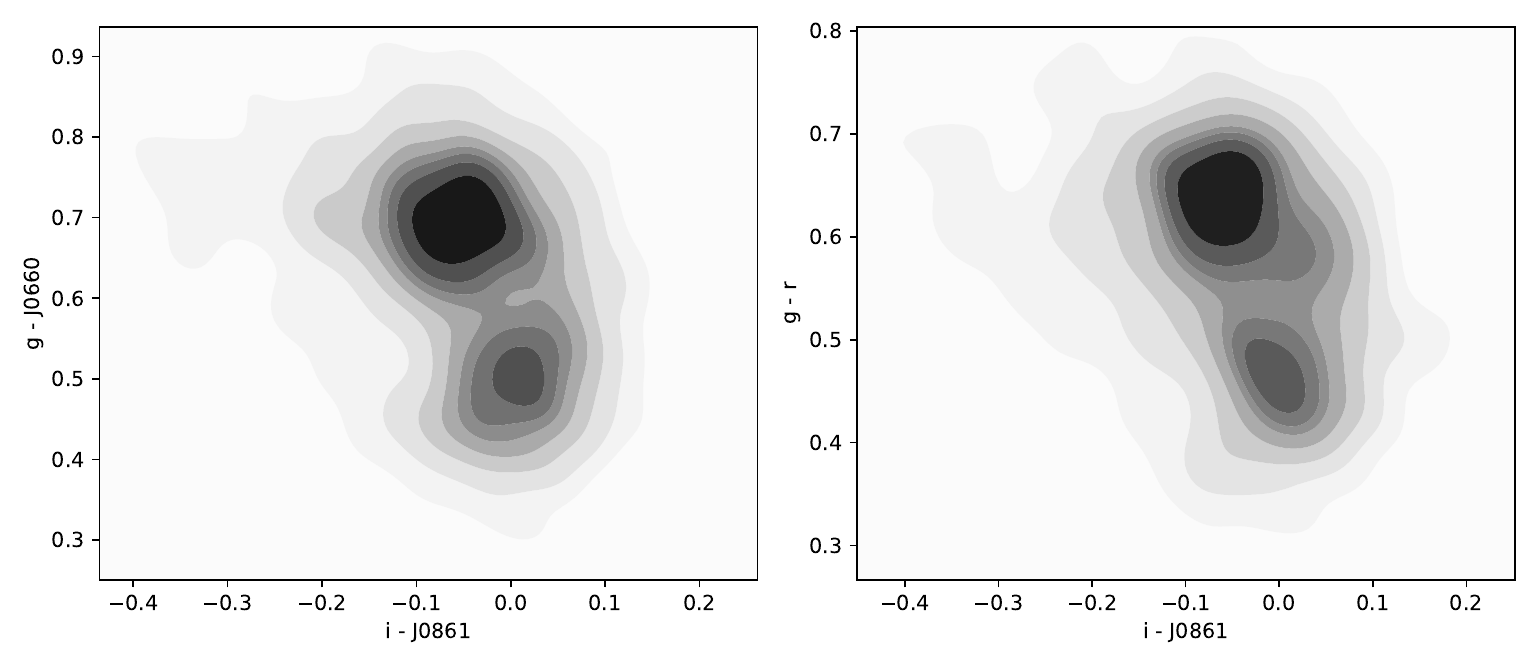}
      \caption{Colour-colour diagram in the form of heat maps. The left panel shows ${g - J0660}$ 
      vs  ${i - J0861}$, while the right panel shows  ${g - r}$ vs  ${i - J0861}$.
      The use of different colours allows to 
      separate between families of different chemical
      composition.
      }
         \label{Fig:sbs_colors}
   \end{figure}

Finally, we show a highlight of our results (a detailed description of the data is provided in Morate et al., submitted): Fig.~\ref{Fig:sbs_colors} shows two colour-colour diagrams, zoomed into their densest regions to avoid outliers.

In Morate et al. (submitted) we present a value added catalogue (VAC) for J-VAR's DR1: the detection catalogue with calibrated magnitudes, RA, DEC, time of observation, among other useful information. We also provide a first look into the taxonomy of the objects and rotational properties of some targets, mostly observed in HF fields.
In conclusion, J-VAR works as a stand-alone survey, but, importantly, it nicely complements J-PLUS, which uses all J-VAR filters.

\subsection{Optical Transients}
\label{sec:transients}

Ranging from novae, supernovae (SNe) up to the progenitors of the most powerful gravitational waves, the transient sky is part of the most challenging puzzles in astrophysics and cosmology.
Although mostly known as responsible for the first evidences of the Dark Energy dominated Universe in the late 1990s, type Ia SNe, for example, are still of paramount importance to cosmology. The so-called "Hubble tension" \citep{PhysRevD.101.043533}, a discrepancy in the measurement of the Hubble constant in different stages of the Universe's evolution, is becoming a decade-long problem and some of the most precise results on which it relies are based on observations of low-redshift type Ia SNe.

From an astrophysical point of view, there seems to be still no universal consensus on the nature of type Ia SNe progenitors and their explosion mechanisms \citep{2023RAA....23h2001L}. The same difficulties arise in finding a single model to explain superluminous SNe explosions \citep{annurev:/content/journals/10.1146/annurev-astro-081817-051819}.

As commented in Sect.~\ref{sec:introduction}, a variety of surveys aim at the study of the
transient sky. In particular, currently ASAS-SN and ZTF which can be considered a progenitor
of LSST.

We took advantage of the repeated visits from J-VAR and applied the image subtraction or differential image technique to look for optical transients within DR1 images.
Our differential image pipeline consists in a sequence of open source packages and scripts to perform image subtraction and produce a catalogue of candidates to be analysed individually. 
The pipeline, optimized for SNe detection, also recovers variable stars, AGN, and asteroids. To reduce false positives, we apply strict shape and multi-band detection criteria, followed by SIMBAD cross-matching. The final step is a visual inspection and confirmation of the candidates.

\begin{figure}
    \centering
    \includegraphics[width=\linewidth]{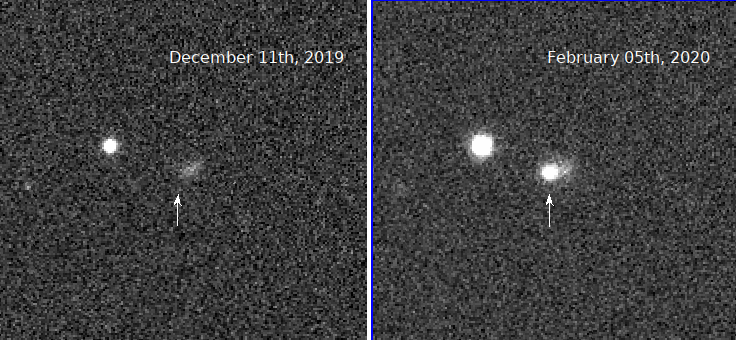}\\
    \includegraphics[width=\linewidth]{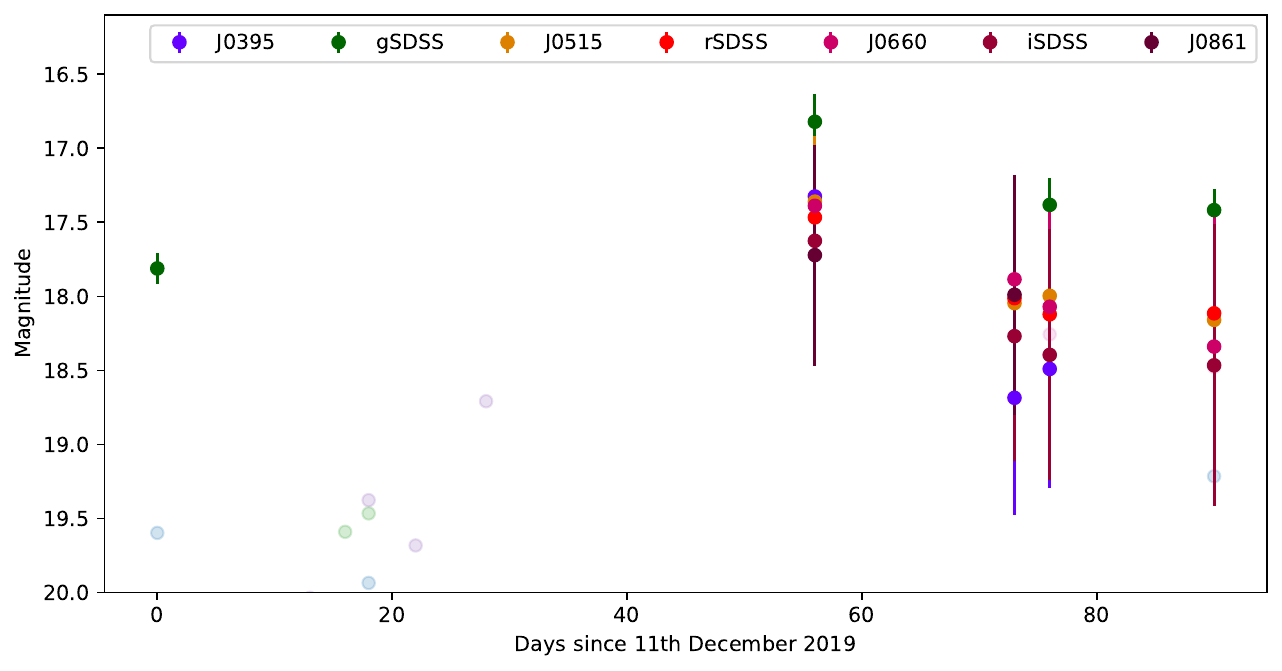}\\
    \includegraphics[width=\linewidth]{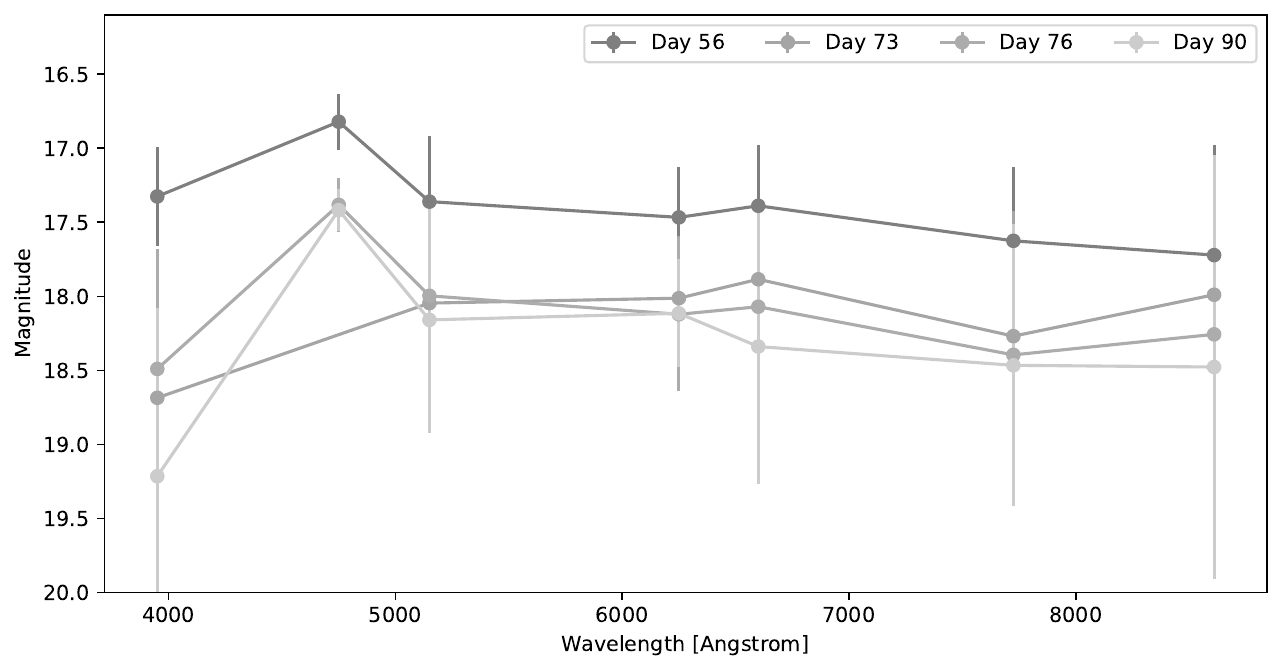}\\
    \caption{{\it Top:} Detection image of SN~2020amv, {\it Middle:} Light curve of 
    SN~2020amv in the seven filters of J-VAR. The shaded points refer to photometric
    errors larger than 1~magnitude, which is equivalent to a non detection. {\it Bottom:}
    Evolution of the spectral energy distribution of SN~2020amv with time. Note the 
    detection of flux in the $J0660$ filter, confirming the nature of the object
    as a type~II supernova.}
    \label{fig:SN2020amv}
\end{figure}

Since our search was performed after the observations of a given field were completed, the pipeline detected some candidates which had been already observed by other surveys, which is the case for supernovae SN\,2017icq (SN Ia), SN\,2019obw (SN Ia), SN\,2020cvy (SN II) and SN\,2020amv (SN II) and transient candidates AT\,2017eke, AT\,2017dzn, AT\,2018fds, AT\,2019roh, AT\,2019ioo and AT\,2020abzu. 
Fig.~\ref{fig:SN2020amv} shows the detection, light curve and spectral energy evolution of the supernova
SN\,2020amv\footnote{\url{https://www.wis-tns.org/object/2020amv}}. This was discovered on 23rd\,January 
2020 and it is detected in J-VAR field 13\,days after discovery. The bottom panel of Fig.~\ref{fig:SN2020amv}
shows an excess in the $J0660$ filter, suggesting the presence of H$\alpha$ in emission, 
characteristic of type\,II supernovae, thus showing the potential of J-VAR to characterise
low-redshift supernovae without the need of spectroscopy.

Our final lists of transient candidates contained a large number of variable sources. We recorded  252 known variable stars and 50 known AGN, along with 236 variable star candidates and 18 AGN candidates.

Finally, we have discovered by visual inspection four transient events which were submitted to the TNS database. Two of them, SN\,2020admb (JVAR20a) and SN\,2024slh (JVAR24b), were successively confirmed as supernovae and classified by the ZTF collaboration as a type Ia supernovae at redshifts $~0.04$ and $0.05$, respectively. The other two, AT\,2021aggv (JVAR21a) and AT\,2024bxn (JVAR24a) are still classified as transient candidates at the TNS database.

\subsection{Variable Stars}
\label{sec:variable_stars}

The J-VAR survey provides multi-band time-domain data that enables the study of various classes of variable stars across a range of timescales, amplitudes, and wavelengths. The combination of its multi-filter approach, flexible cadence, and moderate number of epochs makes it particularly well-suited for the characterization of pulsating stars, 
close or contact
eclipsing binaries, and semi-regular variables.
Moreover, the irregular spacing of the epochs mitigates aliasing effects in period determination, thereby increasing the robustness of variability classification.

The light curves for all the point-sources in common with J-PLUS have been extracted.  This sums up
to 1.3 million sources. 
The light curves have been obtained with ensemble differential photometry (see e.g., \citealt{Everett+Howell2001}; \citealt{Tamuz+2005}). For each
star, at least 15\,comparison stars are selected. In all cases, this is
achieved in a search radius smaller than 9\,arcminutes.
The resulting photometric precision (RMS) is 2\% down to mag $\sim$16 and 5\% to mag $\sim$18
in the broad-band filters.
The extraction and calibration process is described in detail in Pyrzas et al. (submitted).
In total, the J-VAR DR1 catalogue has 1621 objecs 
classified as variables in VSX and 15221 objects classified
as variables in {\it Gaia } DR3. The most frequent stellar
{\it Gaia } DR3 classes are shown in Table~\ref{tab:number_of_variables}.

\begin{figure}
    \centering
    \includegraphics[width=0.5\textwidth]{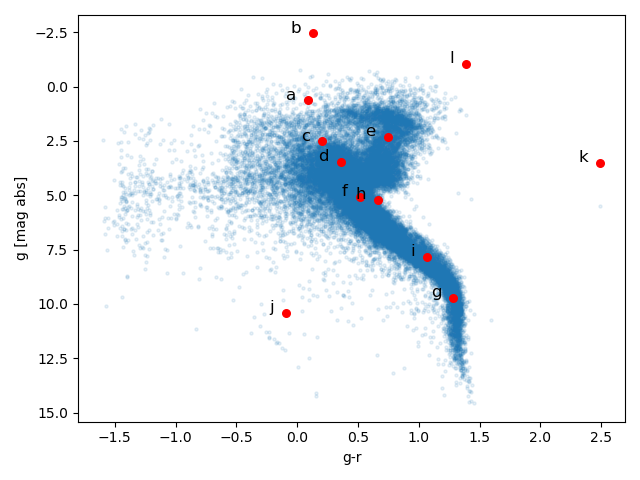}
    \caption{The colour-magnitude diagram of J-VAR sources. The light curves of the 
    objects marked in red are shown in Fig.\ref{fig:LC_samples}. 
    }
    \label{fig:HRvar}
\end{figure}

\begin{figure*}
    \centering
    \begin{tabular}{c c c}
    \includegraphics[width=0.33\linewidth]{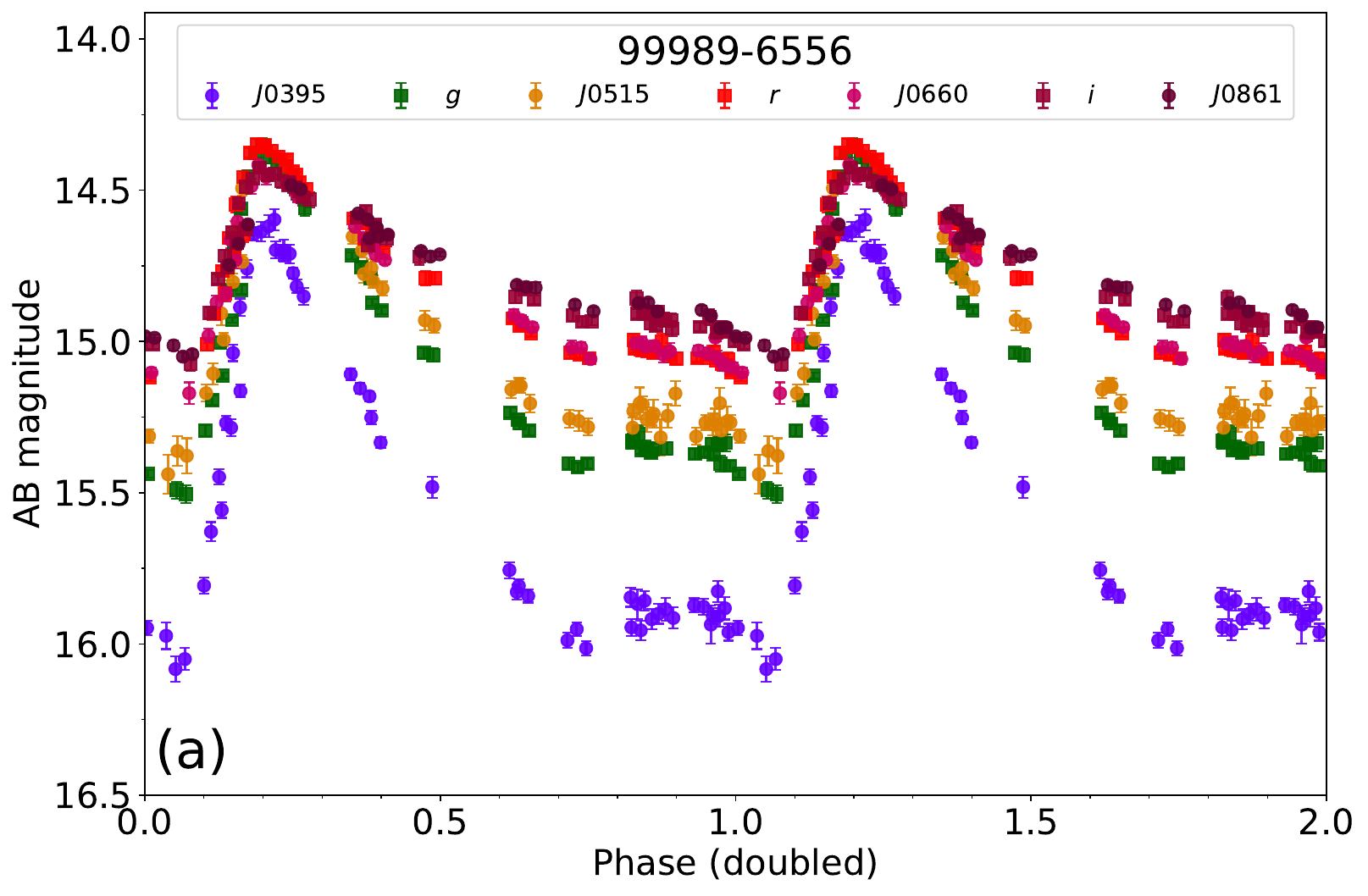}     &  
    \includegraphics[width=0.33\linewidth]{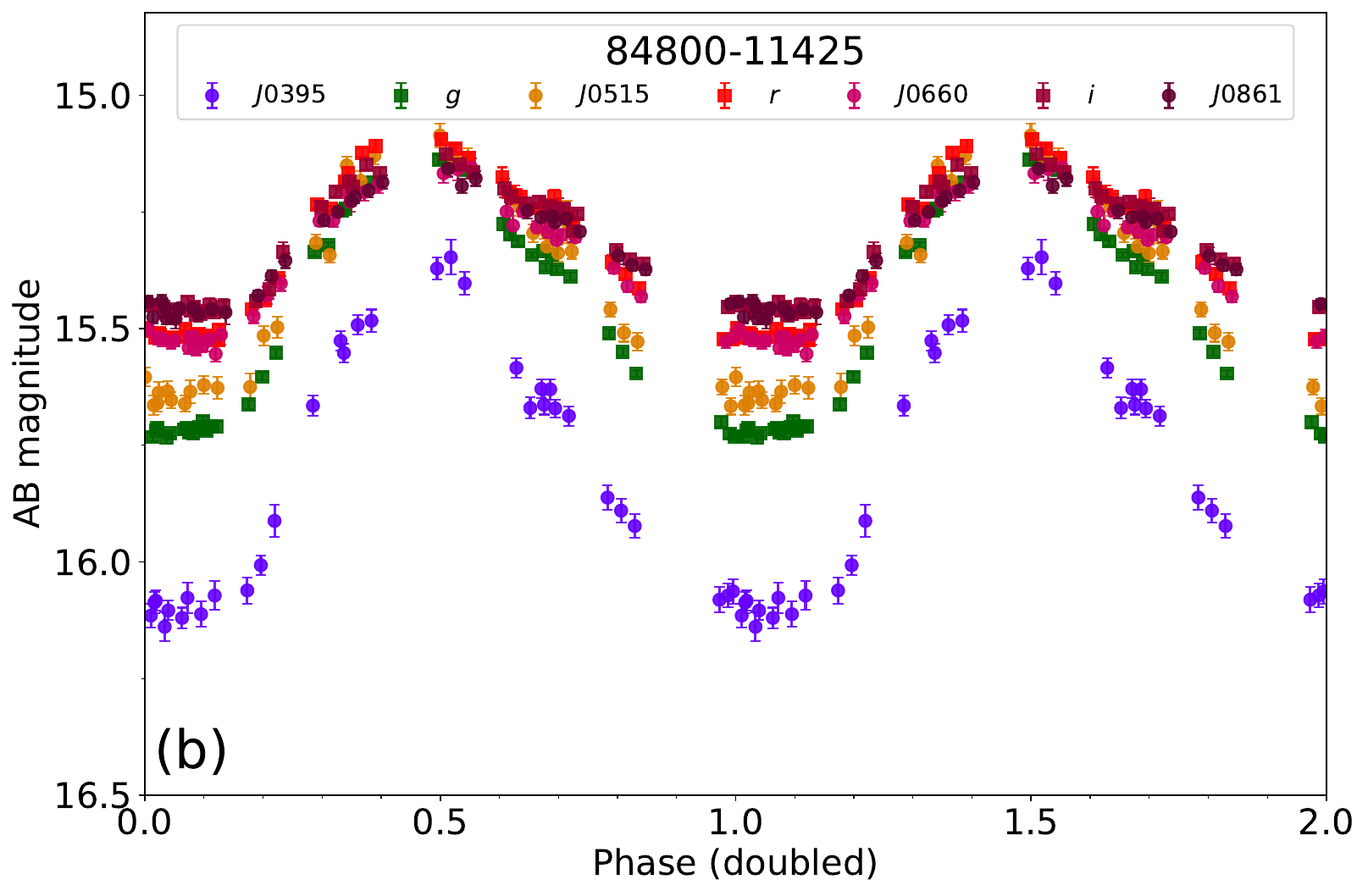}     &
    \includegraphics[width=0.33\linewidth]{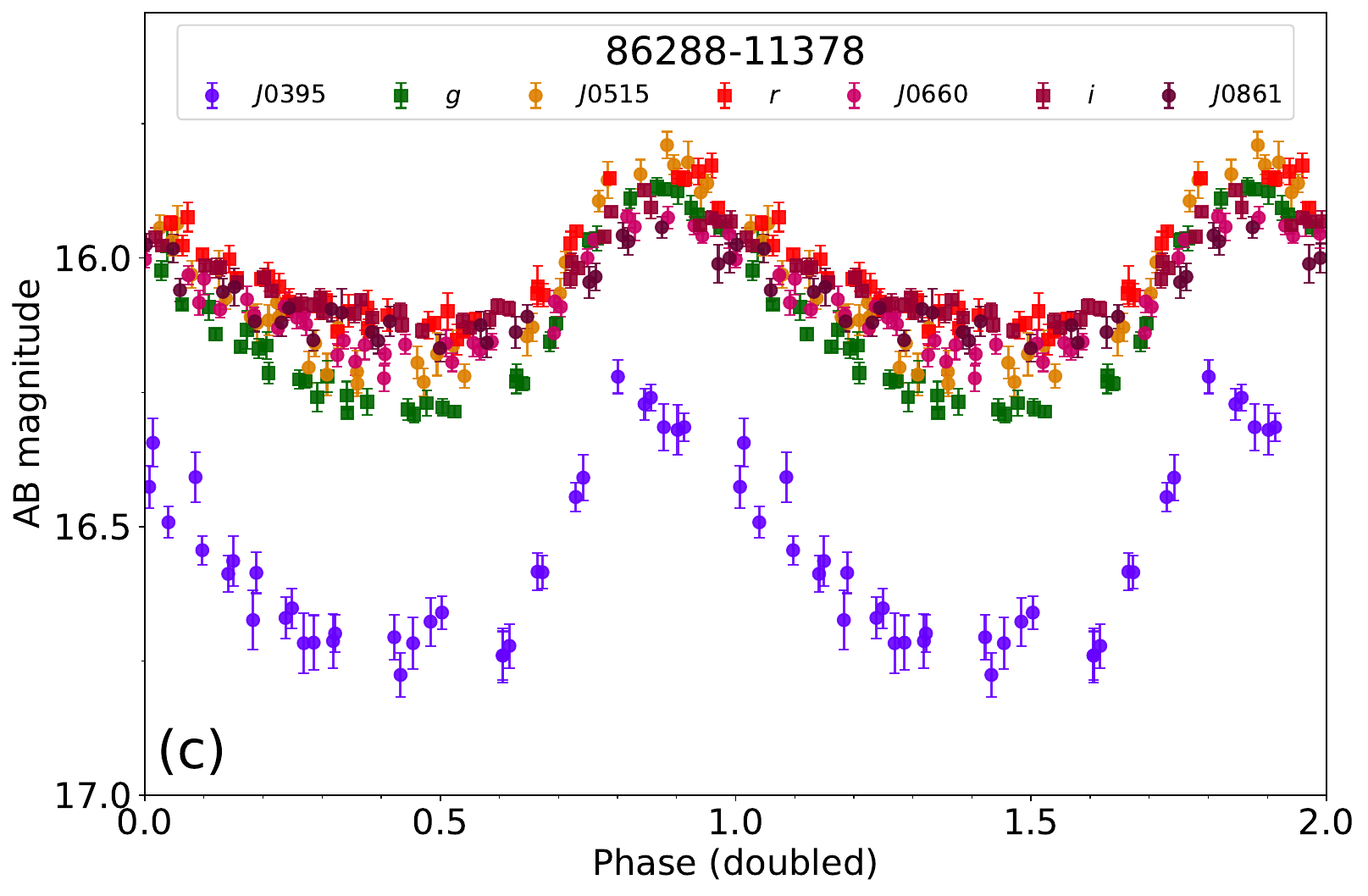} \\
    \includegraphics[width=0.33\linewidth]{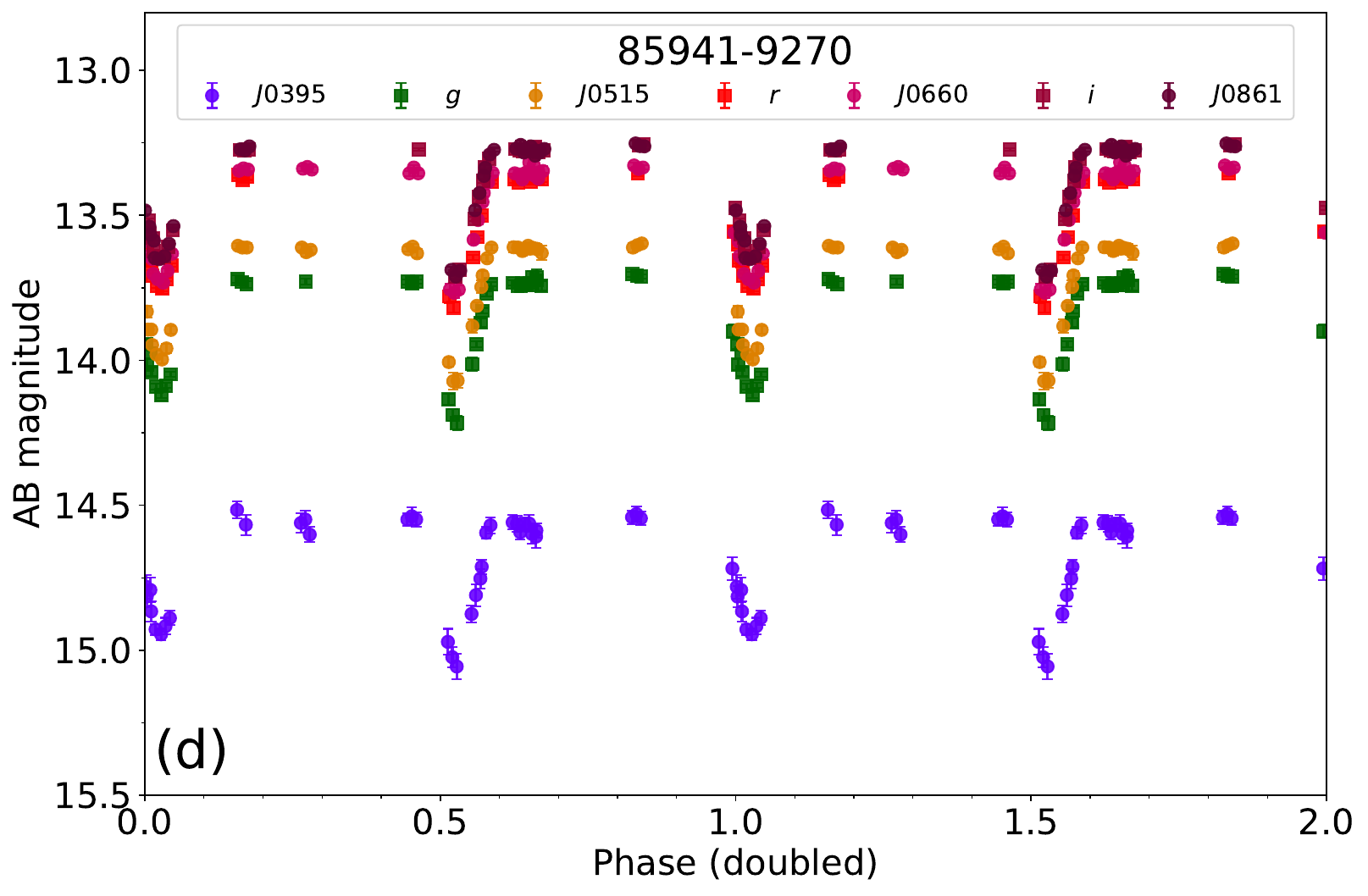}     &  
    \includegraphics[width=0.33\linewidth]{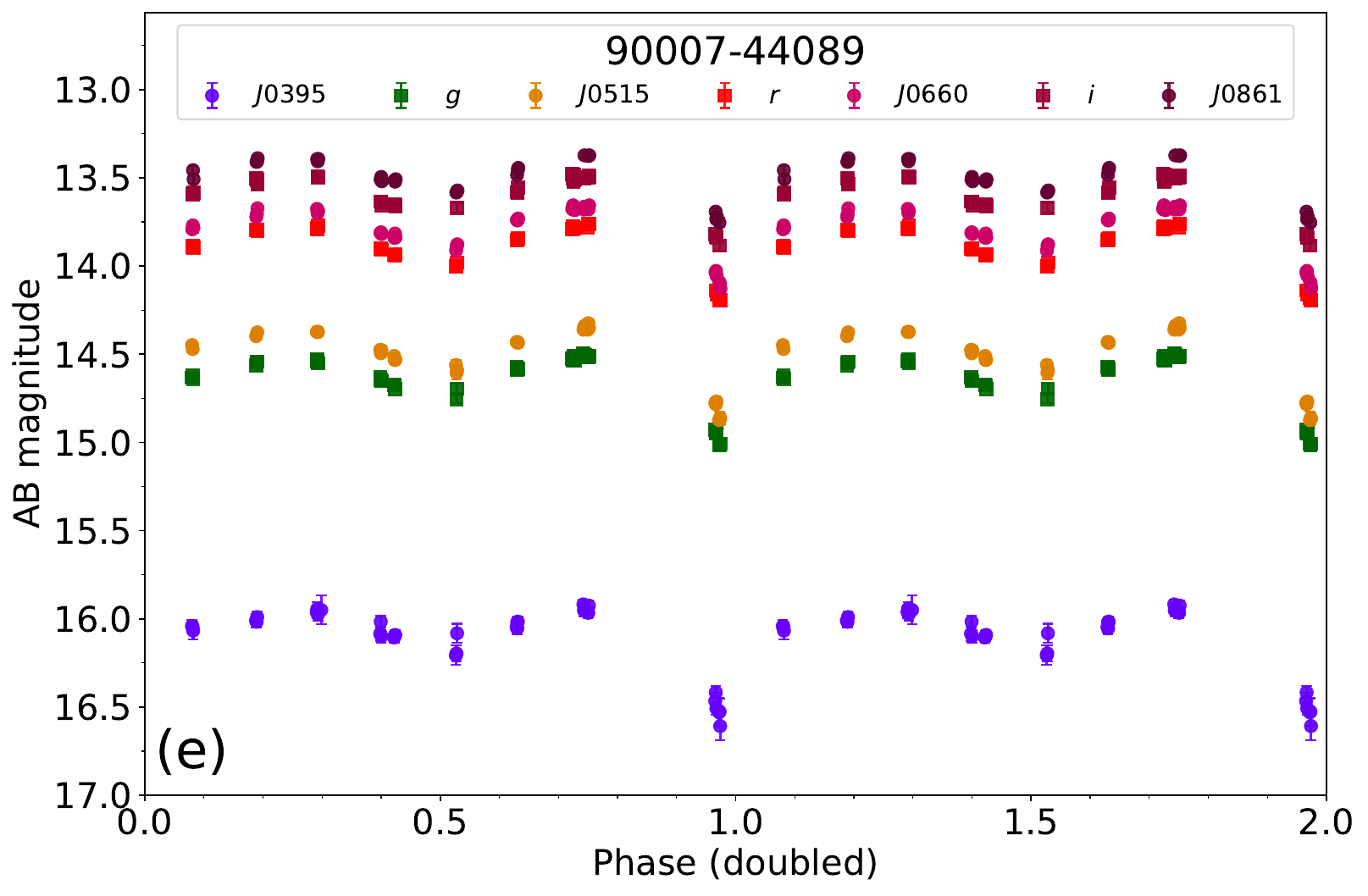}     &
    \includegraphics[width=0.33\linewidth]{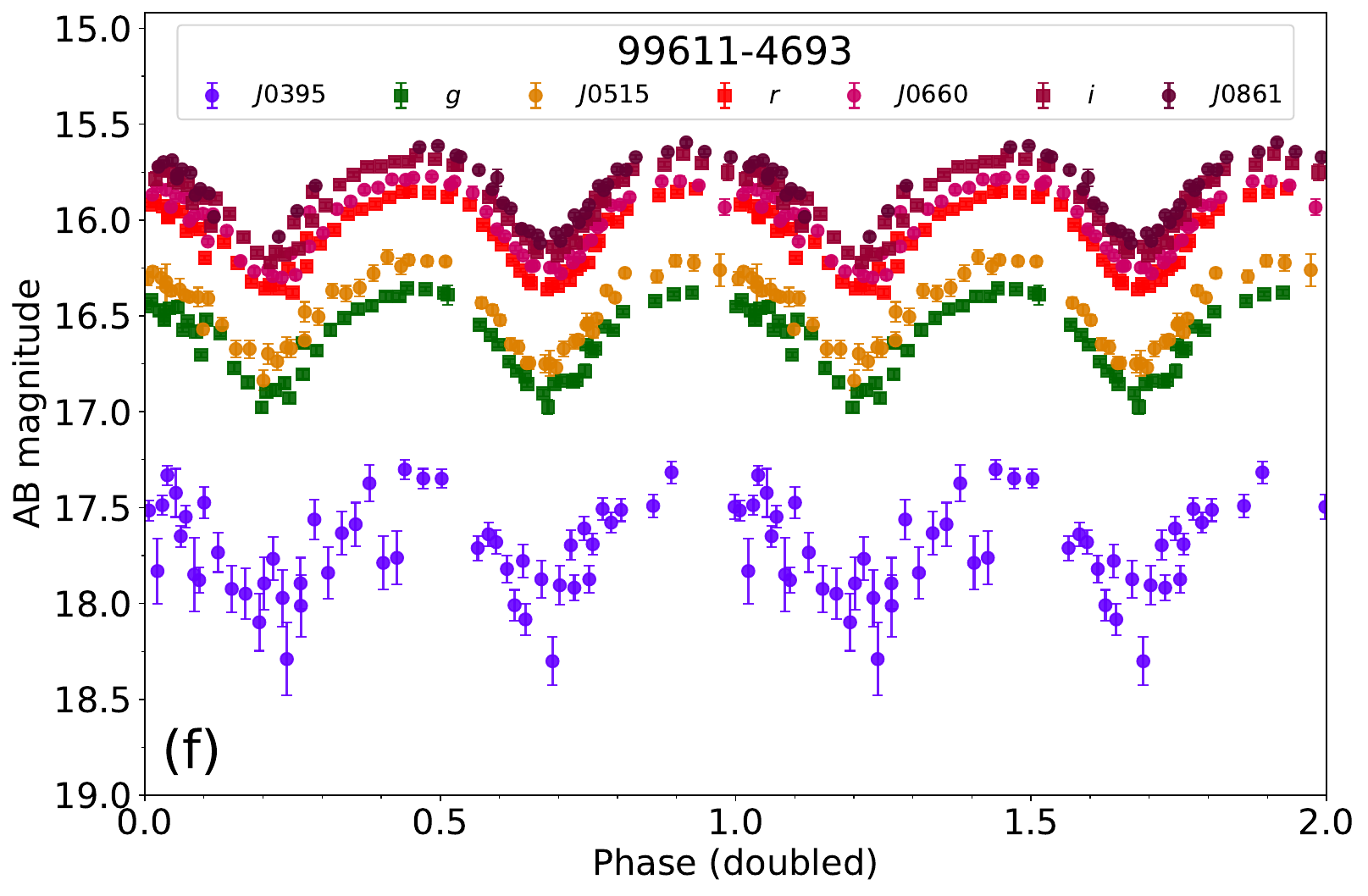}     \\
    \includegraphics[width=0.33\linewidth]{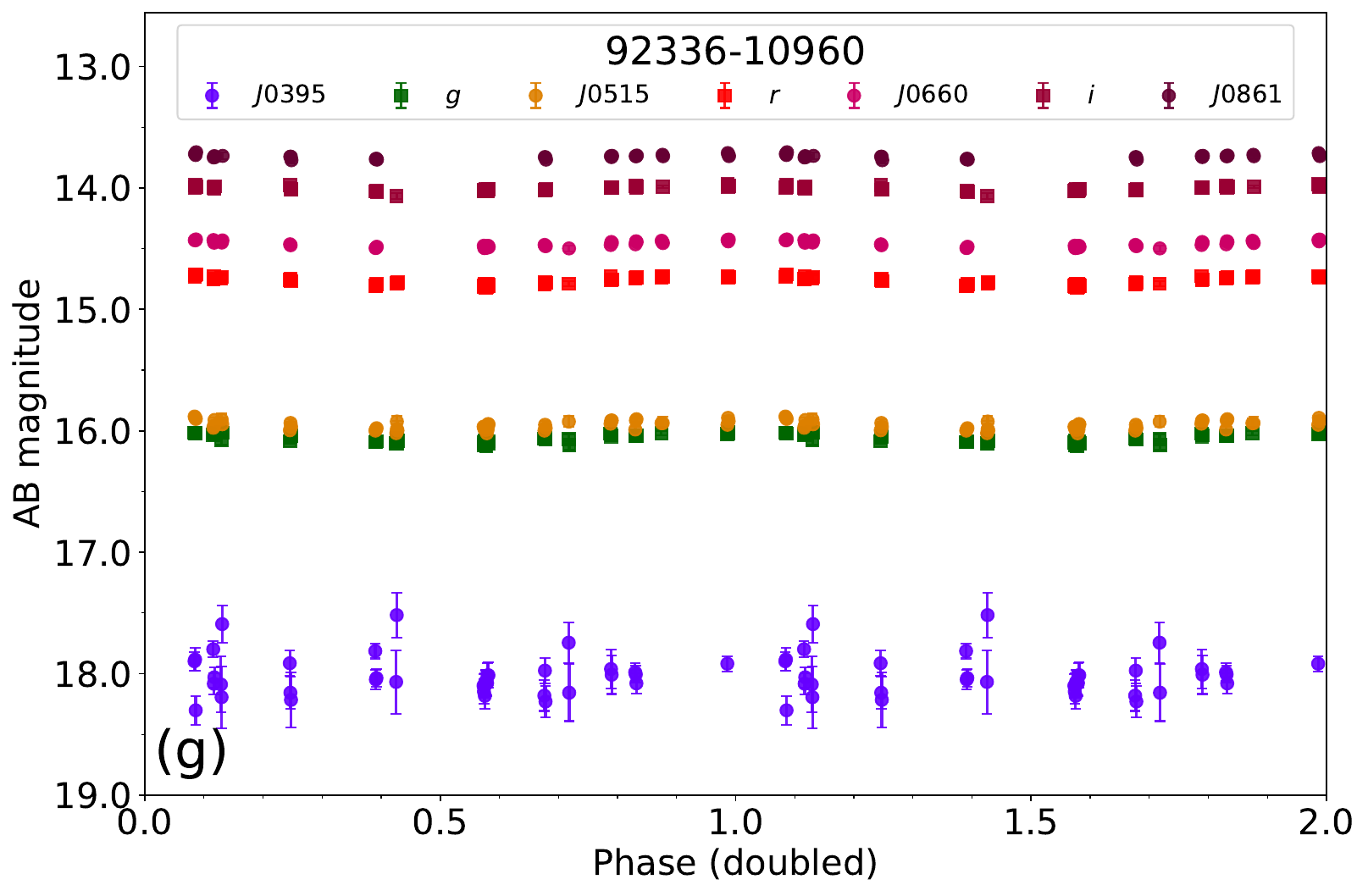} &
    \includegraphics[width=0.33\linewidth]{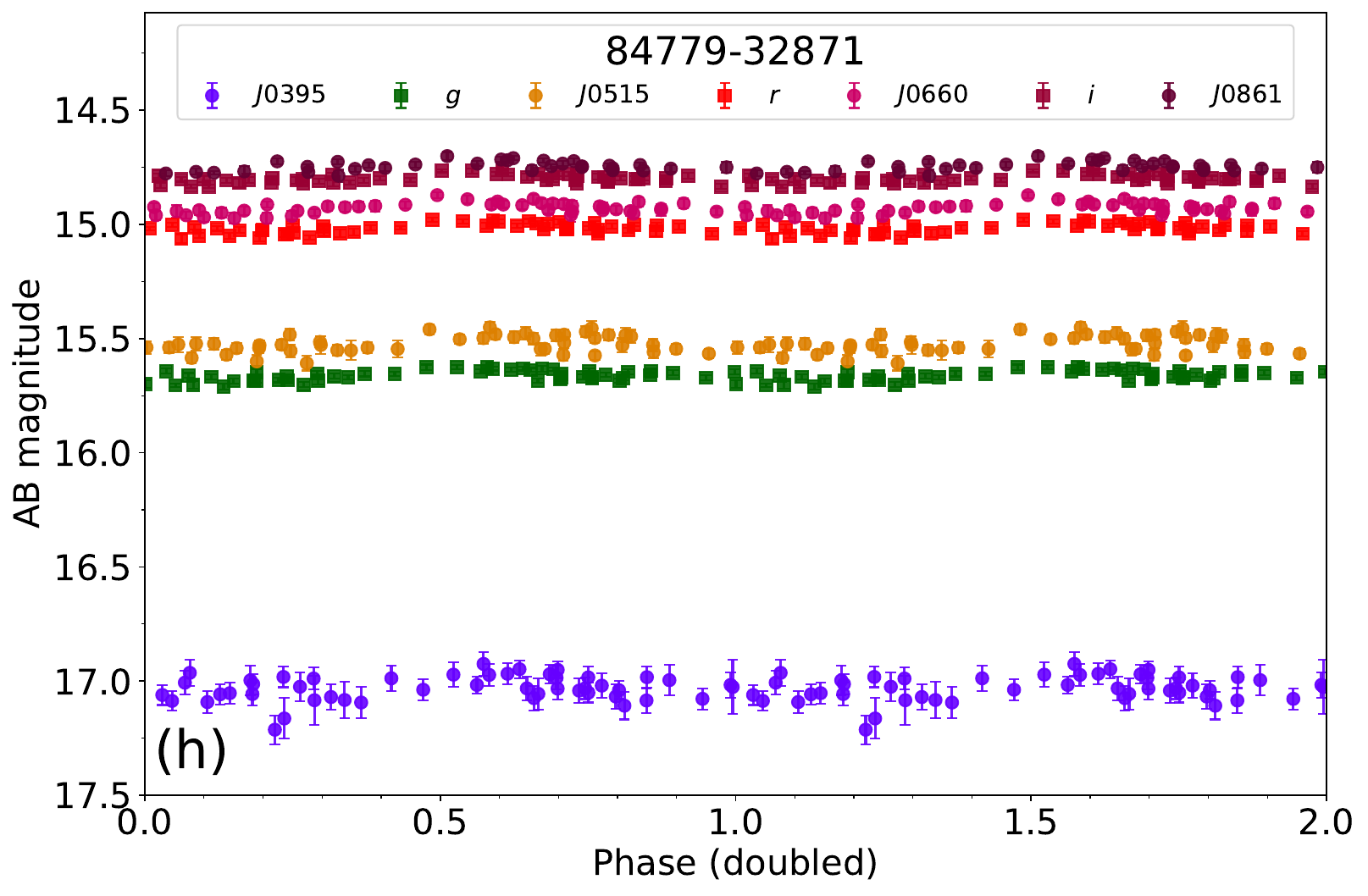}     &  
    \includegraphics[width=0.33\linewidth]{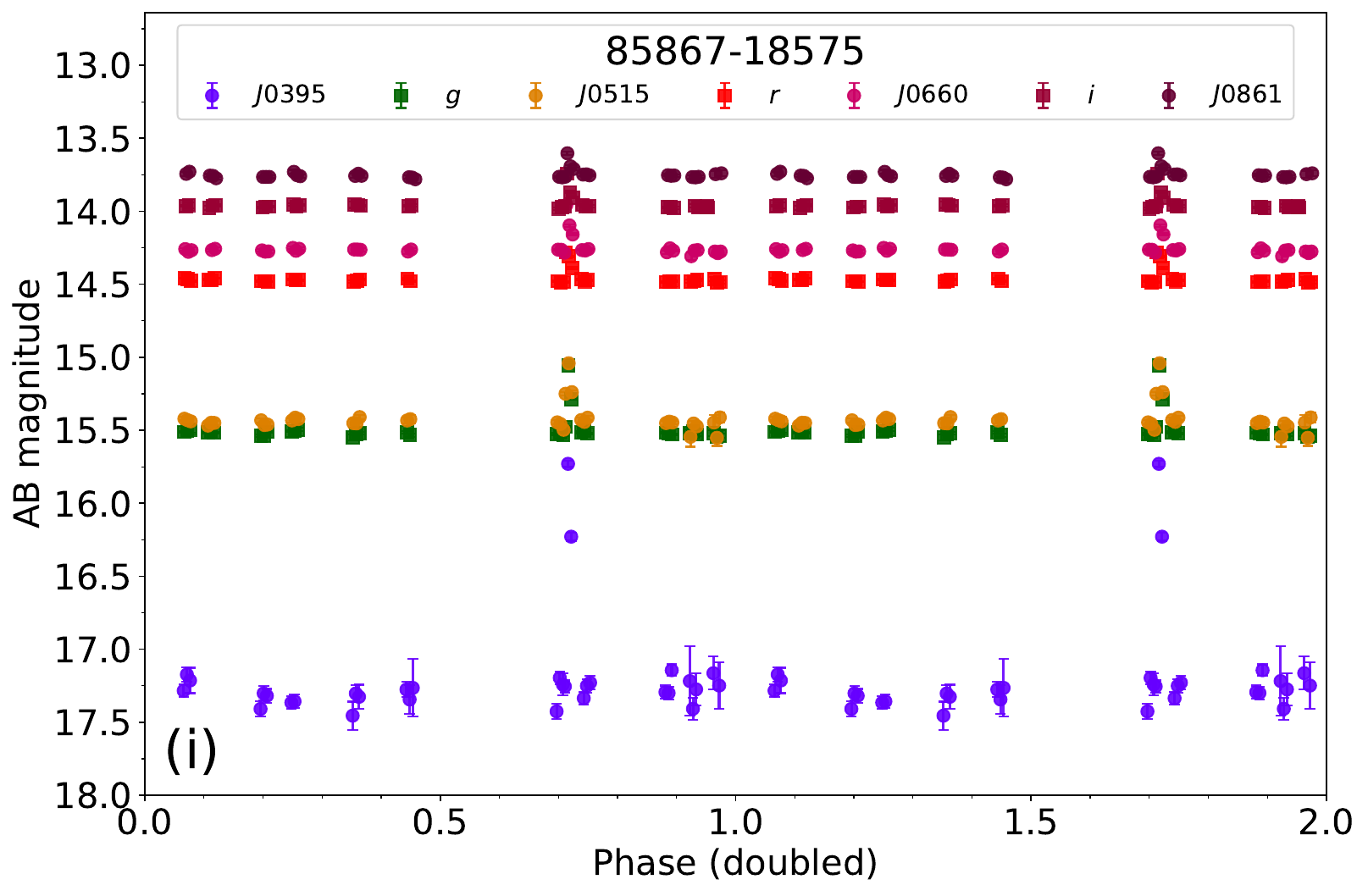}     \\
    \includegraphics[width=0.33\linewidth]{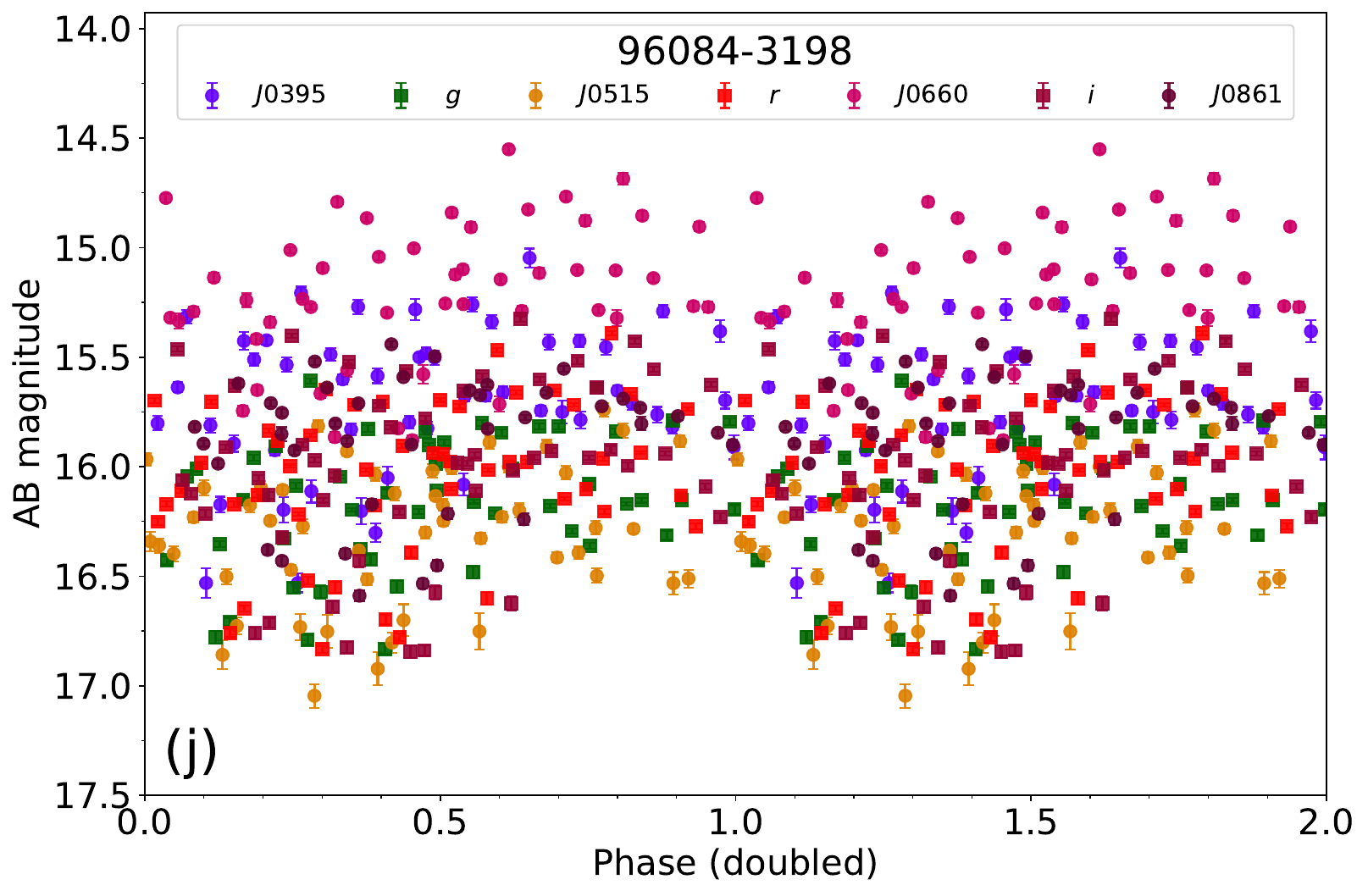}     &
    \includegraphics[width=0.33\linewidth]{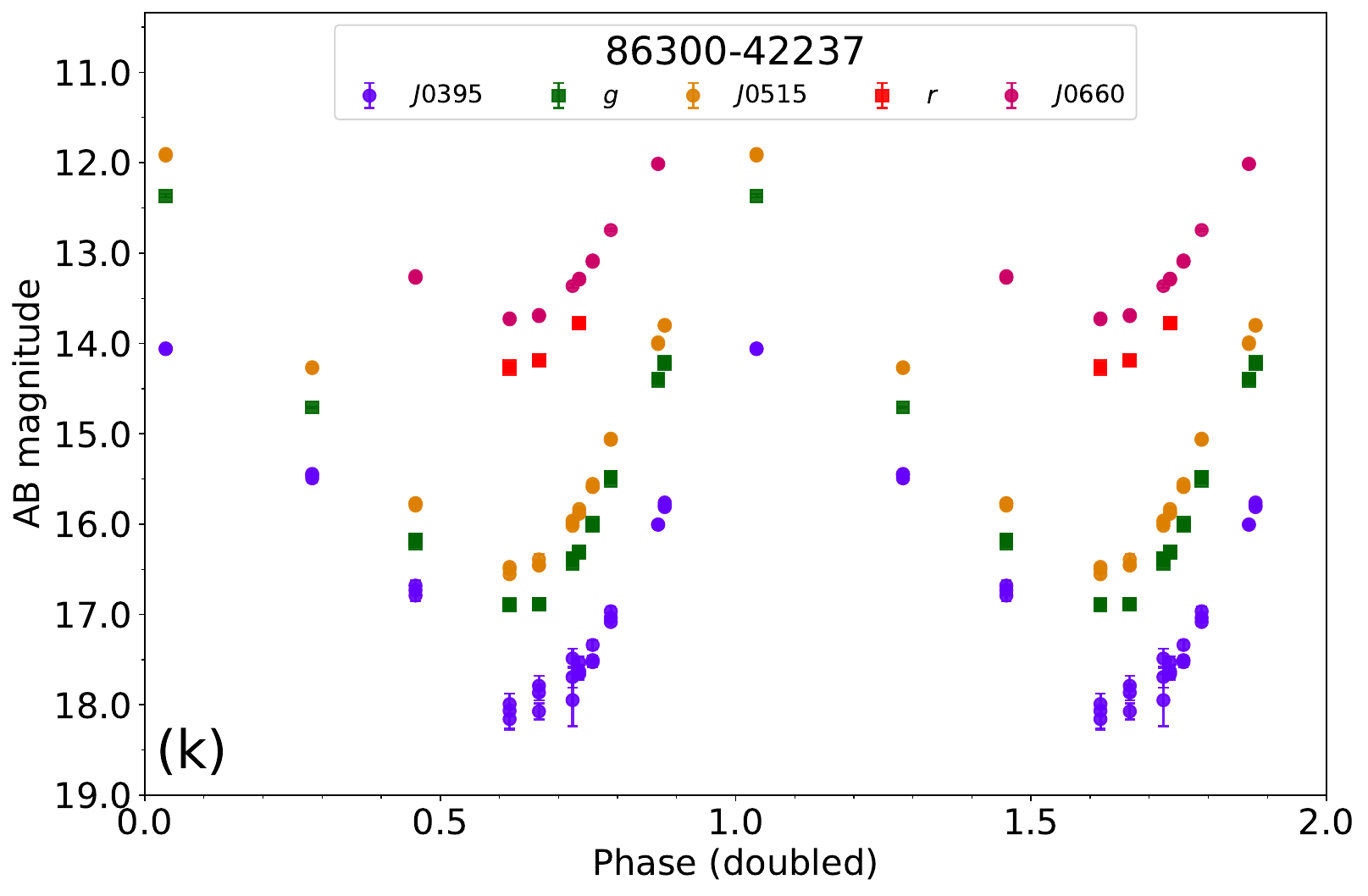} &
    \includegraphics[width=0.33\linewidth]{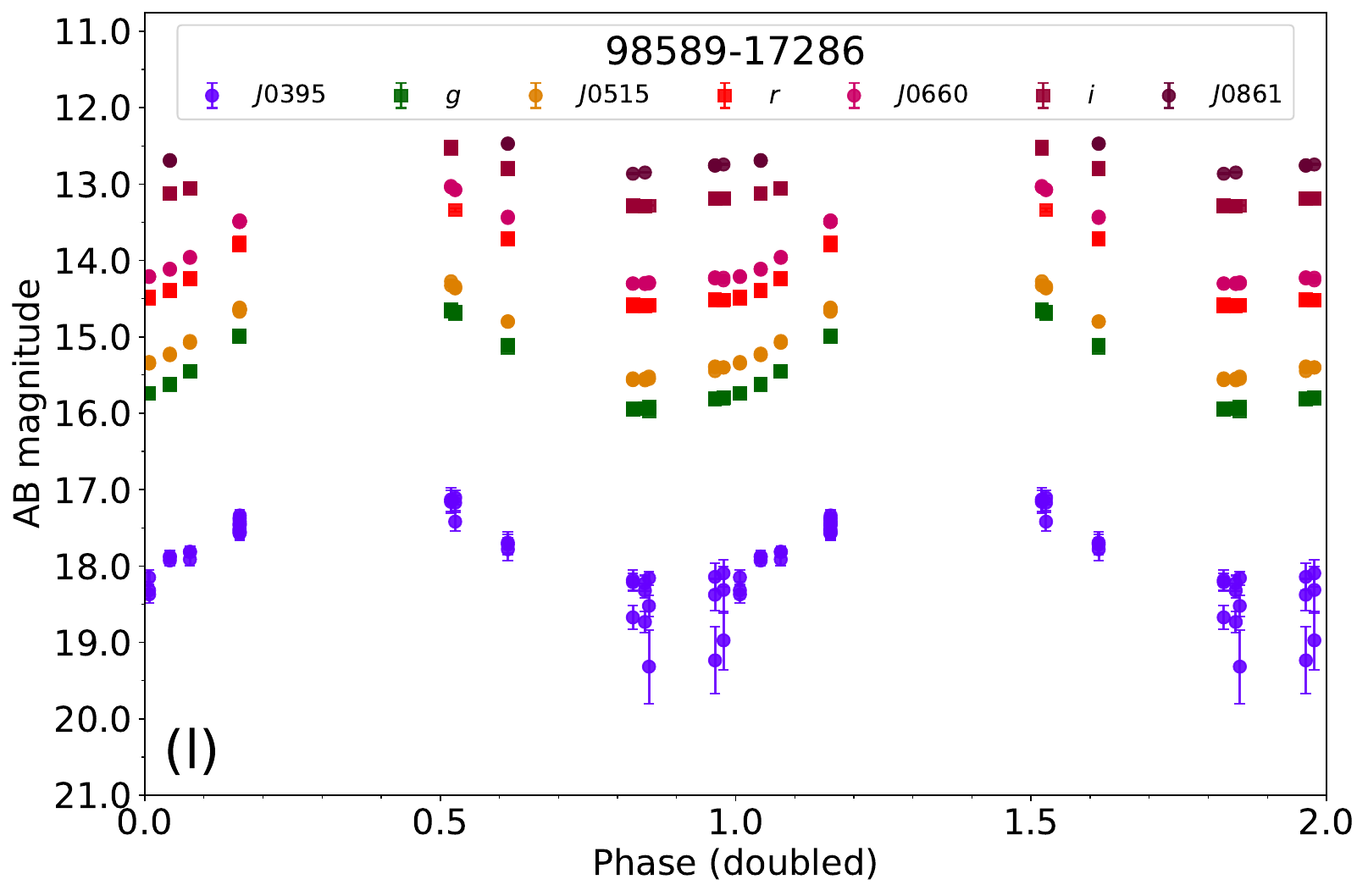} \\
            \end{tabular}
    \caption{Phase-folded light curves of variable stars from J-VAR DR1 with known periods
    , $P$.  Periods are taken either from Kulkarni et al. (submitted, $\star$) or from VSX (\citealt{VSX_Watson2022}; $\dagger$).
    The phase has been duplicated to clearly show the full brightness variation cycle. \textit{From top-left to bottom-right:} \textit{(a)} RRab ($P^{\star}=0.55630\,$d), \textit{(b)} RRc  ($P^{\star}=0.38300\,$d), \textit{(c)} high-amplitude $\delta$ Scuti pulsating stars ($P^{\dagger}=0.04862\,$d), \textit{(d)} EA  ($P^{\dagger}=1.17814\,$d), \textit{(e)} EB  ($P^{\dagger}=8.98140\,$d), and \textit{(f)} EW  ($P^{\dagger}=0.28265\,$d) eclipsing binaries, \textit{(g)} BY Dra  ($P^{\dagger}=8.67666\,$d), \textit{(h)}  RS CVn ($P^{\dagger}=0.17216\,$d), \textit{(i)} RS CVn with a possible flare ($P^{\dagger}=1.64620\,$d), \textit{(j)} CV  ($P^{\dagger}=0.06799\,$d), \textit{(k)} LPV M  ($P^{\dagger}=262.31\,$d), and \textit{(l)} SR   ($P^{\dagger}=143.00\,$d) variables. The \texttt{OBJ\_ID}s are shown in each panel. Light curve generation is described in Pyrzas et al. (submitted).}
    \label{fig:LC_samples}
\end{figure*}

The variety of stellar systems that show changes in luminosity is very rich 
\citep[e.g.][]{Catelan2015}
. The Hertzsprung-Russell diagram in \mbox{Fig.~\ref{fig:HRvar}} shows the locations of the objects whose light curves are shown in 
Fig.~\ref{fig:LC_samples}.
The effective cadence of the observations favours the suitability of the study of some type of variable stars over others. In particular it fits well to the variability time-scales of RR Lyrae stars, Cepheids, high-amplitude $\delta$Sct stars, semi-detached or contact binaries (with short or no interruptions between eclipses), etc. Conversely, the phase coverage is often not good for detached binaries because their eclipses last for a relatively short time in comparison with the whole period, so the probability of capturing both adequately is relatively low. And lastly, long-period variable stars, with periods of hundreds of days, are usually covered by the temporal baseline of J-VAR that typically extends beyond 1 year, although the phase is more irregularly sampled in this case than for shorter period variables mentioned before.

\begin{table}[]
    \caption{Number of most occurring stellar objects (i.e. does not consider AGNs) 
    in {\tt Objects} table based on {\it Gaia} DR3 variability classes.}
    \label{tab:number_of_variables}
    \centering
    \begin{tabular}{c|p{0.4\linewidth}|r}
    \hline
     object  &    number  \\
    \hline\hline
     SOLAR\_LIKE  & Solar-like variability & 4176 \\
            ECL  & Eclipsing binaries & 1165 \\
             RS  & RS Canum Venaticorum  & 1033  \\
DSCT|GDOR|SXPHE  & $\delta$ Scuti / $\gamma$ Doradus/ SX Phoenicis stars  & 901 \\
              S  & Short-timescale  & 494 \\
            LPV  & Long-period variables  & 286 \\
             RR  & RR Lyrae stars  & 315 \\
\hline
    \end{tabular}
\end{table}

\subsubsection{Pulsating stars: RR Lyrae, Cepheids, and $\delta$ Scuti stars}
RR Lyrae stars were a key reference in defining the J-VAR cadence and total number of epochs, as they serve as excellent standard candles for distance determination in the Milky Way and nearby galaxies. The survey’s seven-band photometry allows for refined period determinations, variability classification, and potential metallicity estimates based on color indices. The adopted 11-epoch strategy ensures that light curves sample the pulsation cycle adequately, improving period recovery rates compared to single-band surveys.

In addition to RR Lyrae stars, J-VAR data can also be used to identify and study classical Cepheids and $\delta$ Scuti stars, which provide complementary constraints on stellar evolution and pulsation theory. 
Given that J-VAR primarily targets the Galactic halo, the survey contains only a small number of Cepheids. The multi-band nature of the survey is particularly useful for distinguishing between different pulsation modes and for improving period-luminosity relationships.

\subsubsection{Eclipsing binaries and other periodic variables}
The survey’s multi-epoch observations facilitates the identification and classification of semi-detached (EB) and contact (EW) eclipsing binaries, and in fewer cases, detached (EA) systems. The broad wavelength coverage helps to constrain better the temperature ratios between the components \citep[e.g.][]{kallrath2009},
improving constraints on stellar parameters. Additionally, J-VAR's ability to capture multiple points within an eclipse event in a single observing block increases its potential for detecting and characterizing short-period systems.

Other periodic variables, such as rotationally modulated stars, including RS CVn (RS), chromospherically active rotational binaries, and BY Dra (BY), late-type dwarf stars with starspots and rotational modulation, can also be analysed using J-VAR data. The survey’s multi-band information helps distinguish between RS and BY stars.

\subsubsection{Semi-regular and cataclysmic variables}
The J-VAR cadence also enables the detection of longer-term variations associated with semi-regular and Mira variables, which exhibit large-amplitude, multi-periodic brightness fluctuations. Additionally, the survey’s transient monitoring capabilities may allow for serendipitous detections of eruptive variables, including cataclysmic variables (CVs) undergoing outbursts.

The presence of short intra-night separations in the observing strategy further provides sensitivity to flickering behaviour in CVs and rapidly variable phenomena in compact stellar remnants. Combined with spectroscopic follow-up, these data could contribute to the discovery of rare or previously unknown subclasses of variable stars.

\section{Conclusions and perspectives}
\label{sec:conclusions}

\begin{figure*}
    \centering
    \includegraphics[width=\linewidth]{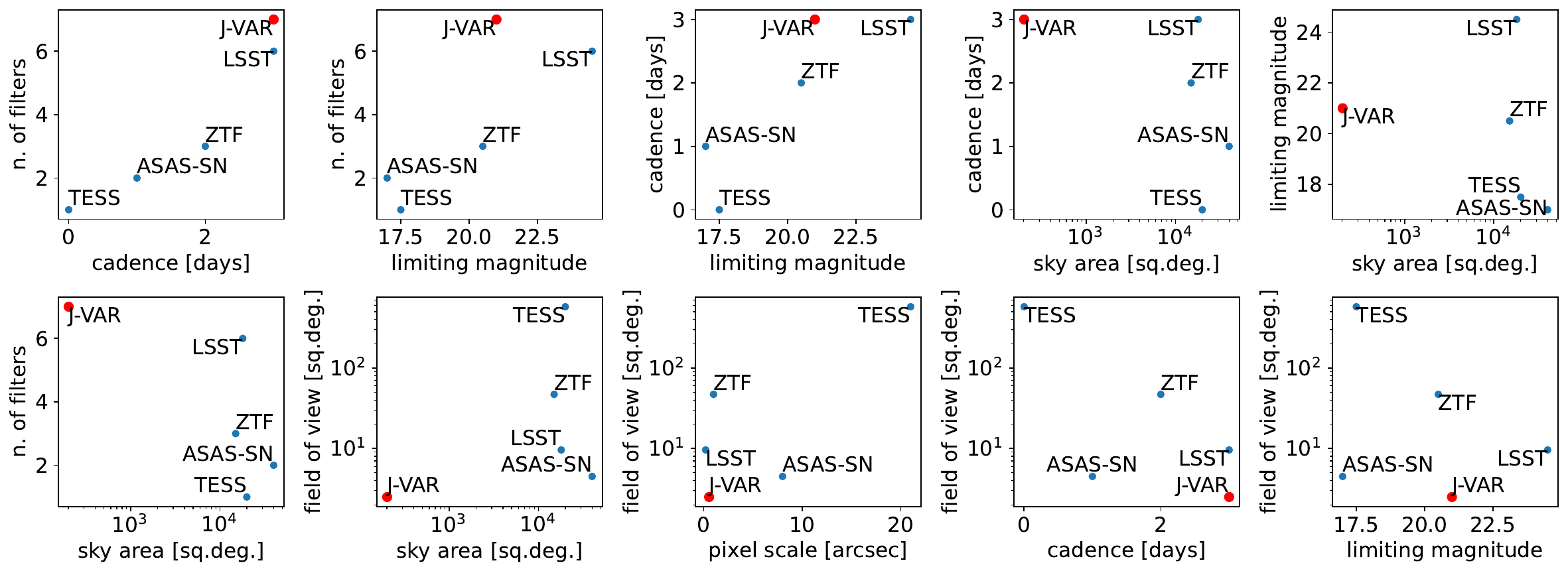}
    \caption{Scatter plots comparing the main characteristics of J-VAR DR1 (highlighted in red) 
    with respect to other
    wide-field photometric surveys.}
    \label{fig:comparison}
\end{figure*}


The parameter space for time-domain wide-field photometric
surveys can be summarized by six factors: (i) survey area -- the amount of sky area where the light curves are distributed, (ii) cadence -- the time interval between two observations of the same targets, (iii) limiting magnitude, (iv) number of filters used to characterize a light curve, (v) pixel scale of the camera, and (vi) field of view\footnote{In this analysis we are not using the etendue 
(the multiplication of the size of the primary mirror by the field of view) or the exposure
time, since they are somewhat embedded in the values that we are considering, in particular limiting
magnitude and cadence}.
The science cases driving the different projects are reflected in how they cover
such parameter space (see Fig.~\ref{fig:comparison}). Since TESS is aiming at transits of extrasolar planets, its
cadence is very short with respect to all other projects. TESS also employs a vast 
field of view and a very large pixel scale. With the noticeable exception of ZTF, 
ground based projects have all comparable fields of view and pixel scales. This is 
mostly due to the need of having an optical system which is adapted to the atmospheric 
conditions of the observing site. The three parameters where there is the largest 
scatter are sky area,
limiting magnitude, and number of filters. Ground-based transient surveys like
ZTF and ASAS-SN cover a very large sky area in a small number of filters as often as
possible. The main difference between these two surveys is the limiting magnitude
(ZTF is carried out with a larger telescope) and the survey area (ZTF is carried out
with one telescope only, thus only covering the Northern celestial hemisphere, 
while ASAS-SN, being a network of telescopes, can survey the
whole sky). The observing strategy of LSST is the result of the trade off between different
projects, which makes the number of filters larger than other surveys and the cadence 
somewhat less frequent, although it has an unrivalled depth.

J-VAR leverages three characteristics of the JAST80 telescope: good image quality, wide field of view, and the narrow band filters. To the best of our knowledge, at the time of beginning the survey (2017), J-VAR was the only wide-field time-domain survey routinely observing with narrow-band filters. This brings J-VAR to occupy a unique position in the phase-space for wide-field time-domain surveys as the cadence
is comparable with the one of LSST, the depth comparable with ZTF, and the filter set is 
unique.

J-VAR is not carried out with a dedicated telescope, unlike other time-domain surveys, and therefore shares observing time with other awarded proposals.
Projects like ASAS-SN and ZTF can observe within a few days all the sky
available from their sites and, therefore, they can, in principle, be extended indefinitely
to increase the length of the light curves. In the case of J-VAR, 
the observation of a field is limited to 11\,epochs. 
The number of completed fields grows at a rate of approximately 58~square degrees 
per year (see  Fig~\ref{fig:evolution}). This has been
roughly constant since the beginning, when J-VAR was carried out as a two-years long
large project at the OAJ, and now when it is a filler among J-PLUS and the
Legacy Surveys at the OAJ. It is likely that J-VAR will continue at the same pace 
until 2027, when the Legacy Surveys will end. At the current rate, J-VAR will reach
1000\,square degrees in 2035, although this could vary in case of changes in the
programs carried out on the JAST80 after 2027.

Adopting a unique strategy leveraging narrow band filters, J-VAR represents an 
unprecedented opportunity for time-domain astrophysics. 
Thanks to the use of narrow-band filters, J-VAR constitutes a bridge between 
spectroscopic time-domain surveys  \citep[e.g., SDSS-V;][]{sdss5}
and wide-field photometric surveys.
J-VAR’s cadence and wavelength coverage make it complementary to large-scale surveys such as {\it Gaia}, ZTF, and LSST. While those surveys provide extensive temporal coverage, J-VAR adds the advantage of quasi simultaneous multi-band observations, aiding in precise classification and improving constraints on variability mechanisms. The multi-filter approach also facilitates cross-matching with existing variability catalogues, enabling further refinement of classification schemes.
The high-frequency mode further extends J-VAR’s scope, enabling intra-night variability studies (asteroids, eclipsing binaries, CVs) not accessible in most wide-field surveys.

J-VAR DR1 contains 1.3 million light curves of 11\,epochs in seven photometric bands with
no selection bias on colour or type. 
By covering a diverse range of variable star types, J-VAR provides a unique dataset for time-domain astrophysics. The combination of multiple bands, strategically spaced epochs, and intra-night sampling opens the door to numerous scientific applications, from refining period-luminosity relations to characterizing new classes of variables. The survey’s contribution to variable star studies will be further enhanced through future data releases and dedicated variability catalogues, ensuring an average of 5000 new multi-band light curves per newly observed
square degree.

\begin{acknowledgements}

We ackowledge support from the Governments of Spain and Arag\'on through their general budgets and the Fondo de Inversiones de Teruel.

The Spanish Ministry of Science and Innovation (MCIN/AEI/10.13039/501100011033) and ERDF, A way of making Europe with grant PID2021-124918NB-C42.

AE acknowledges the financial support from the Spanish Ministry of Science and Innovation and the European Union - NextGenerationEU through the Recovery and Resilience Facility project ICTS-MRR-2021-03-CEFCA.

DM acknowledges the financial support from the European Union - NextGenerationEU through the Recovery and Resilience Facility program Planes Complementarios con las CCAA de Astrof\'{\i}sica y F\'{\i}sica de Altas Energ\'{\i}as - LA4.

SK has received financial support from the Aragonese Government through the Research Groups E16\_23R and the Spanish Ministry of Science and Innovation (MCIN/AEI/10.13039/501100011033 y FEDER, Una manera de hacer Europa) with grant PID2021-124918NB-C41.

AAC acknowledges financial support from the Severo Ochoa grant CEX2021-001131-S funded by
MCIN/AEI/10.13039/501100011033. AAC and DM acknowledge financial support from the Spanish project PID2023-153123NB-I00,
funded by MCIN/AEI.

P.C. acknowledges financial support from the Spanish Virtual Observatory project funded by the Spanish Ministry of Science and Innovation/State Agency of Research MCIN/AEI/10.13039/501100011033 through grant PID2023-146210NB-I00.

F.R.H. acknowledges support from FAPESP grants 2018/21661-9 and 2021/11345-5.

The authors would like to acknowledge the use of the following open-source software packages that were essential in the analysis presented herein: Python, NumPy, Astropy, pandas, matplotlib, and PyVO (including TAP). Their combined functionality enabled efficient data handling, manipulation, visualization, and access to Virtual Observatory services.

\end{acknowledgements}

\bibliographystyle{aa} 
\bibliography{jvar} 

\begin{thebibliography}{47}
\expandafter\ifx\csname natexlab\endcsname\relax\def\natexlab#1{#1}\fi

\bibitem[{{Auvergne} {et~al.}(2009){Auvergne}, {Bodin}, {Boisnard}, {Buey},
  {Chaintreuil}, {Epstein}, {Jouret}, {Lam-Trong}, {Levacher}, {Magnan},
  {Perez}, {Plasson}, {Plesseria}, {Peter}, {Steller}, {Tiph{\`e}ne}, {Baglin},
  {Agogu{\'e}}, {Appourchaux}, {Barbet}, {Beaufort}, {Bellenger}, {Berlin},
  {Bernardi}, {Blouin}, {Boumier}, {Bonneau}, {Briet}, {Butler}, {Cautain},
  {Chiavassa}, {Costes}, {Cuvilho}, {Cunha-Parro}, {de Oliveira Fialho},
  {Decaudin}, {Defise}, {Djalal}, {Docclo}, {Drummond}, {Dupuis}, {Exil},
  {Faur{\'e}}, {Gaboriaud}, {Gamet}, {Gavalda}, {Grolleau}, {Gueguen},
  {Guivarc'h}, {Guterman}, {Hasiba}, {Huntzinger}, {Hustaix}, {Imbert},
  {Jeanville}, {Johlander}, {Jorda}, {Journoud}, {Karioty}, {Kerjean},
  {Lafond}, {Lapeyrere}, {Landiech}, {Larqu{\'e}}, {Laudet}, {Le Merrer},
  {Leporati}, {Leruyet}, {Levieuge}, {Llebaria}, {Martin}, {Mazy}, {Mesnager},
  {Michel}, {Moalic}, {Monjoin}, {Naudet}, {Neukirchner}, {Nguyen-Kim},
  {Ollivier}, {Orcesi}, {Ottacher}, {Oulali}, {Parisot}, {Perruchot},
  {Piacentino}, {Pinheiro da Silva}, {Platzer}, {Pontet}, {Pradines},
  {Quentin}, {Rohbeck}, {Rolland}, {Rollenhagen}, {Romagnan}, {Russ}, {Samadi},
  {Schmidt}, {Schwartz}, {Sebbag}, {Smit}, {Sunter}, {Tello}, {Toulouse},
  {Ulmer}, {Vandermarcq}, {Vergnault}, {Wallner}, {Waultier}, \&
  {Zanatta}}]{corot_ref}
{Auvergne}, M., {Bodin}, P., {Boisnard}, L., {et~al.} 2009, \aap, 506, 411

\bibitem[{{Bellm} {et~al.}(2019){Bellm}, {Kulkarni}, {Graham}, {Dekany},
  {Smith}, {Riddle}, {Masci}, {Helou}, {Prince}, {Adams}, {Barbarino},
  {Barlow}, {Bauer}, {Beck}, {Belicki}, {Biswas}, {Blagorodnova}, {Bodewits},
  {Bolin}, {Brinnel}, {Brooke}, {Bue}, {Bulla}, {Burruss}, {Cenko}, {Chang},
  {Connolly}, {Coughlin}, {Cromer}, {Cunningham}, {De}, {Delacroix}, {Desai},
  {Duev}, {Eadie}, {Farnham}, {Feeney}, {Feindt}, {Flynn}, {Franckowiak},
  {Frederick}, {Fremling}, {Gal-Yam}, {Gezari}, {Giomi}, {Goldstein},
  {Golkhou}, {Goobar}, {Groom}, {Hacopians}, {Hale}, {Henning}, {Ho}, {Hover},
  {Howell}, {Hung}, {Huppenkothen}, {Imel}, {Ip}, {Ivezi{\'c}}, {Jackson},
  {Jones}, {Juric}, {Kasliwal}, {Kaspi}, {Kaye}, {Kelley}, {Kowalski},
  {Kramer}, {Kupfer}, {Landry}, {Laher}, {Lee}, {Lin}, {Lin}, {Lunnan},
  {Giomi}, {Mahabal}, {Mao}, {Miller}, {Monkewitz}, {Murphy}, {Ngeow},
  {Nordin}, {Nugent}, {Ofek}, {Patterson}, {Penprase}, {Porter}, {Rauch},
  {Rebbapragada}, {Reiley}, {Rigault}, {Rodriguez}, {van Roestel}, {Rusholme},
  {van Santen}, {Schulze}, {Shupe}, {Singer}, {Soumagnac}, {Stein}, {Surace},
  {Sollerman}, {Szkody}, {Taddia}, {Terek}, {Van Sistine}, {van Velzen},
  {Vestrand}, {Walters}, {Ward}, {Ye}, {Yu}, {Yan}, \& {Zolkower}}]{ZTF_ref}
{Bellm}, E.~C., {Kulkarni}, S.~R., {Graham}, M.~J., {et~al.} 2019, \pasp, 131,
  018002

\bibitem[{{Benitez} {et~al.}(2014){Benitez}, {Dupke}, {Moles}, {Sodre},
  {Cenarro}, {Marin-Franch}, {Taylor}, {Cristobal}, {Fernandez-Soto}, {Mendes
  de Oliveira}, {Cepa-Nogue}, {Abramo}, {Alcaniz}, {Overzier},
  {Hernandez-Monteagudo}, {Alfaro}, {Kanaan}, {Carvano}, {Reis}, {Martinez
  Gonzalez}, {Ascaso}, {Ballesteros}, {Xavier}, {Varela}, {Ederoclite},
  {Vazquez Ramio}, {Broadhurst}, {Cypriano}, {Angulo}, {Diego}, {Zandivarez},
  {Diaz}, {Melchior}, {Umetsu}, {Spinelli}, {Zitrin}, {Coe}, {Yepes}, {Vielva},
  {Sahni}, {Marcos-Caballero}, {Kitaura}, {Maroto}, {Masip}, {Tsujikawa},
  {Carneiro}, {Gonzalez Nuevo}, {Carvalho}, {Reboucas}, {Carvalho}, {Abdalla},
  {Bernui}, {Pigozzo}, {Ferreira}, {Chandrachani Devi}, {Bengaly}, {Campista},
  {Amorim}, {Asari}, {Bongiovanni}, {Bonoli}, {Bruzual}, {Cardiel}, {Cava},
  {Cid Fernandes}, {Coelho}, {Cortesi}, {Delgado}, {Diaz Garcia}, {Espinosa},
  {Galliano}, {Gonzalez-Serrano}, {Falcon-Barroso}, {Fritz}, {Fernandes},
  {Gorgas}, {Hoyos}, {Jimenez-Teja}, {Lopez-Aguerri}, {Lopez-San Juan},
  {Mateus}, {Molino}, {Novais}, {OMill}, {Oteo}, {Perez-Gonzalez}, {Poggianti},
  {Proctor}, {Ricciardelli}, {Sanchez-Blazquez}, {Storchi-Bergmann}, {Telles},
  {Schoennell}, {Trujillo}, {Vazdekis}, {Viironen}, {Daflon},
  {Aparicio-Villegas}, {Rocha}, {Ribeiro}, {Borges}, {Martins}, {Marcolino},
  {Martinez-Delgado}, {Perez-Torres}, {Siffert}, {Calvao}, {Sako}, {Kessler},
  {Alvarez-Candal}, {De Pra}, {Roig}, {Lazzaro}, {Gorosabel}, {Lopes de
  Oliveira}, {Lima-Neto}, {Irwin}, {Liu}, {Alvarez}, {Balmes}, {Chueca},
  {Costa-Duarte}, {da Costa}, {Dantas}, {Diaz}, {Fabregat}, {Ferrari},
  {Gavela}, {Gracia}, {Gruel}, {Gutierrez}, {Guzman}, {Hernandez-Fernandez},
  {Herranz}, {Hurtado-Gil}, {Jablonsky}, {Laporte}, {Le Tiran}, {Licandro},
  {Lima}, {Martin}, {Martinez}, {Montero}, {Penteado}, {Pereira}, {Peris},
  {Quilis}, {Sanchez-Portal}, {Soja}, {Solano}, {Torra}, \&
  {Valdivielso}}]{jpas}
{Benitez}, N., {Dupke}, R., {Moles}, M., {et~al.} 2014, arXiv e-prints,
  arXiv:1403.5237

\bibitem[{{Bertin} \& {Arnouts}(1996)}]{Bertin1996}
{Bertin}, E. \& {Arnouts}, S. 1996, \aaps, 117, 393

\bibitem[{{Bonoli} {et~al.}(2021){Bonoli}, {Mar{\'\i}n-Franch}, {Varela},
  {V{\'a}zquez Rami{\'o}}, {Abramo}, {Cenarro}, {Dupke}, {V{\'\i}lchez},
  {Crist{\'o}bal-Hornillos}, {Gonz{\'a}lez Delgado},
  {Hern{\'a}ndez-Monteagudo}, {L{\'o}pez-Sanjuan}, {Muniesa}, {Civera},
  {Ederoclite}, {Hern{\'a}n-Caballero}, {Marra}, {Baqui}, {Cortesi},
  {Cypriano}, {Daflon}, {de Amorim}, {D{\'\i}az-Garc{\'\i}a}, {Diego},
  {Mart{\'\i}nez-Solaeche}, {P{\'e}rez}, {Placco}, {Prada}, {Queiroz},
  {Alcaniz}, {Alvarez-Candal}, {Cepa}, {Maroto}, {Roig}, {Siffert}, {Taylor},
  {Benitez}, {Moles}, {Sodr{\'e}}, {Carneiro}, {Mendes de Oliveira}, {Abdalla},
  {Angulo}, {Aparicio Resco}, {Balaguera-Antol{\'\i}nez}, {Ballesteros},
  {Brito-Silva}, {Broadhurst}, {Carrasco}, {Castro}, {Cid Fernandes}, {Coelho},
  {de Melo}, {Doubrawa}, {Fernandez-Soto}, {Ferrari}, {Finoguenov},
  {Garc{\'\i}a-Benito}, {Iglesias-P{\'a}ramo}, {Jim{\'e}nez-Teja}, {Kitaura},
  {Laur}, {Lopes}, {Lucatelli}, {Mart{\'\i}nez}, {Maturi}, {Overzier},
  {Pigozzo}, {Quartin}, {Rodr{\'\i}guez-Mart{\'\i}n}, {Salzano}, {Tamm},
  {Tempel}, {Umetsu}, {Valdivielso}, {von Marttens}, {Zitrin},
  {D{\'\i}az-Mart{\'\i}n}, {L{\'o}pez-Alegre}, {L{\'o}pez-Sainz},
  {Yanes-D{\'\i}az}, {Rueda-Teruel}, {Rueda-Teruel}, {Abril Iba{\~n}ez}, {L
  Ant{\'o}n Bravo}, {Bello Ferrer}, {Bielsa}, {Casino}, {Castillo}, {Chueca},
  {Cuesta}, {Garzar{\'a}n Calderaro}, {Iglesias-Marzoa}, {{\'I}niguez},
  {Lamadrid Gutierrez}, {Lopez-Martinez}, {Lozano-P{\'e}rez}, {Ma{\'\i}cas
  Sacrist{\'a}n}, {Molina-Ib{\'a}{\~n}ez}, {Moreno-Signes}, {Rodr{\'\i}guez
  Llano}, {Royo Navarro}, {Tilve Rua}, {Andrade}, {Alfaro}, {Akras},
  {Arnalte-Mur}, {Ascaso}, {Barbosa}, {Beltr{\'a}n Jim{\'e}nez}, {Benetti},
  {Bengaly}, {Bernui}, {Blanco-Pillado}, {Borges Fernandes}, {Bregman},
  {Bruzual}, {Calderone}, {Carvano}, {Casarini}, {Chaves-Montero},
  {Chies-Santos}, {Coutinho de Carvalho}, {Dimauro}, {Duarte Puertas},
  {Figueruelo}, {Gonz{\'a}lez-Serrano}, {Guerrero}, {Gurung-L{\'o}pez},
  {Herranz}, {Huertas-Company}, {Irwin}, {Izquierdo-Villalba}, {Kanaan},
  {Kehrig}, {Kirkpatrick}, {Lim}, {Lopes}, {Lopes de Oliveira},
  {Marcos-Caballero}, {Mart{\'\i}nez-Delgado}, {Mart{\'\i}nez-Gonz{\'a}lez},
  {Mart{\'\i}nez-Somonte}, {Oliveira}, {Orsi}, {Penna-Lima}, {Reis}, {Spinoso},
  {Tsujikawa}, {Vielva}, {Vitorelli}, {Xia}, {Yuan}, {Arroyo-Polonio},
  {Dantas}, {Galarza}, {Gon{\c{c}}alves}, {Gon{\c{c}}alves}, {Gonzalez},
  {Gonzalez}, {Greisel}, {Jim{\'e}nez-Esteban}, {Landim}, {Lazzaro}, {Magris},
  {Monteiro-Oliveira}, {Pereira}, {Rebou{\c{c}}as}, {Rodriguez-Espinosa},
  {Santos da Costa}, \& {Telles}}]{Bonoli2021}
{Bonoli}, S., {Mar{\'\i}n-Franch}, A., {Varela}, J., {et~al.} 2021, \aap, 653,
  A31

\bibitem[{{Borucki} {et~al.}(2010){Borucki}, {Koch}, {Basri}, {Batalha},
  {Brown}, {Caldwell}, {Caldwell}, {Christensen-Dalsgaard}, {Cochran},
  {DeVore}, {Dunham}, {Dupree}, {Gautier}, {Geary}, {Gilliland}, {Gould},
  {Howell}, {Jenkins}, {Kondo}, {Latham}, {Marcy}, {Meibom}, {Kjeldsen},
  {Lissauer}, {Monet}, {Morrison}, {Sasselov}, {Tarter}, {Boss}, {Brownlee},
  {Owen}, {Buzasi}, {Charbonneau}, {Doyle}, {Fortney}, {Ford}, {Holman},
  {Seager}, {Steffen}, {Welsh}, {Rowe}, {Anderson}, {Buchhave}, {Ciardi},
  {Walkowicz}, {Sherry}, {Horch}, {Isaacson}, {Everett}, {Fischer}, {Torres},
  {Johnson}, {Endl}, {MacQueen}, {Bryson}, {Dotson}, {Haas}, {Kolodziejczak},
  {Van Cleve}, {Chandrasekaran}, {Twicken}, {Quintana}, {Clarke}, {Allen},
  {Li}, {Wu}, {Tenenbaum}, {Verner}, {Bruhweiler}, {Barnes}, \&
  {Prsa}}]{kepler_ref}
{Borucki}, W.~J., {Koch}, D., {Basri}, G., {et~al.} 2010, Science, 327, 977

\bibitem[{Catelan \& Smith(2015)}]{Catelan2015}
Catelan, M. \& Smith, H.~A. 2015, Pulsating stars (John Wiley \& Sons)

\bibitem[{{Cenarro} {et~al.}(2019){Cenarro}, {Moles},
  {Crist{\'o}bal-Hornillos}, {Mar{\'\i}n-Franch}, {Ederoclite}, {Varela},
  {L{\'o}pez-Sanjuan}, {Hern{\'a}ndez-Monteagudo}, {Angulo}, {V{\'a}zquez
  Rami{\'o}}, {Viironen}, {Bonoli}, {Orsi}, {Hurier}, {San Roman}, {Greisel},
  {Vilella-Rojo}, {D{\'\i}az-Garc{\'\i}a}, {Logro{\~n}o-Garc{\'\i}a},
  {Gurung-L{\'o}pez}, {Spinoso}, {Izquierdo-Villalba}, {Aguerri}, {Allende
  Prieto}, {Bonatto}, {Carvano}, {Chies-Santos}, {Daflon}, {Dupke},
  {Falc{\'o}n-Barroso}, {Gon{\c{c}}alves}, {Jim{\'e}nez-Teja}, {Molino},
  {Placco}, {Solano}, {Whitten}, {Abril}, {Ant{\'o}n}, {Bello}, {Bielsa de
  Toledo}, {Castillo-Ram{\'\i}rez}, {Chueca}, {Civera},
  {D{\'\i}az-Mart{\'\i}n}, {Dom{\'\i}nguez-Mart{\'\i}nez},
  {Garzar{\'a}n-Calderaro}, {Hern{\'a}ndez-Fuertes}, {Iglesias-Marzoa},
  {I{\~n}iguez}, {Jim{\'e}nez Ruiz}, {Kruuse}, {Lamadrid}, {Lasso-Cabrera},
  {L{\'o}pez-Alegre}, {L{\'o}pez-Sainz}, {Ma{\'\i}cas}, {Moreno-Signes},
  {Muniesa}, {Rodr{\'\i}guez-Llano}, {Rueda-Teruel}, {Rueda-Teruel},
  {Soriano-Lagu{\'\i}a}, {Tilve}, {Valdivielso}, {Yanes-D{\'\i}az}, {Alcaniz},
  {Mendes de Oliveira}, {Sodr{\'e}}, {Coelho}, {Lopes de Oliveira}, {Tamm},
  {Xavier}, {Abramo}, {Akras}, {Alfaro}, {Alvarez-Candal}, {Ascaso}, {Beasley},
  {Beers}, {Borges Fernandes}, {Bruzual}, {Buzzo}, {Carrasco}, {Cepa},
  {Cortesi}, {Costa-Duarte}, {De Pr{\'a}}, {Favole}, {Galarza}, {Galbany},
  {Garcia}, {Gonz{\'a}lez Delgado}, {Gonz{\'a}lez-Serrano},
  {Guti{\'e}rrez-Soto}, {Hernandez-Jimenez}, {Kanaan}, {Kuncarayakti},
  {Landim}, {Laur}, {Licandro}, {Lima Neto}, {Lyman}, {Ma{\'\i}z
  Apell{\'a}niz}, {Miralda-Escud{\'e}}, {Morate}, {Nogueira-Cavalcante},
  {Novais}, {Oncins}, {Oteo}, {Overzier}, {Pereira}, {Rebassa-Mansergas},
  {Reis}, {Roig}, {Sako}, {Salvador-Rusi{\~n}ol}, {Sampedro},
  {S{\'a}nchez-Bl{\'a}zquez}, {Santos}, {Schmidtobreick}, {Siffert}, {Telles},
  \& {Vilchez}}]{jplus}
{Cenarro}, A.~J., {Moles}, M., {Crist{\'o}bal-Hornillos}, D., {et~al.} 2019,
  \aap, 622, A176

\bibitem[{{Cenarro} {et~al.}(2014){Cenarro}, {Moles}, {Mar{\'\i}n-Franch},
  {Crist{\'o}bal-Hornillos}, {Yanes D{\'\i}az}, {Ederoclite}, {Varela},
  {V{\'a}zquez Rami{\'o}}, {Valdivielso}, {Ben{\'\i}tez}, {Cepa}, {Dupke},
  {Fern{\'a}ndez-Soto}, {Mendes de Oliveira}, {Sodr{\'e}}, {Taylor},
  {Rueda-Teruel}, {Rueda-Teruel}, {Luis-Simoes}, {Chueca}, {Ant{\'o}n},
  {Bello}, {D{\'\i}az-Mart{\'\i}n}, {Guill{\'e}n-Civera},
  {Hern{\'a}ndez-Fuertes}, {Iglesias-Marzoa}, {Jim{\'e}nez-Mej{\'\i}as},
  {Lasso-Cabrera}, {L{\'o}pez-Alegre}, {L{\'o}pez-Sainz},
  {Rodr{\'\i}guez-Hern{\'a}ndez}, {Su{\'a}rez}, {Lamadrid}, {Ma{\'\i}cas},
  {Abril-Iba{\~n}ez}, {Tilve}, \& {Rodr{\'\i}guez-Llano}}]{oaj}
{Cenarro}, A.~J., {Moles}, M., {Mar{\'\i}n-Franch}, A., {et~al.} 2014, in
  Society of Photo-Optical Instrumentation Engineers (SPIE) Conference Series,
  Vol. 9149, Observatory Operations: Strategies, Processes, and Systems V, ed.
  A.~B. {Peck}, C.~R. {Benn}, \& R.~L. {Seaman}, 91491I

\bibitem[{{Clementini} {et~al.}(2023){Clementini}, {Ripepi}, {Garofalo},
  {Molinaro}, {Muraveva}, {Leccia}, {Rimoldini}, {Holl}, {Jevardat de
  Fombelle}, {Sartoretti}, {Marchal}, {Audard}, {Nienartowicz}, {Andrae},
  {Marconi}, {Szabados}, {Evans}, {Lecoeur-Taibi}, {Mowlavi}, {Musella}, \&
  {Eyer}}]{Clementini2023}
{Clementini}, G., {Ripepi}, V., {Garofalo}, A., {et~al.} 2023, \aap, 674, A18

\bibitem[{{del Pino} {et~al.}(2024){del Pino}, {L{\'o}pez-Sanjuan},
  {Hern{\'a}n-Caballero}, {Dom{\'\i}nguez-S{\'a}nchez}, {von Marttens},
  {Fern{\'a}ndez-Ontiveros}, {Coelho}, {Lumbreras-Calle}, {Vega-Ferrero},
  {Jimenez-Esteban}, {Cruz}, {Marra}, {Quartin}, {Galarza}, {Angulo},
  {Cenarro}, {Crist{\'o}bal-Hornillos}, {Dupke}, {Ederoclite},
  {Hern{\'a}ndez-Monteagudo}, {Mar{\'\i}n-Franch}, {Moles}, {Sodr{\'e}},
  {Varela}, \& {V{\'a}zquez Rami{\'o}}}]{delPino2024}
{del Pino}, A., {L{\'o}pez-Sanjuan}, C., {Hern{\'a}n-Caballero}, A., {et~al.}
  2024, \aap, 691, A221

\bibitem[{{DeMeo} \& {Carry}(2013)}]{demeo+carry}
{DeMeo}, F.~E. \& {Carry}, B. 2013, \icarus, 226, 723

\bibitem[{{Drake} {et~al.}(2009){Drake}, {Djorgovski}, {Mahabal}, {Beshore},
  {Larson}, {Graham}, {Williams}, {Christensen}, {Catelan}, {Boattini},
  {Gibbs}, {Hill}, \& {Kowalski}}]{CRTS_ref}
{Drake}, A.~J., {Djorgovski}, S.~G., {Mahabal}, A., {et~al.} 2009, \apj, 696,
  870

\bibitem[{{Everett} \& {Howell}(2001)}]{Everett+Howell2001}
{Everett}, M.~E. \& {Howell}, S.~B. 2001, \pasp, 113, 1428

\bibitem[{{Gaia Collaboration} {et~al.}(2023){Gaia Collaboration}, {Vallenari},
  {Brown}, {Prusti}, {de Bruijne}, {Arenou}, {Babusiaux}, {Biermann},
  {Creevey}, {Ducourant}, {Evans}, {Eyer}, {Guerra}, {Hutton}, {Jordi},
  {Klioner}, {Lammers}, {Lindegren}, {Luri}, {Mignard}, {Panem}, {Pourbaix},
  {Randich}, {Sartoretti}, {Soubiran}, {Tanga}, {Walton}, {Bailer-Jones},
  {Bastian}, {Drimmel}, {Jansen}, {Katz}, {Lattanzi}, {van Leeuwen}, {Bakker},
  {Cacciari}, {Casta{\~n}eda}, {De Angeli}, {Fabricius}, {Fouesneau},
  {Fr{\'e}mat}, {Galluccio}, {Guerrier}, {Heiter}, {Masana}, {Messineo},
  {Mowlavi}, {Nicolas}, {Nienartowicz}, {Pailler}, {Panuzzo}, {Riclet}, {Roux},
  {Seabroke}, {Sordo}, {Th{\'e}venin}, {Gracia-Abril}, {Portell}, {Teyssier},
  {Altmann}, {Andrae}, {Audard}, {Bellas-Velidis}, {Benson}, {Berthier},
  {Blomme}, {Burgess}, {Busonero}, {Busso}, {C{\'a}novas}, {Carry}, {Cellino},
  {Cheek}, {Clementini}, {Damerdji}, {Davidson}, {de Teodoro}, {Nu{\~n}ez
  Campos}, {Delchambre}, {Dell'Oro}, {Esquej}, {Fern{\'a}ndez-Hern{\'a}ndez},
  {Fraile}, {Garabato}, {Garc{\'\i}a-Lario}, {Gosset}, {Haigron}, {Halbwachs},
  {Hambly}, {Harrison}, {Hern{\'a}ndez}, {Hestroffer}, {Hodgkin}, {Holl},
  {Jan{\ss}en}, {Jevardat de Fombelle}, {Jordan}, {Krone-Martins}, {Lanzafame},
  {L{\"o}ffler}, {Marchal}, {Marrese}, {Moitinho}, {Muinonen}, {Osborne},
  {Pancino}, {Pauwels}, {Recio-Blanco}, {Reyl{\'e}}, {Riello}, {Rimoldini},
  {Roegiers}, {Rybizki}, {Sarro}, {Siopis}, {Smith}, {Sozzetti}, {Utrilla},
  {van Leeuwen}, {Abbas}, {{\'A}brah{\'a}m}, {Abreu Aramburu}, {Aerts},
  {Aguado}, {Ajaj}, {Aldea-Montero}, {Altavilla}, {{\'A}lvarez}, {Alves},
  {Anders}, {Anderson}, {Anglada Varela}, {Antoja}, {Baines}, {Baker},
  {Balaguer-N{\'u}{\~n}ez}, {Balbinot}, {Balog}, {Barache}, {Barbato},
  {Barros}, {Barstow}, {Bartolom{\'e}}, {Bassilana}, {Bauchet}, {Becciani},
  {Bellazzini}, {Berihuete}, {Bernet}, {Bertone}, {Bianchi}, {Binnenfeld},
  {Blanco-Cuaresma}, {Blazere}, {Boch}, {Bombrun}, {Bossini}, {Bouquillon},
  {Bragaglia}, {Bramante}, {Breedt}, {Bressan}, {Brouillet}, {Brugaletta},
  {Bucciarelli}, {Burlacu}, {Butkevich}, {Buzzi}, {Caffau}, {Cancelliere},
  {Cantat-Gaudin}, {Carballo}, {Carlucci}, {Carnerero}, {Carrasco},
  {Casamiquela}, {Castellani}, {Castro-Ginard}, {Chaoul}, {Charlot}, {Chemin},
  {Chiaramida}, {Chiavassa}, {Chornay}, {Comoretto}, {Contursi}, {Cooper},
  {Cornez}, {Cowell}, {Crifo}, {Cropper}, {Crosta}, {Crowley}, {Dafonte},
  {Dapergolas}, {David}, {David}, {de Laverny}, {De Luise}, \& {De
  March}}]{GaiaDR3}
{Gaia Collaboration}, {Vallenari}, A., {Brown}, A.~G.~A., {et~al.} 2023, \aap,
  674, A1

\bibitem[{Gal-Yam(2019)}]{annurev:/content/journals/10.1146/annurev-astro-081817-051819}
Gal-Yam, A. 2019, Annual Review of Astronomy and Astrophysics, 57, 305

\bibitem[{{Hodapp} {et~al.}(2004){Hodapp}, {Kaiser}, {Aussel}, {Burgett},
  {Chambers}, {Chun}, {Dombeck}, {Douglas}, {Hafner}, {Heasley}, {Hoblitt},
  {Hude}, {Isani}, {Jedicke}, {Jewitt}, {Laux}, {Luppino}, {Lupton}, {Maberry},
  {Magnier}, {Mannery}, {Monet}, {Morgan}, {Onaka}, {Price}, {Ryan},
  {Siegmund}, {Szapudi}, {Tonry}, {Wainscoat}, \& {Waterson}}]{pansta2004AN}
{Hodapp}, K.~W., {Kaiser}, N., {Aussel}, H., {et~al.} 2004, Astronomische
  Nachrichten, 325, 636

\bibitem[{{Hubble}(1929)}]{hubble}
{Hubble}, E. 1929, PNAS, 15, 168

\bibitem[{{Ivezi{\'c}} {et~al.}(2019{\natexlab{a}}){Ivezi{\'c}}, {Kahn},
  {Tyson}, {Abel}, {Acosta}, {Allsman}, {Alonso}, {AlSayyad}, {Anderson},
  {Andrew}, {Angel}, {Angeli}, {Ansari}, {Antilogus}, {Araujo}, {Armstrong},
  {Arndt}, {Astier}, {Aubourg}, {Auza}, {Axelrod}, {Bard}, {Barr}, {Barrau},
  {Bartlett}, {Bauer}, {Bauman}, {Baumont}, {Bechtol}, {Bechtol}, {Becker},
  {Becla}, {Beldica}, {Bellavia}, {Bianco}, {Biswas}, {Blanc}, {Blazek},
  {Blandford}, {Bloom}, {Bogart}, {Bond}, {Booth}, {Borgland}, {Borne},
  {Bosch}, {Boutigny}, {Brackett}, {Bradshaw}, {Brandt}, {Brown}, {Bullock},
  {Burchat}, {Burke}, {Cagnoli}, {Calabrese}, {Callahan}, {Callen}, {Carlin},
  {Carlson}, {Chandrasekharan}, {Charles-Emerson}, {Chesley}, {Cheu}, {Chiang},
  {Chiang}, {Chirino}, {Chow}, {Ciardi}, {Claver}, {Cohen-Tanugi}, {Cockrum},
  {Coles}, {Connolly}, {Cook}, {Cooray}, {Covey}, {Cribbs}, {Cui}, {Cutri},
  {Daly}, {Daniel}, {Daruich}, {Daubard}, {Daues}, {Dawson}, {Delgado},
  {Dellapenna}, {de Peyster}, {de Val-Borro}, {Digel}, {Doherty}, {Dubois},
  {Dubois-Felsmann}, {Durech}, {Economou}, {Eifler}, {Eracleous}, {Emmons},
  {Fausti Neto}, {Ferguson}, {Figueroa}, {Fisher-Levine}, {Focke}, {Foss},
  {Frank}, {Freemon}, {Gangler}, {Gawiser}, {Geary}, {Gee}, {Geha}, {Gessner},
  {Gibson}, {Gilmore}, {Glanzman}, {Glick}, {Goldina}, {Goldstein}, {Goodenow},
  {Graham}, {Gressler}, {Gris}, {Guy}, {Guyonnet}, {Haller}, {Harris},
  {Hascall}, {Haupt}, {Hernandez}, {Herrmann}, {Hileman}, {Hoblitt}, {Hodgson},
  {Hogan}, {Howard}, {Huang}, {Huffer}, {Ingraham}, {Innes}, {Jacoby}, {Jain},
  {Jammes}, {Jee}, {Jenness}, {Jernigan}, {Jevremovi{\'c}}, {Johns}, {Johnson},
  {Johnson}, {Jones}, {Juramy-Gilles}, {Juri{\'c}}, {Kalirai}, {Kallivayalil},
  {Kalmbach}, {Kantor}, {Karst}, {Kasliwal}, {Kelly}, {Kessler}, {Kinnison},
  {Kirkby}, {Knox}, {Kotov}, {Krabbendam}, {Krughoff}, {Kub{\'a}nek},
  {Kuczewski}, {Kulkarni}, {Ku}, {Kurita}, {Lage}, {Lambert}, {Lange},
  {Langton}, {Le Guillou}, {Levine}, {Liang}, {Lim}, {Lintott}, {Long},
  {Lopez}, {Lotz}, {Lupton}, {Lust}, {MacArthur}, {Mahabal}, {Mandelbaum},
  {Markiewicz}, {Marsh}, {Marshall}, {Marshall}, {May}, {McKercher}, {McQueen},
  {Meyers}, {Migliore}, {Miller}, \& {Mills}}]{lsst}
{Ivezi{\'c}}, {\v{Z}}., {Kahn}, S.~M., {Tyson}, J.~A., {et~al.}
  2019{\natexlab{a}}, \apj, 873, 111

\bibitem[{{Ivezi{\'c}} {et~al.}(2019{\natexlab{b}}){Ivezi{\'c}}, {Kahn},
  {Tyson}, {Abel}, {Acosta}, {Allsman}, {Alonso}, {AlSayyad}, {Anderson},
  {Andrew}, \& et~al.}]{ivezic2019ApJ}
{Ivezi{\'c}}, {\v Z}., {Kahn}, S.~M., {Tyson}, J.~A., {et~al.}
  2019{\natexlab{b}}, \apj, 873, 111

\bibitem[{{Ivezi{\'c}} {et~al.}(2001){Ivezi{\'c}}, {Tabachnik}, {Rafikov},
  {Lupton}, {Quinn}, {Hammergren}, {Eyer}, {Chu}, {Armstrong}, {Fan},
  {Finlator}, {Geballe}, {Gunn}, {Hennessy}, {Knapp}, {Leggett}, {Munn},
  {Pier}, {Rockosi}, {Schneider}, {Strauss}, {Yanny}, {Brinkmann}, {Csabai},
  {Hindsley}, {Kent}, {Lamb}, {Margon}, {McKay}, {Smith}, {Waddel}, {York}, \&
  {SDSS Collaboration}}]{sso_sdss}
{Ivezi{\'c}}, {\v{Z}}., {Tabachnik}, S., {Rafikov}, R., {et~al.} 2001, \aj,
  122, 2749

\bibitem[{Kallrath \& Milone(2009)}]{kallrath2009}
Kallrath, J. \& Milone, E.~F. 2009, Eclipsing Binary Stars: Modeling and
  Analysis, 2nd edn. (New York: Springer-Verlag)

\bibitem[{Knox \& Millea(2020)}]{PhysRevD.101.043533}
Knox, L. \& Millea, M. 2020, Phys. Rev. D, 101, 043533

\bibitem[{{Kollmeier} {et~al.}(2025){Kollmeier}, {Rix}, {Aerts}, {Aird},
  {Alfaro}, {Almeida}, {Anderson}, {Jim{\'e}nez Arranz}, {Arseneau}, {Assef},
  {Aviram}, {Aydar}, {Badenes}, {Bandyopadhyay}, {Barger}, {Barkhouser},
  {Bauer}, {Bender}, {Besser}, {Bhattarai}, {Bilgi}, {Bird}, {Bizyaev},
  {Blanc}, {Blanton}, {Bochanski}, {Bovy}, {Brandon}, {Brandt}, {Brownstein},
  {Buchner}, {Burchett}, {Carlberg}, {Casey}, {Castaneda-Carlos},
  {Chakraborty}, {Chanam{\'e}}, {Chandra}, {Cherinka}, {Chilingarian},
  {Comparat}, {Cosens}, {Covey}, {Crane}, {Crumpler}, {Cunha}, {Cunningham},
  {Dai}, {Darling}, {Davidson}, {Davis}, {De Lee}, {Deacon}, {M{\'e}ndez
  Delgado}, {Demasi}, {Demianenko}, {Derwent}, {D'Onghia}, {Di Mille}, {Dias},
  {Donor}, {Drory}, {Dwelly}, {Egorov}, {Egorova}, {El-Badry}, {Engelman},
  {Eracleous}, {Fan}, {Farr}, {Fries}, {Frinchaboy}, {Froning}, {G{\"a}nsicke},
  {Garc{\'\i}a}, {Gelfand}, {Gentile Fusillo}, {Glover}, {Grabowski}, {Grebel},
  {Green}, {Grier}, {Gupta}, {Gray}, {H{\"a}berle}, {Hall}, {Hammond},
  {Hawkins}, {Harding}, {Heged{\H{u}}s}, {Herbst}, {Hermes}, {Rodr{\'\i}guez
  Hidalgo}, {Hilder}, {Hogg}, {Holtzman}, {Horta}, {Huang}, {Hwang},
  {Ibarra-Medel}, {Imig}, {Inight}, {Jana}, {Ji}, {Jofre}, {Johns}, {Johnson},
  {Johnson}, {Johnston}, {Jones}, {Katkov}, {Koekemoer}, {Kounkel}, {Kreckel},
  {Krishnarao}, {Krumpe}, {Kumari}, {Kupfer}, {Lacerna}, {Laporte}, {Lepine},
  {Li}, {Liu}, {Loebman}, {Long}, {Roman-Lopes}, {Lu}, {Majewski}, {Maoz},
  {McKinnon}, {Medan}, {Merloni}, {Minniti}, {Morrison}, {Myers},
  {M{\'e}sz{\'a}ros}, {Nandra}, {Nayak}, {Ness}, {Nidever}, {O'Brien}, {Oeur},
  {Oravetz}, {Oravetz}, {Otto}, {Adamane Pallathadka}, {Palunas}, {Pan},
  {Pappalardo}, {Pandey}, {Negrete Pe{\~n}aloza}, {Pinsonneault}, {Pogge},
  {Taghizadeh Popp}, {Price-Whelan}, {Pulatova}, {Qiu}, {Ramirez}, {Rankine},
  {Ricci}, {Runnoe}, {Sanchez}, {Salvato}, {Sattler}, {Saydjari}, {Sayres},
  {Schlaufman}, {Schneider}, {Schreiber}, {Schwope}, {Serna}, {Shen},
  {Sif{\'o}n}, {Singh}, {Sinha}, {Smee}, {Song}, {Souto}, {Stassun},
  {Steinmetz}, {Stone-Martinez}, {Stringfellow}, {Stutz}, {Jos{\'e}}, {S{\'a}},
  {nchez-Gallego}, {Tan}, {Tayar}, {Thai}, {Thakar}, {Ting}, {Tkachenko},
  {Tovmasian}, {Trakhtenbrot}, {Fern{\'a}ndez-Trincado}, {Troup}, {Trump},
  {Tuttle}, {van der Marel}, \& {Villanova}}]{sdss5}
{Kollmeier}, J.~A., {Rix}, H.-W., {Aerts}, C., {et~al.} 2025, arXiv e-prints,
  arXiv:2507.06989

\bibitem[{{Leavitt} \& {Pickering}(1912)}]{leavitt}
{Leavitt}, H.~S. \& {Pickering}, E.~C. 1912, "Harvard College Observatory
  Circular", 173, 1

\bibitem[{{Liu} {et~al.}(2023){Liu}, {R{\"o}pke}, \&
  {Han}}]{2023RAA....23h2001L}
{Liu}, Z.-W., {R{\"o}pke}, F.~K., \& {Han}, Z. 2023, Research in Astronomy and
  Astrophysics, 23, 082001

\bibitem[{{L{\'o}pez-Sanjuan} {et~al.}(2019){L{\'o}pez-Sanjuan}, {V{\'a}zquez
  Rami{\'o}}, {Varela}, {Spinoso}, {Angulo}, {Muniesa}, {Viironen},
  {Crist{\'o}bal-Hornillos}, {Cenarro}, {Ederoclite}, {Mar{\'\i}n-Franch},
  {Moles}, {Ascaso}, {Bonoli}, {Chies-Santos}, {Coelho}, {Costa-Duarte},
  {Cortesi}, {D{\'\i}az-Garc{\'\i}a}, {Dupke}, {Galbany},
  {Hern{\'a}ndez-Monteagudo}, {Logro{\~n}o-Garc{\'\i}a}, {Molino}, {Orsi},
  {Placco}, {Sampedro}, {San Roman}, {Vilella-Rojo}, {Whitten}, {Mendes de
  Oliveira}, \& {Sodr{\'e}}}]{Lopez-Sanjuan2019}
{L{\'o}pez-Sanjuan}, C., {V{\'a}zquez Rami{\'o}}, H., {Varela}, J., {et~al.}
  2019, \aap, 622, A177

\bibitem[{{L{\'o}pez-Sanjuan} {et~al.}(2024){L{\'o}pez-Sanjuan}, {V{\'a}zquez
  Rami{\'o}}, {Xiao}, {Yuan}, {Carrasco}, {Varela}, {Crist{\'o}bal-Hornillos},
  {Tremblay}, {Ederoclite}, {Mar{\'\i}n-Franch}, {Cenarro}, {Coelho}, {Daflon},
  {del Pino}, {Dom{\'\i}nguez S{\'a}nchez}, {Fern{\'a}ndez-Ontiveros},
  {Hern{\'a}n-Caballero}, {Jim{\'e}nez-Esteban}, {Alcaniz}, {Angulo}, {Dupke},
  {Hern{\'a}ndez-Monteagudo}, {Moles}, \& {Sodr{\'e}}}]{Lopez-Sanjuan2024}
{L{\'o}pez-Sanjuan}, C., {V{\'a}zquez Rami{\'o}}, H., {Xiao}, K., {et~al.}
  2024, \aap, 683, A29

\bibitem[{{Mahlke} {et~al.}(2018){Mahlke}, {Bouy}, {Altieri}, {Verdoes Kleijn},
  {Carry}, {Bertin}, {de Jong}, {Kuijken}, {McFarland}, \&
  {Valentijn}}]{Mahlke+2018}
{Mahlke}, M., {Bouy}, H., {Altieri}, B., {et~al.} 2018, \aap, 610, A21

\bibitem[{{Mahlke} {et~al.}(2019){Mahlke}, {Solano}, {Bouy}, {Carry}, {Verdoes
  Kleijn}, \& {Bertin}}]{soss-pip2019}
{Mahlke}, M., {Solano}, E., {Bouy}, H., {et~al.} 2019, Astronomy and Computing,
  28, 100289

\bibitem[{{Ma{\'\i}z Apell{\'a}niz} {et~al.}(2021){Ma{\'\i}z Apell{\'a}niz},
  {Alfaro}, {Barb{\'a}}, {Holgado}, {V{\'a}zquez Rami{\'o}}, {Varela},
  {Ederoclite}, {Lorenzo-Guti{\'e}rrez}, {Garc{\'\i}a-Lario}, {Garc{\'\i}a
  Escudero}, {Garc{\'\i}a}, \& {Coelho}}]{galante}
{Ma{\'\i}z Apell{\'a}niz}, J., {Alfaro}, E.~J., {Barb{\'a}}, R.~H., {et~al.}
  2021, \mnras, 506, 3138

\bibitem[{{Mar\'{\i}n-Franch} {et~al.}(2015){Mar\'{\i}n-Franch}, {Taylor},
  {Cenarro}, {Cristobal-Hornillos}, \& {Moles}}]{Marin-Franch2015}
{Mar\'{\i}n-Franch}, A., {Taylor}, K., {Cenarro}, J., {Cristobal-Hornillos},
  D., \& {Moles}, M. 2015, in IAU General Assembly, Vol.~29, 2257381

\bibitem[{{Mar\'{\i}n-Franch} {et~al.}(2012){Mar\'{\i}n-Franch}, {Taylor},
  {Cepa}, {Laporte}, {Cenarro}, {Chueca}, {Cristobal-Hornillos}, {Ederoclite},
  {Gruel}, {Hern{\'a}ndez-Fuertes}, {L{\'o}pez-Sainz}, {Luis-Simoes}, {Moles},
  {Rueda-Teruel}, {Rueda-Teruel}, {Varela}, {Yanes-D{\'\i}az}, {Benitez},
  {Dupke}, {Fern{\'a}ndez-Soto}, {Mendes de Oliveira}, {Sims}, {Sodr{\'e}}, \&
  {Toerne}}]{Marin-Franch2012}
{Mar\'{\i}n-Franch}, A., {Taylor}, K., {Cepa}, J., {et~al.} 2012, in Society of
  Photo-Optical Instrumentation Engineers (SPIE) Conference Series, Vol. 8446,
  Ground-based and Airborne Instrumentation for Astronomy IV, ed. I.~S.
  {McLean}, S.~K. {Ramsay}, \& H.~{Takami}, 84466H

\bibitem[{{Mayor} \& {Queloz}(1995)}]{51Pegb}
{Mayor}, M. \& {Queloz}, D. 1995, \nat, 378, 355

\bibitem[{{Morate} {et~al.}(2021){Morate}, {Marcio Carvano}, {Alvarez-Candal},
  {De Pr{\'a}}, {Licandro}, {Galarza}, {Mahlke}, {Solano-M{\'a}rquez},
  {Cenarro}, {Crist{\'o}bal-Hornillos}, {Hern{\'a}ndez-Monteagudo},
  {L{\'o}pez-Sanjuan}, {Mar{\'\i}n-Franch}, {Moles}, {Varela}, {V{\'a}zquez
  Rami{\'o}}, {Alcaniz}, {Dupke}, {Ederoclite}, {Sodr{\'e}}, {Angulo},
  {Jim{\'e}nez-Esteban}, {Siffert}, \& {J-PLUS Collaboration}}]{morate2021}
{Morate}, D., {Marcio Carvano}, J., {Alvarez-Candal}, A., {et~al.} 2021, \aap,
  655, A47

\bibitem[{{Oke} \& {Gunn}(1983)}]{abmag_definition}
{Oke}, J.~B. \& {Gunn}, J.~E. 1983, \apj, 266, 713

\bibitem[{{Popescu} {et~al.}(2016){Popescu}, {Licandro}, {Morate}, {de
  Le{\'o}n}, {Nedelcu}, {Rebolo}, {McMahon}, {Gonzalez-Solares}, \&
  {Irwin}}]{Popescu+2016}
{Popescu}, M., {Licandro}, J., {Morate}, D., {et~al.} 2016, \aap, 591, A115

\bibitem[{{Ricker} {et~al.}(2015){Ricker}, {Winn}, {Vanderspek}, {Latham},
  {Bakos}, {Bean}, {Berta-Thompson}, {Brown}, {Buchhave}, {Butler}, {Butler},
  {Chaplin}, {Charbonneau}, {Christensen-Dalsgaard}, {Clampin}, {Deming},
  {Doty}, {De Lee}, {Dressing}, {Dunham}, {Endl}, {Fressin}, {Ge}, {Henning},
  {Holman}, {Howard}, {Ida}, {Jenkins}, {Jernigan}, {Johnson}, {Kaltenegger},
  {Kawai}, {Kjeldsen}, {Laughlin}, {Levine}, {Lin}, {Lissauer}, {MacQueen},
  {Marcy}, {McCullough}, {Morton}, {Narita}, {Paegert}, {Palle}, {Pepe},
  {Pepper}, {Quirrenbach}, {Rinehart}, {Sasselov}, {Sato}, {Seager},
  {Sozzetti}, {Stassun}, {Sullivan}, {Szentgyorgyi}, {Torres}, {Udry}, \&
  {Villasenor}}]{tess_ref}
{Ricker}, G.~R., {Winn}, J.~N., {Vanderspek}, R., {et~al.} 2015, Journal of
  Astronomical Telescopes, Instruments, and Systems, 1, 014003

\bibitem[{{Sesar} {et~al.}(2017){Sesar}, {Hernitschek}, {Mitrovi{\'c}},
  {Ivezi{\'c}}, {Rix}, {Cohen}, {Bernard}, {Grebel}, {Martin}, {Schlafly},
  {Burgett}, {Draper}, {Flewelling}, {Kaiser}, {Kudritzki}, {Magnier},
  {Metcalfe}, {Tonry}, \& {Waters}}]{Sesar2017}
{Sesar}, B., {Hernitschek}, N., {Mitrovi{\'c}}, S., {et~al.} 2017, \aj, 153,
  204

\bibitem[{{Shappee} {et~al.}(2014){Shappee}, {Prieto}, {Grupe}, {Kochanek},
  {Stanek}, {De Rosa}, {Mathur}, {Zu}, {Peterson}, {Pogge}, {Komossa}, {Im},
  {Jencson}, {Holoien}, {Basu}, {Beacom}, {Szczygie{\l}}, {Brimacombe},
  {Adams}, {Campillay}, {Choi}, {Contreras}, {Dietrich}, {Dubberley},
  {Elphick}, {Foale}, {Giustini}, {Gonzalez}, {Hawkins}, {Howell}, {Hsiao},
  {Koss}, {Leighly}, {Morrell}, {Mudd}, {Mullins}, {Nugent}, {Parrent},
  {Phillips}, {Pojmanski}, {Rosing}, {Ross}, {Sand}, {Terndrup}, {Valenti},
  {Walker}, \& {Yoon}}]{ASASSN_ref}
{Shappee}, B.~J., {Prieto}, J.~L., {Grupe}, D., {et~al.} 2014, \apj, 788, 48

\bibitem[{{Soszy{\'n}ski} {et~al.}(2011){Soszy{\'n}ski}, {Dziembowski},
  {Udalski}, {Poleski}, {Szyma{\'n}ski}, {Kubiak}, {Pietrzy{\'n}ski},
  {Wyrzykowski}, {Ulaczyk}, {Koz{\l}owski}, \& {Pietrukowicz}}]{Soszynski2011}
{Soszy{\'n}ski}, I., {Dziembowski}, W.~A., {Udalski}, A., {et~al.} 2011,
  \actaa, 61, 1

\bibitem[{{Soszy{\'n}ski} {et~al.}(2010){Soszy{\'n}ski}, {Udalski},
  {Szyma{\'n}ski}, {Kubiak}, {Pietrzy{\'n}ski}, {Wyrzykowski}, {Ulaczyk}, \&
  {Poleski}}]{Soszynski2010}
{Soszy{\'n}ski}, I., {Udalski}, A., {Szyma{\'n}ski}, M.~K., {et~al.} 2010,
  \actaa, 60, 165

\bibitem[{{Sykes} {et~al.}(2000){Sykes}, {Cutri}, {Fowler}, {Tholen},
  {Skrutskie}, {Price}, \& {Tedesco}}]{sso_2mass}
{Sykes}, M.~V., {Cutri}, R.~M., {Fowler}, J.~W., {et~al.} 2000, \icarus, 146,
  161

\bibitem[{{Tamuz} {et~al.}(2005){Tamuz}, {Mazeh}, \& {Zucker}}]{Tamuz+2005}
{Tamuz}, O., {Mazeh}, T., \& {Zucker}, S. 2005, \mnras, 356, 1466

\bibitem[{{Taylor}(2005)}]{topcat}
{Taylor}, M.~B. 2005, in Astronomical Society of the Pacific Conference Series,
  Vol. 347, Astronomical Data Analysis Software and Systems XIV, ed.
  P.~{Shopbell}, M.~{Britton}, \& R.~{Ebert}, 29

\bibitem[{{Tonry} {et~al.}(2018){Tonry}, {Denneau}, {Heinze}, {Stalder},
  {Smith}, {Smartt}, {Stubbs}, {Weiland}, \& {Rest}}]{atlas2018PASP}
{Tonry}, J.~L., {Denneau}, L., {Heinze}, A.~N., {et~al.} 2018, \pasp, 130,
  064505

\bibitem[{{Watson} {et~al.}(2022){Watson}, {Henden}, \&
  {Price}}]{VSX_Watson2022}
{Watson}, C., {Henden}, A.~A., \& {Price}, A. 2022, {VizieR Online Data
  Catalog: AAVSO International Variable Star Index VSX (Watson+, 2006-)},
  VizieR On-line Data Catalog: B/vsx. Originally published in:
  2006SASS...25...47W

\end{thebibliography}

\begin{appendix}

\section{Pointings in J-VAR DR1}

The coordinates of the pointings of J-PLUS\,DR1 are shown in Table\,\ref{tab:pointings}. These 
are the coordinates that the telescope used to point for the first exposure of each observing
sequence.

\longtab[1]{
\begin{longtable}{lcc|cc|c|cc}
\caption{\label{tab:pointings} 
List of pointings of J-VAR DR1. The first three columns contain the 
identifier of the field, the right ascension and the declination (in degrees). The fourth and fifth
column refer to the time of the first and last observation of the field in ``JVN'' (in modified 
Julian date). The sixth column
is ``True'' if a field has high frequency observations and ``False'' otherwise. If ``True'', the last
two columns show the time of the first and last observation of this sequence (in modified Julian date).
}\\
\hline
field &    ra &  dec & earliest MJD & latest MJD &  has HF & earliest MJD HF & latest MJD HF \\
\hline
\hline
\endfirsthead
\caption{continued}
\hline
field &    ra &  dec & earliest MJD & latest MJD &  has HF & earliest MJD HF & latest MJD HF \\
\hline
\endhead
\hline
\endfoot
\hline
\hline
\endlastfoot
JVAR00056 & 187.0911 & 20.6962 & 59621.042367 & 59722.940706 & True & 59717.875677 & 59717.973275 \\
JVAR00316 & 256.5094 & 23.481 & 58230.099057 & 58666.92478 & True & 60031.176817 & 60062.036111 \\
JVAR00317 & 258.0164 & 23.481 & 58230.129392 & 58749.835845 & True & 59783.879983 & 59783.989491 \\
JVAR00421 & 256.5934 & 24.873 & 58236.111892 & 58756.823368 & True & 59014.015631 & 59014.111672 \\
JVAR00540 & 119.8148 & 27.6581 & 58901.886291 & 59286.931817 & True & 59891.168177 & 59891.228559 \\
JVAR00596 & 207.1414 & 27.658 & 58143.139323 & 58497.196262 & True & 59599.209716 & 60053.125885 \\
JVAR00674 & 168.9654 & 29.05 & 58555.105307 & 58636.922118 & True & 59595.132459 & 59595.23022 \\
JVAR00739 & 112.2042 & 30.4428 & 58143.049844 & 58845.957986 & True & 60039.84684 & 60039.933617 \\
JVAR00746 & 123.4108 & 30.443 & 59165.156962 & 59249.990613 & False &  &  \\
JVAR00749 & 128.2137 & 30.443 & 59173.095434 & 59287.906134 & True & 60041.87272 & 60041.959531 \\
JVAR00752 & 133.0154 & 30.4428 & 58828.058027 & 58921.895787 & False &  &  \\
JVAR00753 & 134.6166 & 30.443 & 58828.08445 & 58898.876273 & True & 60044.872106 & 60044.958848 \\
JVAR00757 & 141.0197 & 30.4428 & 58494.058281 & 58772.191424 & True & 58865.123623 & 59944.187581 \\
JVAR00762 & 149.0241 & 30.4428 & 58851.03349 & 59262.941447 & True & 59290.888466 & 59291.023669 \\
JVAR00764 & 152.2257 & 30.4428 & 58846.194265 & 59630.087222 & True & 60023.924977 & 60024.011939 \\
JVAR00840 & 273.8913 & 30.443 & 58242.04195 & 58666.95309 & True & 59718.086314 & 59718.144126 \\
JVAR00841 & 275.4923 & 30.443 & 58242.072344 & 58725.966076 & False &  &  \\
JVAR00842 & 277.0928 & 30.4428 & 58242.101175 & 59366.047824 & True & 59013.054983 & 59013.126389 \\
JVAR00844 & 280.2944 & 30.443 & 58242.131615 & 58702.004595 & True & 59712.105955 & 59867.860949 \\
JVAR00846 & 283.4963 & 30.443 & 58362.878108 & 59366.077488 & False &  &  \\
JVAR00860 & 126.9228 & 31.8352 & 59611.893223 & 59722.902396 & True & 60045.857338 & 60045.949693 \\
JVAR00863 & 131.7944 & 31.835 & 58080.076453 & 58624.923773 & False &  &  \\
JVAR00864 & 133.4184 & 31.8352 & 58850.943513 & 59279.060648 & True & 60042.854398 & 60042.941152 \\
JVAR00867 & 138.2901 & 31.8352 & 59620.951696 & 59698.866829 & True & 60058.836285 & 60058.923073 \\
JVAR00877 & 154.5291 & 31.8352 & 58846.135226 & 59279.092847 & True & 59948.148119 & 59948.24581 \\
JVAR00909 & 206.4934 & 31.835 & 58550.174149 & 58608.918241 & True & 59717.976047 & 59718.074954 \\
JVAR00941 & 258.4574 & 31.835 & 58240.052888 & 58922.1689 & False &  &  \\
JVAR01066 & 283.8711 & 33.2276 & 59729.008397 & 59843.875579 & True & 59782.970017 & 59783.067396 \\
JVAR01276 & 279.5964 & 36.0123 & 59677.136765 & 59749.032245 & False &  &  \\
JVAR01280 & 105.8673 & 37.4047 & 59499.128524 & 59591.896678 & True & 59506.110318 & 59506.207998 \\
JVAR01477 & 270.1453 & 38.7971 & 59679.08849 & 59731.987593 & True & 60065.116447 & 60065.159149 \\
JVAR01486 & 105.901 & 40.1895 & 58208.858142 & 59165.065313 & True & 58894.797043 & 58894.919722 \\
JVAR01489 & 111.3042 & 40.1895 & 58081.986823 & 59205.105243 & True & 59563.912992 & 59564.016736 \\
JVAR01514 & 156.3319 & 40.1895 & 58210.936719 & 59200.197072 & True & 58895.041302 & 58895.209491 \\
JVAR01515 & 158.1334 & 40.19 & 58210.972899 & 58621.913657 & True & 60028.916563 & 60029.005064 \\
JVAR01516 & 159.9342 & 40.19 & 58211.001395 & 59975.087396 & True & 59949.234635 & 59951.264664 \\
JVAR01522 & 170.7408 & 40.1895 & 58851.122494 & 59015.944363 & True & 58857.121198 & 58857.254595 \\
JVAR01527 & 179.7463 & 40.1895 & 58494.09673 & 58636.952095 & True & 58530.020966 & 58564.203843 \\
JVAR01561 & 240.9837 & 40.1895 & 59616.177425 & 59719.933704 & False &  &  \\
JVAR01578 & 271.6026 & 40.1895 & 59679.118351 & 59878.793079 & False &  &  \\
JVAR01724 & 181.0813 & 42.9742 & 58494.133027 & 58621.972535 & True & 58531.015492 & 58531.161991 \\
JVAR01725 & 182.9599 & 42.9742 & 58851.150584 & 59204.196285 & True & 58859.962633 & 58860.113854 \\
JVAR01734 & 199.8667 & 42.9742 & 58494.163108 & 58603.930231 & True & 58532.013119 & 58532.218912 \\
JVAR01755 & 239.316 & 42.9742 & 58494.197969 & 58626.053727 & True & 58533.04305 & 58533.165486 \\
JVAR01756 & 241.1945 & 42.9742 & 58862.221684 & 59021.917083 & True & 59007.029948 & 59741.015729 \\
JVAR01777 & 280.6438 & 42.9742 & 59323.0982 & 59410.138356 & True & 59391.934057 & 59829.841557 \\
JVAR01778 & 282.5227 & 42.9742 & 59024.96695 & 59407.953056 & True & 59454.929647 & 59454.954954 \\
JVAR01871 & 280.8233 & 44.3666 & 59369.080862 & 59419.91015 & True & 59500.820399 & 59745.025885 \\
JVAR01872 & 282.7449 & 44.3666 & 59025.919554 & 59375.016782 & True & 59392.03548 & 59393.038189 \\
JVAR01873 & 284.6663 & 44.3666 & 59396.104439 & 59435.910104 & True & 59505.842946 & 59836.960671 \\
JVAR02023 & 223.025 & 47.1514 & 58851.206811 & 58999.920648 & True & 58898.113929 & 58898.233137 \\
JVAR02024 & 225.0426 & 47.1514 & 58862.164057 & 59016.014722 & True & 59299.116117 & 59393.007199 \\
JVAR02140 & 284.1486 & 48.5437 & 59396.076175 & 59435.939329 & True & 59838.859693 & 59838.957269 \\
JVAR02320 & 133.2358 & 52.7209 & 58143.167841 & 59957.139132 & False &  &  \\
JVAR02322 & 137.7534 & 52.7209 & 58239.93489 & 59937.236933 & True & 59513.129902 & 59513.166916 \\
JVAR02355 & 212.2952 & 52.7209 & 57902.106117 & 59996.214722 & True & 59014.936638 & 59714.996065 \\
JVAR02356 & 214.5538 & 52.7209 & 57904.012575 & 59996.125741 & True & 60040.028449 & 60040.116198 \\
JVAR02357 & 216.8127 & 52.7209 & 57906.008443 & 58324.933866 & True & 58893.168247 & 59013.050694 \\
JVAR02359 & 221.3304 & 52.721 & 57942.986626 & 58615.94125 & True & 60033.070903 & 60033.160492 \\
JVAR02360 & 223.5894 & 52.721 & 57943.019242 & 58670.926632 & False &  &  \\
JVAR02402 & 138.8162 & 54.1133 & 58148.110492 & 59165.124016 & True & 60015.986632 & 60016.092407 \\
JVAR02434 & 213.4442 & 54.1132 & 57900.016406 & 58622.032882 & True & 58887.096383 & 58887.193924 \\
JVAR02435 & 215.7763 & 54.113 & 57900.041499 & 58619.108843 & True & 59390.876742 & 59390.991123 \\
JVAR02436 & 218.1083 & 54.1131 & 57900.066534 & 58619.137523 & True & 59605.222841 & 59743.976597 \\
JVAR02437 & 220.4404 & 54.113 & 57900.092645 & 59022.989977 & False &  &  \\
JVAR02438 & 222.7724 & 54.1133 & 57900.117703 & 58637.084606 & True & 59301.019659 & 59301.117292 \\
JVAR02439 & 225.1047 & 54.1133 & 57902.07967 & 59285.082998 & True & 59301.121047 & 60044.129439 \\
JVAR02503 & 197.8546 & 55.5056 & 58851.178721 & 59285.017928 & True & 58860.117367 & 58885.142645 \\
JVAR02508 & 209.9134 & 55.506 & 57909.911186 & 58294.96794 & True & 58653.909635 & 59712.975307 \\
JVAR02509 & 212.3253 & 55.506 & 57909.940584 & 58294.998021 & True & 58650.077506 & 59713.98864 \\
JVAR02510 & 214.737 & 55.5059 & 57909.968374 & 58563.150764 & True & 60040.973322 & 60061.032876 \\
JVAR02511 & 217.1483 & 55.506 & 57901.970365 & 58622.061574 & True & 58653.98349 & 60045.141771 \\
JVAR02512 & 219.5603 & 55.506 & 57899.965203 & 58643.934826 & True & 58648.937598 & 58649.1264 \\
JVAR02513 & 221.9724 & 55.506 & 57899.990758 & 58658.920984 & True & 60018.152106 & 60018.208281 \\
JVAR02514 & 224.3844 & 55.506 & 57902.008212 & 59996.154942 & False &  &  \\
JVAR02515 & 226.7964 & 55.5056 & 57902.029832 & 59999.172477 & False &  &  \\
JVAR02520 & 238.8553 & 55.506 & 57944.99886 & 58722.964074 & True & 60041.065278 & 60041.152807 \\
JVAR02622 & 132.2325 & 58.2904 & 58093.187407 & 58772.135498 & True & 59599.108918 & 59599.206516 \\
JVAR02840 & 188.7493 & 62.468 & 58235.923594 & 58622.002789 & True & 58654.028686 & 58896.229792 \\
JVAR03384 & 3.4766 & -2.3038 & 58702.096348 & 59169.970197 & True & 59455.104855 & 59511.892095 \\
JVAR03420 & 10.4249 & -0.9114 & 58725.137147 & 59199.865255 & True & 59513.966649 & 59514.064178 \\
JVAR03479 & 6.2563 & 1.8733 & 59801.101869 & 59975.795475 & True & 59956.764323 & 59956.787407 \\
JVAR03481 & 9.0367 & 1.8733 & 59146.850145 & 59498.940475 & True & 59510.902008 & 59510.999792 \\
JVAR03488 & 18.7682 & 1.8733 & 58746.063756 & 59884.933854 & True & 58856.761997 & 59839.191944 \\
JVAR03489 & 20.1584 & 1.8733 & 59146.879369 & 59569.797998 & True & 59949.743015 & 59949.840532 \\
JVAR03490 & 21.5487 & 1.873 & 58746.01511 & 59884.90647 & False &  &  \\
JVAR03504 & 41.0114 & 1.8733 & 59821.082737 & 59956.822813 & True & 59909.982355 & 59948.981574 \\
JVAR03539 & 3.4834 & 4.658 & 58079.75901 & 59818.136863 & True & 59563.765041 & 59563.869051 \\
JVAR03557 & 28.5617 & 4.6581 & 59146.907506 & 59499.029977 & True & 59880.931591 & 59881.035104 \\
JVAR03566 & 41.101 & 4.6581 & 59467.165828 & 59884.046771 & True & 59501.061256 & 59513.022083 \\
JVAR03589 & 30.0145 & 6.0507 & 58750.074531 & 58859.950116 & True & 59597.767888 & 59597.86566 \\
JVAR03590 & 31.4104 & 6.0505 & 58764.105064 & 59174.985023 & True & 59512.037737 & 59512.159826 \\
JVAR03615 & 24.4934 & 7.4428 & 58079.850666 & 58781.08191 & True & 59183.917367 & 59183.993744 \\
JVAR03624 & 37.0898 & 7.4428 & 58781.113571 & 59611.801516 & True & 59593.804068 & 59867.200012 \\
JVAR03653 & 332.1144 & 10.228 & 58337.082216 & 58751.990266 & True & 59455.004392 & 59455.102025 \\
JVAR03660 & 341.9802 & 10.2278 & 58671.048929 & 58842.826134 & True & 59183.792251 & 59839.090856 \\
JVAR03661 & 343.3896 & 10.228 & 58671.077286 & 58841.819062 & True & 59510.78151 & 59510.87934 \\
JVAR03705 & 343.4482 & 11.62 & 59163.771522 & 59909.78081 & True & 59512.817876 & 59512.915417 \\
JVAR04246 & 35.1077 & 28.328 & 59146.936858 & 59870.033356 & True & 59502.135642 & 59503.083854 \\
JVAR04376 & 34.6054 & 32.506 & 58079.925098 & 58769.099699 & True & 59882.014959 & 59882.11853 \\
JVAR04377 & 36.2407 & 32.5056 & 58769.103443 & 59977.804282 & True & 59596.816973 & 59596.916956 \\
\end{longtable}
}

\section{Tables of the Database}

This section contains the description of the tables in the database. Following J-PLUS, measurements in different filters are shown as arrays. The order of filters is GSDSS, RSDSS, ISDSS, J0395, J0515, J0660, J0861. The labels of the filters have been capitalised to indicate that they are a label.

\begin{table*}[h!]
\label{tab:objects}
\caption{Content of the {\tt Objects} table  of the database}
\centering
\begin{tabular}{|l|l|l|p{0.7\textwidth}|}
\hline
Column Name & Data Type & Unit & Description \\ \hline\hline
obj\_id & object & None & The J-PLUS object ID, with format \{REF\_TILE\}-\{NUMBER\}; used for internal cross-matching with OBJ\_ID in the Photometry Tables \\ \hline
ra & float64 & deg & The RA of the object in the J-PLUS reference tile \\ \hline
dec & float64 & deg & The DEC of the object in the J-PLUS reference tile \\ \hline
mean\_ra & float64 & deg & the mean RA position of all object detections in the J-VAR images \\ \hline
rms\_ra & float64 & deg & the RMS (actually NMAD) of MEAN\_RA \\ \hline
mean\_dec & float64 & deg & the mean RA position of all object detections in the J-VAR images \\ \hline
rms\_dec & float64 & deg & the RMS (actually NMAD) of MEAN\_DEC \\ \hline
mag\_auto & object & mag & array of values of the J-PLUS photometry MAG\_AUTO \\ \hline
mag\_auto\_err & object & None & array of values of the ERRORS of J-PLUS photometry MAG\_AUTO \\ \hline
mag\_aper\_6 & object & mag & array of values of the J-PLUS photometry MAG\_APER\_6 (6 arcsec aperture with aperture correction factor applied) \\ \hline
mag\_aper\_6\_err & object & None & array of values of the ERRORS of J-PLUS photometry MAG\_APER\_6 \\ \hline
mag\_aper\_3 & object & mag & array of values of the J-PLUS photometry MAG\_APER\_3 (3 arcsec aperture with aperture correction factor applied) \\ \hline
mag\_aper\_3\_err & object & None & array of values of the ERRORS of J-PLUS photometry MAG\_APER\_3 \\ \hline
flags & object & None & array of Sextractor flags of the J-PLUS photometry \\ \hline
mask\_flags & object & None & array of the flags obtained after applying masks on J-PLUS tiles during UPAD processing \\ \hline
single\_det & object & None & array indicating whether an object is detected on the J-PLUS SingleMode Sextractor catalogue (0=no, 1=yes) \\ \hline
cls & float32 & None & ClassStar assigned by Sextractor during photometry on the J-PLUS tile \\ \hline
mps & float32 & None & Probability of being a star using the morphological method to classify the object (López-Sanjuan et al. 2019b) \\ \hline
sgc & float32 & None & Probability of being a star by combining the available priors and classifications (López-Sanjuan et al. 2019b) \\ \hline
bannjos & object & None & Object classification via the BANNJOS algorithm (del Pino et al. in prep); classes are 'STAR','QSO','APS' (= Ambiguous Point Source) \\ \hline
fwhm & float32 & None & Full Width at Half Maximum of the object on the J-PLUS reference (rSDSS) tile \\ \hline
morph\_flag & int32 & None & Additional Flag (0 or 1) to indicate potential issues with an object's morphology. A value of "1" is obtained when CLS < 0.3 or MPS < 0.3 or FWHM > 3 and it might be worth checking the object out \\ \hline
snr & object & None & Signal-To-Noise ratio of the J-PLUS photometry \\ \hline
hpix11 & int32 & None & Healpix index ORDER 11 (NESTED schema) of object position \\ \hline
ebv & float32 & None & E(B-V) colour excess estimated from the Schlafly \& Finkbeiner (2011) recalibration of the Schlegel, Finkbeiner \& Davis (1998) dust map at infinity \\ \hline
ebv\_err & float32 & None & Error in the above E(B-V) value \\ \hline
vsx\_var & int32 & None & Flag to indicate whether or not the object is a known variable included in the VSX catalogue; 0=no, 1=yes \\ \hline
vsx\_type & object & None & The VSX variable type classification of known variables; value is 'N/A' for all other objects \\ \hline
vsx\_period & float64 & None & The period (if available, otherwise period=0) of known variables; value is also 0 for all other objects \\ \hline
gaia\_sid & int64 & None & The {\it Gaia} DR3 source\_id of the object; crossmatch radius is 1.1 arcsec, ID=0 if no match is found \\ \hline
gaia\_xmulti & int32 & None & Flag indicating the presence of multiple objects within the 1.1 arcsec crossmatch radius. Values are 0 if no, 1 if yes, -1 for unmatched objects. In case of multiple {\it Gaia} objects, the source\_id assigned is the one corresponding to the smallest angular separation from the J-VAR object \\ \hline
gaia\_prlx & float64 & None & {\it Gaia} DR3 object parallax in mas; 'nan' if no parallax is available \\ \hline
gaia\_prlx\_err & float64 & None & {\it Gaia} DR3 object parallax error in mas; 'nan' if no parallax is available \\ \hline
gaia\_gmag & float64 & None & The mean {\it Gaia} DR3 G magnitude (phot\_g\_mean\_mag in gaia\_source) \\ \hline
gaia\_var & int32 & None & Flag to indicate whether or not the object is classified as variable (phot\_variable\_flag in gaia\_source). Values are 0 if no (GAIA=NOT\_AVAILABLE), 1 if yes (GAIA=VARIABLE), -1 for unmatched objects \\ \hline
gaia\_class & object & None & The {\it Gaia} classification of the variable (best\_class\_name in vari\_classifier\_result). There are {\it Gaia} objects flagged as 'VARIABLE' but     without classification (in\_vari\_classification\_result == False in vari\_summary). These objects are assigned a 'NO\_CLASS' value, while unmatched objects are assigned an 'N/A' value \\ \hline
gaia\_score & float32 & None & The score of the {\it Gaia} classification algorithm (best\_class\_score in vari\_classifier\_result). 'VARIABLE' objects with 'NO\_CLASS' get a score=0, unmatched objects get a score=-1 \\ \hline
\end{tabular}
\end{table*}

\begin{table*}[h!]
\caption{Description of the columns of the {\tt Light Curves} Table  of the database}
\label{tab:lightcurves}
\centering
\begin{tabular}{|l|l|l|p{0.6\textwidth}|}
\hline
Column Name & Data Type & Unit & Description \\ \hline\hline
obj\_id & object & None & The J-PLUS object ID, with format {REF\_TILE}-{NUMBER} \\ \hline 
ra & float64 & deg & The RA of the object in the J-PLUS reference tile \\ \hline
dec & float64 & deg & The DEC of the object in the J-PLUS reference tile \\ \hline
hpix11 & int32 & None & Healpix index ORDER 11 (NESTED schema) of object position \\ \hline
field & int32 & None & Number of the J-VAR field \\ \hline
filter & object & None & Name of the filter \\ \hline
mag & object & mag & The array containing the magnitudes \\ \hline
mag\_err & object & None & The array containing the magnitude errors. \\ \hline
flags & object & None & array of Sextractor flags of the J-PLUS photometry; follows FORDER \\ \hline
fwhm & object & None & The array containing the FWHM value of the object in each image. \\ \hline
corr\_curves & object & None & The array containing the curves used to calibrate each star's light curve \\ \hline 
mjd & object & None & Corresponding MJD(UTC) value of the UTC timestamps. \\ \hline
\end{tabular}
\end{table*}

\begin{table*}[h!]
\caption{Description of the columns of the {\tt Time Stamps} table of the database}
\label{tab:timestamps}
\centering
\begin{tabular}{|l|l|l|p{0.6\textwidth}|}
\hline
Column Name & Data Type & Unit & Description \\ \hline\hline
field & int32 & None & Number of the J-VAR field \\ \hline
filter & object & None & Name of the filter \\ \hline
utc & object & None & List with the dates and times of each of the images used for each measure of the object in a a field and filter. \\ \hline
mjd & object & None & Corresponding MJD(UTC) value of the UTC timestamps. \\ \hline
imtype & object & None & a flag to keep track of the different images. \\ \hline
%
\end{tabular}
\end{table*}

\begin{table*}[h!]
\caption{Description of the columns of the {\tt Photometry} table  of the database}
\label{tab:photometry}
\centering
\begin{tabular}{|l|l|l|p{0.6\textwidth}|}
\hline
Column Name & Data Type & Unit & Description \\ \hline\hline
field & int32 & None & Number of the J-VAR field \\ \hline
filter & object & None & Name of the filter \\ \hline
obj\_id & object & None & The J-PLUS object ID, with format {REF\_TILE}-{NUMBER} \\ \hline 
cnum & int32 & None & The number of comparison stars used to correct the object's light curve \\ \hline
srad & float32 & None & Search radius to encounter the above comparison stars \\ \hline 
mag & object & mag & The array containing the magnitudes \\ \hline
mag\_err & object & None & The array containing the magnitude errors. \\ \hline
flags & object & None & The array containing the photometric flags. \\ \hline 
fwhm & object & None & The array containing the FWHM value of the object in each image. \\ \hline
corr\_curves & object & None & The array containing the curves used to calibrate each star's light curve \\ \hline 
\end{tabular}
\end{table*}

\begin{table}[h!]
\caption{Description of the columns of the {\tt SSOs\_DETECTIONS} table of the database.
}
\label{tab:ssos_detections}
\centering
\begin{tabular}{|l|l|l|l|}
\hline
Column Name & Data Type & Unit & Description \\ \hline\hline
source\_number & int32 & None & None \\ \hline
catalog\_number & int32 & None & None \\ \hline
ra & float64 & deg & None \\ \hline
dec & float64 & deg & None \\ \hline
epoch & float32 & None & None \\ \hline
mag & float32 & None & None \\ \hline
magerr & float32 & None & None \\ \hline
mag\_aper & float32 & None & None \\ \hline
mag\_aper\_err & float32 & None & None \\ \hline
flux\_aper & float32 & None & None \\ \hline
flux\_aper\_err & float32 & None & None \\ \hline
flux & float32 & None & None \\ \hline
fluxerr & float32 & None & None \\ \hline
pm & float32 & None & None \\ \hline
pmerr & float32 & None & None \\ \hline
pmra & float32 & None & None \\ \hline
pmra\_err & float32 & None & None \\ \hline
pmdec & float32 & None & None \\ \hline
pmdec\_err & float32 & None & None \\ \hline
mid\_exposure\_mjd & float32 & None & None \\ \hline
date\_obs & object & None & None \\ \hline
exptime & float32 & None & None \\ \hline
matched & bool & None & None \\ \hline
skybot\_number & float32 & None & None \\ \hline
skybot\_name & object & None & None \\ \hline
skybot\_class & object & None & None \\ \hline
skybot\_mag & float32 & None & None \\ \hline
skybot\_ra & float32 & None & None \\ \hline
skybot\_dec & float32 & None & None \\ \hline
skybot\_pmra & float32 & None & None \\ \hline
skybot\_pmdec & float32 & None & None \\ \hline
skybot\_deltara & float32 & None & None \\ \hline
skybot\_deltadec & float32 & None & None \\ \hline
object & object & None & None \\ \hline
filter & object & None & None \\ \hline
ra\_image & float32 & None & None \\ \hline
dec\_image & float32 & None & None \\ \hline
image\_filename & object & None & None \\ \hline
extension & float32 & None & None \\ \hline
xwin\_image & float32 & None & None \\ \hline
ywin\_image & float32 & None & None \\ \hline
awin\_image & float32 & None & None \\ \hline
errawin\_image & float32 & None & None \\ \hline
bwin\_image & float32 & None & None \\ \hline
errbwin\_image & float32 & None & None \\ \hline
thetawin\_image & float32 & None & None \\ \hline
errthetawin\_image & float32 & None & None \\ \hline
erra\_world & float32 & None & None \\ \hline
errb\_world & float32 & None & None \\ \hline
errtheta\_world & float32 & None & None \\ \hline
flags\_extraction & float32 & None & None \\ \hline
flags\_scamp & float32 & None & None \\ \hline
flags\_ima & float32 & None & None \\ \hline
flags\_ssos & float32 & None & None \\ \hline
appid\_object & object & None & None \\ \hline
mag\_cal & float32 & None & None \\ \hline
mag\_cal\_err & float32 & None & None \\ \hline
\end{tabular}
\end{table}

\begin{table}[h!]
\caption{Description of the columns of the{\tt SSO\_magnitudes} table of the database.}
\label{tab:ssos_magnitudes}
\centering
\begin{tabular}{|l|l|l|l|}
\hline
Column Name & Data Type & Unit & Description \\ \hline
ast\_id & float32 & None & ID of the SSO \\ \hline
ast\_name & object & None & MPC name of the SSO \\ \hline
mag\_j0395 & float32 & None & None \\ \hline
err\_j0395 & float32 & None & None \\ \hline
mjd\_j0395 & float32 & None & None \\ \hline
mag\_gsdss & float32 & None & None \\ \hline
err\_gsdss & float32 & None & None \\ \hline
mjd\_gsdss & float32 & None & None \\ \hline
mag\_j0515 & float32 & None & None \\ \hline
err\_j0515 & float32 & None & None \\ \hline
mjd\_j0515 & float32 & None & None \\ \hline
mag\_rsdss & float32 & None & None \\ \hline
err\_rsdss & float32 & None & None \\ \hline
mjd\_rsdss & float32 & None & None \\ \hline
mag\_j0660 & float32 & None & None \\ \hline
err\_j0660 & float32 & None & None \\ \hline
mjd\_j0660 & float32 & None & None \\ \hline
mag\_isdss & float32 & None & None \\ \hline
err\_isdss & float32 & None & None \\ \hline
mjd\_isdss & float32 & None & None \\ \hline
mag\_j0861 & float32 & None & None \\ \hline
err\_j0861 & float32 & None & None \\ \hline
mjd\_j0861 & float32 & None & None \\ \hline
n\_filters & int32 & None & None \\ \hline
\end{tabular}

\end{table}

\end{appendix}

\end{document}